\documentclass[a4paper,10pt]{article}
\usepackage[utf8]{inputenc}
\usepackage{color}
\usepackage{amssymb}
\usepackage{amsmath}
\usepackage{graphicx}
\usepackage{mathtools}
\usepackage{ragged2e}
\usepackage{hyperref}
\hypersetup{
    colorlinks=true,
    linkcolor=blue,
    filecolor=cyan,
    urlcolor=blue,
    citecolor=blue
}
\usepackage{cleveref}
\usepackage[margin=2.5cm]{geometry}
\usepackage{cite}
\usepackage{epsfig}
\usepackage{float}
\usepackage{cancel}
\usepackage{amsfonts}
\usepackage{enumitem}
\usepackage[font={footnotesize,normal}]{caption}
\usepackage{tikz}
\usetikzlibrary{shapes}

\begin{document}
 
\begin{flushright}
APCTP Pre2022 - 023
\end{flushright}

\begin{center}

\Large {\bf  Holographic confining/deconfining gauge theories and 
entanglement measures with a magnetic field}

\vspace{8mm}

\renewcommand\thefootnote{\mbox{$\fnsymbol{footnote}$}}
Parul Jain,${}^{1}$\footnote{parul.jain@apctp.org}
Siddhi Swarupa Jena,${}^{2}$\footnote{519ph2015@nitrkl.ac.in} Subhash Mahapatra${}^2$\footnote{mahapatrasub@nitrkl.ac.in}

\vspace{4mm}

${}^1${\small \sl Asia Pacific Center for Theoretical Physics} \\
{\small \sl Pohang 37673, Republic of Korea}
\vskip 0.2cm

${}^2${\small \sl Department of Physics and Astronomy} \\
{\small \sl National Institute of Technology Rourkela} \\
{\small \sl Rourkela - 769008, India}
\vskip 0.2cm

\end{center}




\begin{abstract}
\noindent
\justify
We study various holographic pure and mixed-state entanglement measures in the confined/deconfined phases of a bottom-up AdS/QCD model in the presence of a background magnetic field. We analyze the entanglement entropy, entanglement wedge cross-section, mutual information, and entanglement negativity and investigate how a background magnetic field leaves its imprints on the entanglement structure of these measures. Due to the anisotropy introduced by the magnetic field, we find that the behavior of these measures depends nontrivially on the relative orientation of the strip with respect to the field. In the confining phase, the entanglement entropy and negativity undergo a phase transition at the same critical strip length, the magnitude of which increases/decreases for parallel/perpendicular orientation of the magnetic field. The entanglement wedge cross-section similarly displays discontinuous behavior each time a phase transition between different entangling surfaces occurs, while further exhibiting anisotropic features with a magnetic field. We further find that the magnetic field also introduces substantial changes in the entanglement measures of the deconfined phase; however, these changes remain qualitatively similar for all orientations of the magnetic field. We further study the inequality involving the entanglement wedge and mutual information and find that the former always exceeds half of the latter everywhere in the parameter space of the confined/deconfined phases.

\end{abstract}

\tableofcontents
\addtocounter{page}{1}
\pagebreak
\section{Introduction}
\label{sec1}

The gauge/gravity duality or holography is an elegant theoretical framework that provides an interesting connection between quantum field theory and gravity \cite{Maldacena:1997re,Gubser:1998bc,Witten:1998qj}. In its approximate form, the duality maps a classical theory of gravity in anti-de Sitter (AdS) to a strongly coupled quantum field theory living at the boundary of the AdS space in one lower dimension. The duality has been used to understand various aspects of strongly coupled field theories using classical gravitational tools, and by now there is plenty of evidence that numerous nonperturbative and novel aspects of strongly coupled field theories can be probed using this duality. In recent years, its applications have been found in various domains of physics ranging from condensed matter to black holes. Two of the most promising areas where the compelling ideas of the duality can be applied to obtain important physical results are quantum information and quantum chromodynamics (QCD). In this paper, following up on the seminal work that combined these two areas \cite{Klebanov:2007ws,Nishioka:2006gr}, we further examine how the concept of pure and mixed-state entanglement measures endows the QCD phase diagram in the presence of a crucial and anisotropic parameter: the magnetic field.

Quantum information science in recent years has emerged as a powerful tool to investigate diverse aspects in theoretical physics.  One of the key ingredients of quantum information is entanglement, which essentially means how different parts of the system are correlated. One of the most commonly used entanglement measures is entanglement entropy.  Aspects related to entanglement entropy have been used to study quantum phases \cite{Vidal:2002rm,Osborne:2002zz}, black hole entropy \cite{Bombelli:1986rw,Srednicki:1993im}, quantum communication \cite{Hoi,Karpov}, etc. Perhaps, one of the most striking developments appeared in the context of gauge/gravity duality, where a remarkably successful conjecture for the entanglement entropy was suggested \cite{Ryu:2006bv,Ryu:2006ef}. In this proposal, the entanglement entropy of the boundary theory is related to the area of a certain boundary homologous minimal surface. The proposal geometrizes the concept of entanglement entropy and therefore provides a unique stage in which spacetime geometry, quantum field theories, and quantum information measures can be combined in a single framework. Indeed, in recent years this proposal has been used to probe and investigate various physical problems, such as quantum error-correcting codes and tensor networks \cite{Pastawski:2015qua,Hayden:2016cfa}, large-$N$ phase transitions \cite{Johnson:2013dka,Dey:2015ytd,Dey:2014voa}, quantum gravity \cite{VanRaamsdonk:2010pw,Balasubramanian:2013lsa}, confinement and deconfinement transitions \cite{Klebanov:2007ws,Nishioka:2006gr},  quench dynamics \cite{Balasubramanian:2011ur,Liu:2013iza,Dey:2015poa}, etc.

The entanglement entropy, however, apart from containing UV divergences, is not a good measure of entanglement for the mixed and multipartite states. For such states, various entanglement measures, such as entanglement of formation, (logarithmic) entanglement negativity,  entanglement of purification, etc., have been proposed in the quantum information literature \cite{Vidal:2002zz,Terhal,Horodecki1996,Horodecki,Peres:1996dw,Eisert}. These quantities generally are extremely hard to compute in strongly coupled field theories and only a handful of systems are known where these can be computed explicitly. From the gauge/gravity duality point of view, a few suggestions for these measures have appeared. This includes the entanglement of purification suggestion of \cite{Takayanagi:2017knl,Nguyen:2017yqw}, where the purification was suggested to be dual to the minimal cross-section area of the entanglement wedge $E_W$. Similarly, there have been two separate suggestions for the entanglement negativity. In the first suggestion, the negativity is given by the area of an extremal cosmic brane that is suspended on the boundary of the entanglement wedge  \cite{Kudler-Flam:2018qjo,Kusuki:2019zsp}, whereas in the second suggestion, it is given by certain combinations of the minimal areas of codimension-two  surfaces \cite{Chaturvedi:2016rft,Chaturvedi:2016rcn,Jain:2017aqk,Jain:2017xsu,Jain:2017uhe,Jain:2018bai,Malvimat:2018txq,Malvimat:2018izs,Basak:2020bot,Malvimat:2018cfe}. Interestingly, these holographic quantities, like the entanglement entropy, are again given by the areas of certain bulk surfaces; however, unlike the entanglement entropy, they do not contain UV divergences and are finite by construction.

Let us also mention that $E_W$ has appeared in the holographic proposal of many information-theoretic quantities. This includes the above-mentioned entanglement of purification proposal \cite{Takayanagi:2017knl,Nguyen:2017yqw}, the reflected entropy proposal \cite{Dutta:2019gen}, and the odd entropy proposal \cite{Tamaoka:2018ned}. It also closely appears in the entanglement negativity proposal of \cite{Kudler-Flam:2018qjo,Kusuki:2019zsp}. Moreover, these different proposals of $E_W$ do not always coincide with each other, leading to uncertainty regarding its correct holographic interpretation \cite{Akers:2019gcv}. Therefore, it appears that more caution is required when associating an information-theoretic measure with $E_W$.  In spite of the correct interpretational issues of $E_W$, a great deal of progress has been made in exploring and understanding its properties in various physical situations; see \cite{Jain:2020rbb,Caputa:2018xuf,Umemoto:2018jpc,Bao:2018gck,Espindola:2018ozt,Jeong:2019xdr,Jokela:2019ebz,Ghodrati:2022hbb,Bhattacharyya:2019tsi,Camargo:2021aiq,
Banuls:2022iwk,Asadi:2022mvo,Ali-Akbari:2021zsm,Vasli:2022kfu,Liu:2021rks,Liu:2020blk,Saha:2021kwq,Chowdhury:2021idy} for more details. In this work, we also take this viewpoint and investigate the properties of $E_W$ in QCD-like holographic confined/deconfined phases in the presence of a background magnetic field, to probe its orientation- and anisotropic-dependent properties, and to see whether it provides any novel signature for confinement, without dwelling on its interpretational issues.

On the other hand, QCD is a well-tested quantum field theory of strong interactions capable of describing the subatomic physics of quarks and gluons. At low temperature and chemical potential the hadrons are bound together in a confined phase, whereas at high temperature and chemical potential these hadrons are librated and undergo a phase transition to a deconfined quark-gluon plasma (QGP) phase. Probing QCD properties in the parameter space of temperature, chemical potential, etc. is a nontrivial task and is of great importance. Unfortunately, this remains challenging in a large part of the QCD parameter space. Analytical approaches are difficult because of the strong coupling, whereas numerical-based approaches of lattice QCD are inherently Euclidean in nature. Therefore, the sparse availability of nonperturbative techniques and the failure of traditional perturbative methods have limited our understanding of QCD at strong coupling. Here, the idea of holographic duality again comes in handy and provides an elegant framework within which the strongly coupled region of QCD can be probed.  Indeed, one of the main and original motivations of holography was to better understand gauge theories such as QCD at strong coupling. In particular, building a dual gravity model capable of describing real QCD features reasonably well and from which testable predictions and aspects can be obtained is importance, to both complement and support other takes on the same problem, coming from, e.g., Dyson-Schwinger or functional renormalization group equations,  lattice QCD, effective QCD models, etc. By now, investigation using the holographic QCD framework have been done for both string theory inspired top-down and phenomenological bottom-up models, and many QCD-like properties have been reproduced, let us refer to \cite{Casalderrey-Solana:2011dxg,Gubser:2009md,Jarvinen:2021jbd,Gursoy:2010fj} for detailed reviews.

Recently,  there have been further suggestions that another parameter might play an important role in the QCD phase structure. In particular, there are suggestions that a very strong magnetic field, of the order of $eB\sim0.3~GeV^2$, might be generated in noncentral relativistic heavy-ion collisions and can leave important imprints on QCD properties \cite{Skokov:2009qp,Bzdak:2011yy,Voronyuk:2011jd,Deng:2012pc,DElia:2010abb,DElia:2021tfb,Tuchin:2013ie}. Though the produced large magnetic field decays fast after the collision, it remains sufficiently high near the deconfinement temperature and is therefore expected to modify QCD properties \cite{Tuchin:2013apa,McLerran:2013hla}. Indeed, the produced magnetic field has been shown to not only play a destructive role in the chiral and  deconfinement transition temperatures (also known as inverse magnetic catalysis) \cite{Bali:2011qj,Bali:2012zg,Ilgenfritz:2013ara,Bruckmann:2013oba,Fukushima:2012kc,Ferreira:2014kpa,Mueller:2015fka,Bali:2013esa,Fraga:2012fs,Ayala:2014iba,Ayala:2014gwa,Fraga:2012ev}, but also cause suppression/enhancement of the string tension in a direction parallel/transverse to the magnetic field \cite{Bonati:2014ksa,Bonati:2016kxj,DElia:2021tfb}. Similarly, it was also suggested that it can influence the charge dynamics in QCD, thereby yielding anomaly-induced novel transport phenomena such as the chiral magnetic effect\cite{Fukushima:2008xe,Kharzeev:2007jp,Kharzeev:2015znc}. In the context of gauge/gravity duality as well, a lot of work has been done to construct holographic models to mimic magnetised QCD as closely as possible.  For a related discussion on the interplay between the magnetic field and QCD observables in holography, see \cite{Johnson:2008vna,Callebaut:2011ab,Callebaut:2013ria,Dudal:2015wfn,Dudal:2014jfa,Dudal:2018rki,Gursoy:2017wzz,Jokela:2013qya,Gursoy:2016ofp,Li:2016gfn,Critelli:2016cvq,Giataganas:2017koz,Gursoy:2020kjd,Gursoy:2018ydr,
Rodrigues:2017cha,Rodrigues:2018pep,Bohra:2019ebj,Bohra:2020qom,Dudal:2021jav,Arefeva:2022avn,Arefeva:2020vae,Arefeva:2018cli,Arefeva:2018hyo,Jena:2022nzw,Ballon-Bayona:2022uyy,Arefeva:2020bjk,Ballon-Bayona:2020xtf,He:2020fdi,Zhu:2019igg,
Braga:2020hhs,Mamo:2015dea,Avila:2018sqf,Giataganas:2012zy}. Exploring QCD in the extreme external conditions of high temperature and magnetic field is not only a concern of theoretically challenging exercises, but of direct possible relevance for current particle accelerator driven research programs \cite{STAR:2021mii},  as well as the study of dense neutron stars \cite{Duncan:1992hi}, early Universe physics \cite{Grasso:2000wj}, gravitational-wave physics \cite{Ecker:2019xrw}, etc; let us refer to the review works \cite{Kharzeev:2012ph,Miransky:2015ava}.

Thus, it is clear that the magnetic field appears as an influential parameter in QCD-related physics. Therefore, it is important to investigate how this magnetic field influences information-theoretic measures in QCD phases and, in particular, whether it introduces any anisotropic features in the entanglement structure of QCD phases.

Unfortunately, getting any reliable information on the entanglement measures in interacting field theories is rather difficult. This is primarily due to severe technical difficulties presented both in analytical as well as in numerical calculations. For these reasons, the study of entanglement measures in QCD-like theories is quite limited. With the exception of a few lattice-related works \cite{Buividovich:2008kq,Buividovich:2008gq,Itou:2015cyu,Rabenstein:2018bri}, most studies have been based on holographic proposals. Moreover, these studies were mainly restricted to entanglement entropy.  In \cite{Klebanov:2007ws,Nishioka:2006gr}, the authors first studied the holographic entanglement entropy in the top-down confining phases and observed a phase transition from a connected to a disconnected minimal surface as the size of the subsystem varied. This phase transition was accompanied by a change in the order of the entanglement entropy reflecting (de)confinement. Similar nonanalytic behavior of the entanglement entropy was later observed in lattice-related studies \cite{Buividovich:2008kq,Buividovich:2008gq}. This idea was then tested in many other confining models, both top-down and bottom-up, and similar results were found \cite{Dudal:2016joz,Dudal:2018ztm,Mahapatra:2019uql,Kola1403,Ben-Ami:2014gsa,Fujita0806,Lewkowycz,Georgiou:2015pia,Kim,Ghodrati,Ali-Akbari:2017vtb,Knaute:2017lll,Anber:2018ohz,Arefeva:2020uec,Slepov:2019guc,
Liu:2019npm,Fujita:2020qvp,Fu:2020oep,Jokela:2020wgs,DiNunno:2021eyf,Ghodrati:2021ozc,Yadav:2021hmy}.

The discussion of mixed-state entanglement measures in QCD-like theories is relatively new. A short discussion appeared in \cite{Jokela:2019ebz}, where $E_W$ in a limited confining model was discussed. A thorough discussion of $E_W$ in various top-down and bottom-up confining models was later presented in \cite{Jain:2020rbb}; see also \cite{Ghodrati:2022hbb}. However, the negativity calculation only appeared in \cite{Jain:2020rbb}, and that too was restricted to the confined phase.

Until now, most studies related to probing confinement/deconfinement physics using the pure and mixed-state entanglement measures have been performed in the absence of background electromagnetic fields, in particular, magnetic fields. However, as mentioned before, the magnetic field does play an important role in QCD-related physics and, therefore, can influence the entanglement structure of QCD phases. Indeed, in the presence of a magnetic field, there are several possibilities to align the entangling surfaces. For instance, we can align them parallel or perpendicular to the magnetic field. We can hence certainly expect to find anisotropic signatures in the entanglement measures. This is interesting considering that the standard order parameter, i.e., the Polyakov loop, does depend on the magnitude of the magnetic field, but is insensitive to its direction. As such, it is again clear that further probing confinement/deconfinement physics in an anisotropic setting is important, both from theoretical
as well as phenomenological perspectives \cite{Bali:2011qj,Bruckmann:2013oba}. For the record, let us mention that the effect of a magnetic field on the entanglement entropy in the soft-wall AdS/QCD model was discussed in \cite{Dudal:2016joz}, while no such study has been performed for the entanglement wedge, negativity, and mutual information. For discussions related to the anisotropic entanglement entropy in different contexts, see \cite{Arefeva:2020uec,Chu:2019uoh,Cartwright:2021hpv}.

In this work, we aim to fill this gap and perform a comprehensive investigation of mixed-state entanglement measures, including both $E_W$ and negativity, in the confined and finite-temperature deconfined QCD phases, in the presence of a background magnetic field. For this purpose, we consider the dynamical bottom-up holographic QCD model of \cite{Bohra:2019ebj,Bohra:2020qom}, where a closed-form analytic solution of the Einstein-Maxwell-dilaton gravity system in the presence of a background magnetic field was obtained, thereby greatly simplifying the relevant numerical calculations, and it was shown to exhibit many desirable anisotropic QCD features. We briefly highlight this holographic model and its properties in the next section. For the entanglement entropy, we consider a strip subsystem of length $\ell$ in a direction either parallel or perpendicular to the magnetic field. In both cases, the entanglement entropy undergoes a phase transition from a connected surface to a disconnected surface at some critical strip length $\ell_{crit}$ in the confined phase. Interestingly, the magnitude of this critical strip length increases/decreases for a parallel/perpendicular magnetic field. This provides an important magnetic field induced signature of anisotropy in the entanglement structure. With two equal-size disjoint strips, separated by a distance $x$, four different types of minimal area surfaces $\{S_A,S_B,S_C,S_D\}$ appear, leading to an interesting phase diagram. This two-strip phase diagram is again greatly modified in the presence of a magnetic field, while further exhibiting anisotropic features. The mutual information turns out to be nonzero only in the $S_B$ and $S_C$ phases and is always a monotonic function of $\ell$ and $x$. Similarly, the entanglement wedge cross-section $E_W$ is also nonzero only in the $S_B$ and $S_C$ phases. Interestingly, unlike the mutual information, $E_W$ goes to zero discontinuously for large values of $x$ and $\ell$ and exhibits a nonanalytic behavior while going from the $S_B$ to $S_C$ phase. In particular, going from the $S_B$ to $S_C$ phase, the entanglement wedge cross-section increases at the $S_B/S_C$ transition line. Interestingly, this increment in the area of the entanglement wedge at the $S_B/S_C$ transition line decreases/increases for a parallel/perpendicular magnetic field, yielding a new anisotropic feature in the entanglement structure. We further find that $E_W$ always exceeds half of the mutual information, i.e., the holographically suggested inequality \cite{Takayanagi:2017knl} is always satisfied for both parallel and perpendicular cases. Similarly, the entanglement negativity exhibits many interesting features in the confined phase. For a single-strip subsystem, the negativity turns out to be just $3/2$ times the entanglement entropy. This suggests that the entanglement negativity also undergoes an order change, from $\mathcal{O}(N^2)$ to $\mathcal{O}(N^0)$, at $\ell_{crit}$, and that the magnitude of $\ell_{crit}$ increases/decreases with a parallel/perpendicular magnetic field. Moreover, for two strips, the negativity behaves smoothly across various phase transition lines and there is no discontinuity in its structure. However, unlike the mutual information and entanglement wedge, the negativity can be nonzero in some parts of the $S_A$ phase. The negativity further displays anisotropic features in parallel and perpendicular directions.

The entanglement structure of the deconfined phase is slightly simpler compared to the confined phase. In particular, there is no connected/disconnected transition and the entanglement entropy is now always given by the connected surface. This implies that it is always of order $\mathcal{O}(N^2)$. Accordingly, with two strips, there are only $S_A$ and $S_B$ phases, and the mutual information and entanglement wedge are nonzero only in the $S_B$ phase, whereas the entanglement negativity is nonzero in both the $S_A$ and $S_B$ phases. The mutual information vanishes continuously in the $S_A$ phase, whereas the entanglement wedge vanishes discontinuously. Moreover, the parameter space of the $S_B$ phase is found to increase for both orientations of the magnetic field, suggesting a larger phase space for the nontrivial entanglement wedge in the presence of a magnetic field. Although the magnetic field does introduce substantial changes in the entanglement measures, these changes remain qualitatively the same in both parallel and perpendicular cases, suggesting a limited anisotropic effect of the magnetic field in the deconfined phase.

Before performing explicit calculations, let us also mention that here we model the magnetic field as a constant external field to get first insights into the entanglement structure of QCD phases. This simplistic assumption can be justified for two reasons: (i) it has been suggested that after a fast initial decrease, the generated $B$ is almost frozen for the rest of the lifetime of the plasma, giving more credit to the assumption of a constant $B$ field, and (ii) from a technical point of view, it allows us to have better control over most of the calculations and is therefore quite common in holographic magnetized QCD model building.

The paper is organized as follows. We give an introduction to the bulk gravitational theory in Sec. \ref{sec2}, and briefly talk about the various entanglement measures that we consider for our calculations in Sec. \ref{sec3}. We study the various entanglement measures in the presence of a background magnetic field (both parallel and perpendicular orientations) in the confining phase in Sec. \ref{TAdS} and in the deconfining phase in Sec. \ref{adsblackholephase}. Finally, we end the paper with discussions and conclusions in Sec. \ref{conclusion}.

\section{Einstein-Maxwell-dilaton gravity with a magnetic field}
\label{sec2}
In this section, we describe the relevant details of the magnetised holographic QCD model presented in \cite{Bohra:2019ebj}. The corresponding five-dimensional Einstein-Maxwell-dilaton gravitational action is given by
\begin{eqnarray}
S_{EM} =  -\frac{1}{16 \pi G_{(5)}} \int_{\mathcal{M}} \mathrm{d^5}x \sqrt{-g}  \ \left[R - \frac{f(\phi)}{4}F_{MN}F^{MN} -\frac{1}{2}\partial_{M}\phi \partial^{M}\phi -V(\phi)\right]\,,
\label{actionEF}
\end{eqnarray}
wherein $R$ is the Ricci scalar of the five-dimensional manifold $\mathcal{M}$, $F_{MN}$ is the field-strength tensor for the $U(1)$ gauge field $A_M$ through which a constant background magnetic field will be introduced, $\phi$ represents the dilaton field, and $f(\phi)$ is the gauge kinetic function which denotes the coupling between the $U(1)$ and dilaton fields. The potential for the dilaton field is given by $V(\phi)$ and $G_{(5)}$ is the five-dimensional Newton's constant. Interestingly, with the following Ans\"atze for the metric $g_{MN}$, field-strength tensor $F_{MN}$, and dilaton field $\phi$,
\begin{eqnarray}
& & ds^2=\frac{L^2 e^{2A(z)}}{z^2}\biggl[-g(z)dt^2 + \frac{dz^2}{g(z)} + dy_{1}^2+ e^{B^2 z^2} \biggl( dy_{2}^2 + dy_{3}^2 \biggr) \biggr]\,, \nonumber \\
& & \phi=\phi(z), ~~  F_{MN}=B dy_{2}\wedge dy_{3}\,,
\label{ansatz}
\end{eqnarray}
the Einstein, Maxwell, and dilaton field equations coming from the action (\ref{actionEF}) can be completely solved in closed form in terms of a single parameter $a$,
\begin{eqnarray}
A(z)&=& -a z^2 \,,
\label{asol} \\
g(z) &=& 1-\frac{e^{z^2 \left(3 a-B^2\right)} \left(3 a z^2-B^2 z^2-1\right)+1}{e^{z_h^2 \left(3
   a-B^2\right)} \left(3 a z_h^2-B^2 z_h^2-1\right)+1} \,,
\label{gsol} \\
\phi(z) &=& \int \, dz \sqrt{-\frac{2}{z} \left(3 z A''(z)-3 z A'(z)^2+6 A'(z)+2 B^4 z^3+2 B^2 z\right)} + K_5 \,,
\label{phisol} \\
f(z) &=& g(z)e^{2 A(z)+2 B^2 z^2} \left(-\frac{6 A'(z)}{z}-4 B^2+\frac{4}{z^2}\right)-\frac{2 e^{2 A(z)+2 B^2 z^2} g'(z)}{z} \,,
\label{fsol} \\
V(z) &=& g'(z) \left(-3 z^2 A'(z)-B^2 z^3+3 z\right) e^{-2 A(z)} - g(z)\left(12 + 9 B^2 z^3 A'(z) \right) e^{-2 A(z)} \nonumber \\
& & +g(z) \left(-9 z^2 A'(z)^2-3 z^2 A''(z)+18 z
   A'(z)-2 B^4 z^4+8 B^2 z^2\right)e^{-2 A(z)} \,,
\label{Vsol}
\end{eqnarray}
wherein the AdS radius $L$ has been set to one and $z$ is the usual holographic radial coordinate. The above solution is obtained by using the boundary condition $g(z=z_h)=0$, corresponding to a black hole with a horizon at $z=z_h$. The magnetised black hole solution has the temperature and entropy
\begin{eqnarray}
& & T = \frac{z_{h}^{3} e^{-3A(z_h)-B^2 z_{h}^{2}}}{4 \pi \int_0^{z_h} \, d\xi \ \xi^3 e^{-B^2 \xi^2 -3A(\xi) } } \,,   \nonumber \\
& & S_{BH} = \frac{V_3 e^{3 A(z_h)+B^2 z_{h}^{2}}}{4 G_{(5)} z_{h}^3 } \,,
\label{BHtemp}
\end{eqnarray}
where $V_3$ is the volume of the three-dimensional spatial plane.

There also exists another solution to the field equations, corresponding to the thermal-AdS solution (without a horizon). This no-black-hole solution corresponds to $g(z)=1$ and can be obtained by taking the limit $z_h\rightarrow \infty$ in the above equations. The coordinate $z$ therefore runs from $z=0$ (asymptotic boundary) to $z=z_h$ (for the black hole) or to $z=\infty$ (for thermal-AdS). Importantly, both the thermal-AdS and black hole solutions asymptote to AdS at the boundary $z=0$, but can have a nontrivial structure in the bulk. The constant $K_5$ appearing in Eq.~(\ref{phisol}) is fixed by demanding that $\phi |_{z=0}\rightarrow 0$ to get an asymptotically AdS spacetime. Note that in these solutions a constant background magnetic field $B$ is chosen in the $y_1$ direction, which breaks the $SO(3)$ invariance of the boundary spatial coordinates $\{y_1, y_2, y_3\}$.

Apart from its analytic simplicity, this holographic model also exhibits many desirable anisotropic QCD properties. A few salient features of this model are the following:

\begin{itemize}
\item A Hawking/Page-type phase transition appears between the thermal-AdS and black hole solutions. In particular, the black hole phase is favored at high temperatures, whereas the thermal-AdS phase is favored at low temperatures. Accordingly, there is a phase transition between these two solutions. However, since $B$ explicitly appears in the temperature expression, now the transition temperature is a $B$-dependent quantity. The behavior of the transition temperature as a function of $B$ for various values of $a$ is shown in Fig.~\ref{BvsTcritvsaphasetransition}.
\begin{figure}[h!]
\centering
\includegraphics[width=2.8in,height=2.3in]{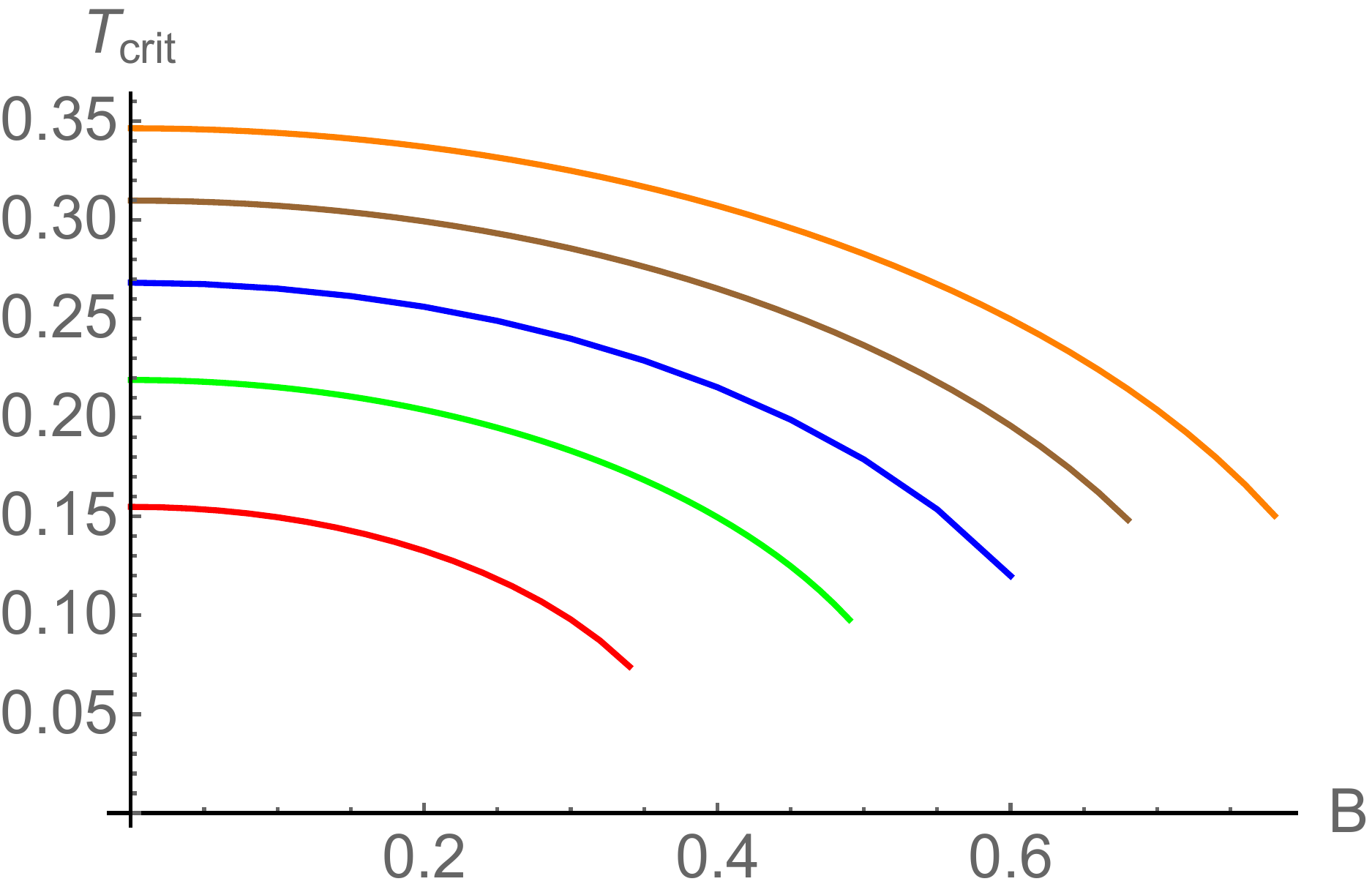}
\caption{ \small Deconfinement transition temperature in terms of magnetic field for various values of $a$. Red, green, blue, brown, and orange
curves correspond to $a=0.05$, $0.10$, $0.15$, $0.20$, and $0.25$, respectively. In units of GeV.}
\label{BvsTcritvsaphasetransition}
\end{figure}
\item These thermal-AdS and black hole phases were further shown to be dual to confined and deconfined phases, respectively, in the dual boundary theory. Since the transition temperature decreases with $B$, this provided a holographic model for inverse magnetic catalysis in the deconfinement sector \cite{Bohra:2019ebj}.

\item The parameter $a$ is the only free parameter in this model, and  Eqs.~(\ref{asol})-(\ref{Vsol}) form a self-consistent solution of the magnetised Einstein-Maxwell-dilaton gravity for any choice of $a$. Nonetheless, in the context of AdS/QCD model building, it is appropriate to fix its value by taking inputs from the dual boundary QCD theory. For instance, by demanding the confined/deconfined (or the dual Hawking/Page)  phase transition temperature in the pure glue sector to be around $270~MeV$, as is reported in lattice QCD \cite{Fromm:2011qi}, one fixes the value of the parameter $a$ to be $0.15~GeV^2$ \cite{Dudal:2017max}. This also fixes the largest attainable magnitude of $B$, by requiring the real-valuedness of the dilaton field, to be around $B\backsimeq 0.6~\text{GeV}$. However, it is important to note that the inverse magnetic behavior is a general result of this model that remains true for other values of $a$ as well, as is shown in Fig.~\ref{BvsTcritvsaphasetransition}.

\item Interestingly, the string tension was further found to decrease/increase with magnetic field in longitudinal/transverse directions. These results are in good agreement with state-of-the-art lattice findings \cite{Bonati:2014ksa,Bonati:2016kxj}.

\item Similarly, the chiral critical temperature again goes down with the magnetic field, indicating inverse magnetic catalysis behavior in the chiral sector. In particular, the chiral condensate magnitude increases with $B$ in the confined phase, whereas it exhibits nonmonotonic thermal features for all $B$ in the deconfined phase.  These chiral results also agree qualitatively well with lattice QCD findings, where similar features have been observed in the chiral sector.

\item The boundary vector-meson mass spectrum also exhibits linear Regge behavior.

\item As far as the stability of the model is concerned, the mass of the dilaton field $\phi$ satisfies the Breitenlohner-Freedman bound for stability in AdS space \cite{Breitenlohner:1982bm}, and the dilaton potential $V$ is bounded from above by its UV boundary value, thereby satisfying the Gubser stability criterion for a well-defined boundary theory \cite{Gubser:2000nd}. Similarly, the null energy condition of the matter field is always satisfied and constructed geometries -- both black hole and thermal-AdS spacetime -- asymptote to AdS at the boundary $z\rightarrow0$.
\end{itemize}
We therefore see that the dual boundary theory of the model (\ref{actionEF}) indeed exhibits many desirable anisotropic QCD features with a magnetic field. Therefore, it is reasonable to use this model to find the anisotropic imprints of a magnetic field on the entanglement structure of QCD phases by studying various entanglement measures.

\section{Entanglement measures}
\label{sec3}
In this section, we briefly talk about various entanglement measures that have gravity duals. To probe the entanglement structure of confined/deconfined QCD phases and make the discussion complete and as general as possible, we concentrate on both pure and mixed-state measures. This includes the (i) entanglement entropy, (ii) mutual information, (iii) entanglement wedge cross-section, and (iv) entanglement negativity.
\subsection{Holographic entanglement entropy}
We begin with the discussion of entanglement entropy. It is a good measure of entanglement for the pure states and in the usual quantum systems it is given by
\begin{equation}
S(A) = -\mathrm{Tr}_A\rho_A\mathrm{ln}\,\rho_A\ \,,
\label{vne}
\end{equation}
where $\rho_A$ is the reduced density matrix of subsystem $A$, obtained by tracing out the degrees of freedom of the rest of the system. In quantum field theories, one can use the replica trick to calculate the entanglement entropy \cite{Calabrese:2009qy}. Holographically, the entanglement entropy can be computed using the Ryu-Takayanagi prescription \cite{Ryu:2006bv,Ryu:2006ef},
\begin{equation}
S(A) = \frac{\mathcal{A}(\Gamma_A^{\text{min}})}{4G_{(d+1)}}\ ,
\label{hee}
\end{equation}
wherein $G_{(d+1)}$ denotes the $(d+1)$-dimensional Newton's constant and $\mathcal{A}(\Gamma_A^{\text{min}})$ represents the area of the $(d-1)$-dimensional minimal surface $\Gamma$ with the condition that the boundary $\partial A$ of the subsystem $A$ is homologous to $\partial\Gamma$. The above equation can also be written in the following way:
\begin{equation}
S(A) = \frac{1}{4G_{(d+1)}}\int_{\Gamma} d^{d-1}\sigma\sqrt{\mathcal{G}^{d-1}_{\mathrm{ind}}}\ ,
\label{heestring}
\end{equation}
wherein the induced metric on the surface $\Gamma$ is given by $\mathcal{G}^{d-1}_{\mathrm{ind}}$, which further needs to be minimized according to the prescription of \cite{Ryu:2006bv,Ryu:2006ef,Klebanov:2007ws,Nishioka:2006gr}. For the record, we have $d=4$ in our cases of interest.

Notice that, with a background magnetic field, we have choices to align the subsystem (or the entangling surface) with respect to the magnetic field. In particular, we can now have two interesting scenarios: (i) align the entangling surface parallel to the magnetic field, and (ii) align it perpendicular to the magnetic field. The relative orientation of the entangling surface can leave anisotropic imprints of the magnetic field on various entanglement measures. Indeed, as we will see shortly, since most of the holographic entanglement measures depend nontrivially on the bulk spacetime metric, which in turn depends nontrivially on the magnetic field, it is therefore reasonable to expect that the magnetic field might generate anisotropic features in the entanglement measures.

\subsection{Holographic mutual information}
We next move on to discuss the mutual information, which serves as a measure of entanglement for disjoint intervals. For two subsystems ($A_1$ and $A_2$), it reflects the amount of shared information between $A_1$ and $A_2$, and in the case of two disjoint intervals on the boundary it is given as \cite{Hayden:2011ag,Headrick:2010zt}
\begin{equation}\label{MIdef}
I(A_1,A_2)=S(A_1) + S(A_2) - S(A_1 \cup A_2) \ ,
\end{equation}
wherein $S(A_1)$, $S(A_2)$, and $S(A_1 \cup A_2)$ represent the entanglement entropies pertaining to $A_1, A_2$,  and $A_1\cup A_2$, respectively. From the above equation (\ref{MIdef}), we can see that the mutual information vanishes in the case of uncorrelated systems, whereas it is nonzero for correlated systems. Moreover, the subadditivity property of the entanglement entropy further implies that the mutual information is non-negative, which in turn signifies the fact that $I(A_1, A_2)$ serves as an upper bound on the correlation between $A_1$ and $A_2$. In the holographic context, the mutual information of the boundary system can be evaluated by computing the entanglement entropies $\{S(A_1), S(A_2), S(A_1 \cup A_2) \}$ individually from the Ryu-Takayanagi prescription. Interestingly, unlike the entanglement entropy, the holographic mutual information does not contain any UV divergences and is UV finite in nature. Therefore, it provides a cut-off or regularization-independent information.  For further information related to mutual information, see \cite{Fischler:2012uv,Kundu:2016dyk,Casini:2015woa,Balasubramanian:2018qqx,Cardy:2013nua,Larkoski:2014pca,Agon:2022efa}. For more on mutual information and two disjoint interval entanglement phase structure in top-down and bottom-up QCD models, see \cite{Mahapatra:2019uql,Ben-Ami:2014gsa}.

\subsection{Entanglement wedge cross-section}
It is well known that entanglement entropy serves as a good measure of entanglement in the case of pure states, but not so in the case of mixed states. Since entanglement entropy is known to exhibit interesting features in QCD phases, it is compelling to ask how the mixed-state measures behave in these phases. When dealing with mixed states, it turns out that the minimal area of the entanglement wedge cross-section can be considered an appropriate measure holographically \footnote{For more information on the entanglement wedge cross-section and its properties and application in various context, see \cite{Jain:2020rbb,Caputa:2018xuf,Umemoto:2018jpc,Bao:2018gck,Espindola:2018ozt,Jeong:2019xdr,Jokela:2019ebz,Ghodrati:2022hbb,Bhattacharyya:2019tsi,Camargo:2021aiq,
Banuls:2022iwk,Asadi:2022mvo,Ali-Akbari:2021zsm,Vasli:2022kfu,Liu:2021rks,Liu:2020blk,Saha:2021kwq,Chowdhury:2021idy}.}.

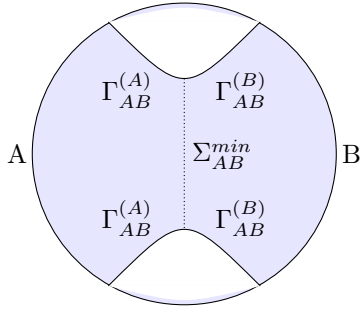
\begin{figure}[h]
\begin{center}
\begin{tikzpicture}
\filldraw[blue!10!, draw=black] (0,0) circle (2);
\draw[densely dotted] (0,1) -- (0,-1);
\filldraw[white!100!, draw=black] (-1,1.75) .. controls (0,0.75) .. (1,1.75);
\fill[white!100!] (-1,1.75) .. controls (0,2) .. (1,1.75);
\filldraw[white!100!, draw=black] (-1,-1.75) .. controls (0,-0.75) .. (1,-1.75);
\fill[white!100!] (-1,-1.75) .. controls (0,-2) .. (1,-1.75);
\node()at (-2.2,0){A};
\node()at (0.5,0){$\Sigma^{min}_{AB}$};
\node()at (-0.75,0.85){$\Gamma^{(A)}_{AB}$};
\node()at (-0.75,-0.85){$\Gamma^{(A)}_{AB}$};
\node()at (0.75,0.85){$\Gamma^{(B)}_{AB}$};
\node()at (0.75,-0.85){$\Gamma^{(B)}_{AB}$};
\node()at (2.2,0){B};
\end{tikzpicture}
\caption{The region in blue is the entanglement wedge $M_{AB}$ corresponding to a pure state. For a thermal state, there would additionally be a black hole in $M_{AB}$. The dotted surface is $\Sigma_{AB}$, which
divides $M_{AB}$ into two parts.}
\label{Entanglementwedgecrosspic}
\end{center}
\end{figure}

In order to calculate the entanglement wedge cross-section holographically, we follow the method suggested in \cite{Takayanagi:2017knl,Nguyen:2017yqw}. On the $d$-dimensional boundary, we consider two nonoverlapping subsystems $A$ and $B$. The minimal surfaces in the $(d+1)$-dimensional bulk corresponding to $A$, $B$, and $AB=A\cup B$ are given as $\Gamma^{min}_A$, $\Gamma^{min}_B$, and $\Gamma^{min}_{AB}$, respectively. The entanglement wedge $M_{AB}$, which is $d+1$-dimensional ($d$-dimensional, if the static case is considered), is then defined as a region in the bulk which shares its boundary with  $A$, $B$, and $\Gamma^{min}_{AB}$, implying
\begin{equation}
\partial M_{AB}=A\cup B\cup \Gamma^{min}_{AB}.
\label{EWboundary}
\end{equation}
It is important to see that if the size of subsystems $A$ and $B$ is very small or if they are too far apart, then the wedge $M_{AB}$ will be of disconnected nature. We can further
divide $\Gamma_{AB}^{min}$ as
\begin{equation}
\Gamma_{AB}^{min} = \Gamma_{AB}^{(A)}\cup \Gamma_{AB}^{(B)}\,,
\label{RTdiv}
\end{equation}
and define
\begin{eqnarray}
& & \tilde{\Gamma}_{A} = A\cup \Gamma_{AB}^{(A)}\,,  \nonumber \\
& & \tilde{\Gamma}_{B} = B\cup \Gamma_{AB}^{(B)}\ .
\label{RTA}
\end{eqnarray}
From the above Eqs.~\eqref{RTdiv} and \eqref{RTA},
we get the following condition for the wedge boundary $\partial M_{AB}$:
\begin{equation}
\partial M_{AB}=\tilde{\Gamma}_{A}\cup \tilde{\Gamma}_{B}.
\end{equation}
$\Sigma_{AB}^{min}$ is then defined as a minimum surface whose boundary conditions are
\begin{equation}
\begin{split}
&(i)\ \partial\Sigma_{AB}^{min}=\partial\tilde{\Gamma}_{A}=\partial\tilde{\Gamma}_{B}\ ,\\
&(ii)\ \Sigma_{AB}^{min}\ \mathrm{is\ homologous\ to}
\ \tilde{\Gamma}_{A}\ \mathrm{inside}\ M_{AB}.
\end{split}
\end{equation}
Using the area of $\Sigma_{AB}^{min}$, which is denoted by
$\mathcal{A}(\Sigma_{AB}^{min})$, one can now define the entanglement wedge cross-section as
\begin{equation}\label{EW}
E_{W}(\rho_{AB}) = \min_{\Gamma_{AB}^{(A)}\subset\Gamma_{AB}^{min}}\left[\frac{\mathcal{A}(\Sigma_{AB}^{min})}{4G_{(d+1)}}\right].
\end{equation}
To put it in words, $E_{W}(\rho_{AB})$ is given by the minimal area of the division of the entanglement wedge $M_{AB}$ which connects subsystems $A$ and $B$. A pictorial representation of the entanglement wedge in the connected space, i.e., in the thermal-AdS spacetime, is shown in Fig.~\ref{Entanglementwedgecrosspic}.

Let us stress here once again that in recent years several entanglement measures have been suggested to be holographically dual to the entanglement wedge cross-section. This includes the entanglement of purification \cite{Takayanagi:2017knl,Nguyen:2017yqw}, reflected entropy \cite{Dutta:2019gen}, and odd entropy \cite{Tamaoka:2018ned}. Unfortunately, these different interpretations do not exactly coincide with each other, leading to uncertainty regarding its correct holographic interpretation. In this work, we do not dwell on the boundary interpretation issues of the entanglement wedge cross-section and mainly concentrate on its properties in the confined/deconfined phases of QCD in the presence of a background magnetic field. Indeed, as we will shortly see, the entanglement wedge cross-section does provide valuable information as far as the entanglement structure in the confined phase is concerned.

\subsection{Holographic entanglement negativity}
Apart from the entanglement wedge cross-section, another quantity that can be taken as a suitable measure of mixed-state entanglement is entanglement negativity. In usual quantum systems this is defined as
\cite{Vidal:2002zz,Horodecki1996}
\begin{equation}\label{EN}
\mathcal{N} = \frac{\|\rho^{T_2}\|-1}{2}\ ,
\end{equation}
where $\rho^{T_2}$ denotes the partial transpose of the reduced density matrix and $\|\rho^{T_2}\|$ denotes its trace norm. One can further define its close cousin, the logarithmic negativity, as
\begin{equation}\label{logen}
\mathcal{E} = \ln \|\rho^{T_2}\| = \ln \mathrm{Tr}|\rho^{T_2}|.
\end{equation}
The logarithmic entanglement negativity serves as an upper bound to the amount of distillable entanglement and has been previously calculated in many-body systems and field theories \cite{Calabrese:2012ew,Calabrese:2012nk,Alba,Calabrese:2013mi,Ruggiero:2016aqr,Hoogeveen:2014bqa,Blondeau-Fournier:2015yoa,Castelnovo,Lee,Eisler,Wen:2015qwa,Wen:2016bla,
Rangamani:2014ywa}. In gauge/gravity duality, two seemingly different (yet equivalent) holographic proposals for the entanglement negativity are available in the market. This includes the proposal of \cite{Kudler-Flam:2018qjo,Kusuki:2019zsp}, in which the logarithmic negativity is given by the area of an extremal cosmic brane that terminates on the boundary of the entanglement wedge, and the proposal of \cite{Chaturvedi:2016rft,Chaturvedi:2016rcn,Jain:2017aqk,Jain:2017xsu,Jain:2017uhe,Jain:2018bai,Malvimat:2018txq,Malvimat:2018izs,Malvimat:2018cfe,Basak:2020bot}, in which the logarithmic negativity is given by certain combinations of the areas of codimension-two minimal bulk surfaces. Both proposals have seemingly different mathematical definitions; however, they both reproduce independent known results for the negativity in conformal field theories and have been tested in diverse physical situations.  In this work, we mainly deal with the latter proposal for two reasons: (i) the former proposal is practically similar to the computation of the entanglement wedge (which we will anyhow compute), and (ii) it is computationally slightly easier to compute the negativity from the latter proposal, as opposed to the former proposal, which requires nontrivial and cumbersome cosmic brane backreaction calculation. Therefore, it might not only be complementary but also more informative if the latter proposal is adopted for the entanglement
negativity calculation. Indeed, as we will see shortly, the latter proposal also provides an interesting and model-independent result for the negativity in all holographic confining/deconfining theories, which can be tested in independent lattice calculations, hence providing an intriguing platform for a nontrivial verification of the proposal.

In order to calculate the holographic logarithmic negativity in the case of a single interval, we follow \cite{Chaturvedi:2016rft,Chaturvedi:2016rcn} and consider a $d$-dimensional boundary system composed of $A$ and its compliment $A^c$. We now consider two additional finite intervals $B_1$ and $B_2$ adjacent to $A$, implying $B=B_1\cup B_2$; see the left part of Fig.~\ref{entanglenegativitypicture}. In terms of the entanglement entropy (Eq.~(\ref{hee})), the holographic logarithmic negativity is then suggested as
\begin{equation}
\label{HEEsc1}
\mathcal{E} = \lim_{B\rightarrow A^c}\frac{3}{4}\left[2S(A)+S(B_1)+S(B_2)-S(A\cup B_1)-S(A\cup B_2)\right]\,.
\end{equation}
It is important to note that in Eq.~(\ref{HEEsc1}) both $B_1$ and $B_2$ have to be taken to infinity so that $B=B_1\cup B_2=A^c$.

In the case of two disjoint intervals $A_1$ (of length $\ell_1$) and $A_2$ (of length $\ell_2$) separated by a distance $x$ (see the right panel of Fig.~\ref{entanglenegativitypicture}), the holographic logarithmic negativity is similarly suggested as
\begin{eqnarray}
\mathcal{E} = \frac{3}{4}\left[S(A_1 \cup A_x) + S(A_x \cup A_2) - S(A_1 \cup A_2 \cup A_x)-S(A_x) \right] \,.
\label{HEEsc1twostrip}
\end{eqnarray}
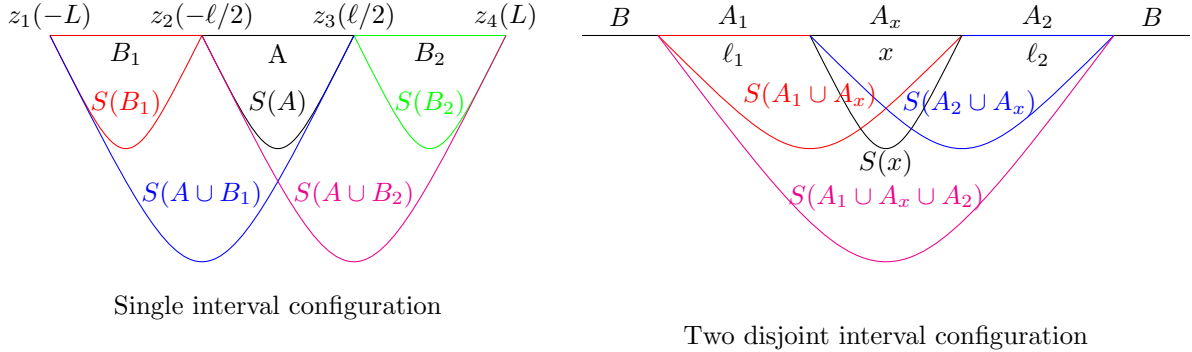
\begin{figure}[ht]
\begin{tikzpicture}
\draw[red] (-20,0) -- (-18,0);
\draw (-18,0) -- (-16,0);
\draw[green] (-16,0) -- (-14,0);
\draw[red] (-20,0) .. controls (-19,-2) .. (-18,0);
\draw (-18,0) .. controls (-17,-2) .. (-16,0);
\draw[green] (-16,0) .. controls (-15,-2) .. (-14,0);
\draw[blue] (-20,0) .. controls (-18,-4) .. (-16,0);
\draw[magenta] (-18,0) .. controls (-16,-4) .. (-14,0);
\node()at (-17,-0.25){A};
\node()at (-19,-0.25){$B_1$};
\node()at (-15,-0.25){$B_2$};
\node()at (-20,0.25){$z_1(-L)$};
\node()at (-18,0.25){$z_2(-\ell/2)$};
\node()at (-16,0.25){$z_3(\ell/2)$};
\node()at (-14,0.25){$z_4(L)$};
\node()at (-17,-0.9){$S(A)$};
\node[red]()at (-19,-0.9){$S(B_1)$};
\node[green]()at (-15,-0.9){$S(B_2)$};
\node[blue]()at (-18,-2.1){$S(A\cup B_1)$};
\node[magenta]()at (-16,-2.1){$S(A\cup B_2)$};
\node()at (-17,-3.6){Single interval configuration};
\draw (-13,0) -- (-12,0);
\draw[red] (-12,0) -- (-10,0);
\draw (-10,0) -- (-8,0);
\draw[blue] (-8,0) -- (-6,0);
\draw (-6,0) -- (-5,0);
\draw[red] (-12,0) .. controls (-10,-2) .. (-8,0);
\draw (-10,0) .. controls (-9,-2) .. (-8,0);
\draw[blue] (-10,0) .. controls (-8,-2) .. (-6,0);
\draw[magenta] (-12,0) .. controls (-9,-4) .. (-6,0);
\node()at (-9,0.25){$A_x$};
\node()at (-9,-0.25){$x$};
\node()at (-11,-0.25){$\ell_1$};
\node()at (-7,-0.25){$\ell_2$};
\node()at (-7,0.25){$A_2$};
\node()at (-12.5,0.25){$B$};
\node()at (-11,0.25){$A_1$};
\node()at (-5.5,0.25){$B$};
\node()at (-9,-1.7){$S(x)$};
\node[red]()at (-10,-0.8){$S(A_1\cup A_x)$};
\node[blue]()at (-7.9,-0.9){$S(A_2\cup A_x)$};
\node[magenta]()at (-9,-2.14){$S(A_1\cup A_x\cup A_2)$};
\node()at (-9,-4){Two disjoint interval configuration};
\end{tikzpicture}
\caption{Illustration of the various bulk minimal surfaces that contribute to the holographic logarithmic entanglement negativity.}
\label{entanglenegativitypicture}
\end{figure}

\section{Confining phase}
\label{TAdS}
In this section, we calculate the previously mentioned four entanglement measures in the confining phase, which is dual to the thermal AdS background, in the presence of a background magnetic field $B$. To compute these measures, we confine ourselves to the simplest situation where the entangling surface is a strip of length $\ell$. However, this entangling strip can be placed parallel or perpendicular to the magnetic field, giving us orientation dependence of these measures.

\subsection{Holographic entanglement entropy}
\subsubsection{Strip in the parallel direction}
We begin by looking at the holographic entanglement entropy for a single interval and consider the boundary subsystem with the domain $\{-\ell^{\parallel}/2\leq y_1 \leq \ell^{\parallel}/2$, $0\leq y_2 \leq \ell_{y_2}, 0\leq y_3 \leq \ell_{y_3} \}$. Here, the strip is placed parallel to the magnetic field in the $y_1$ direction. In the thermal AdS background, it turns out that there are two surfaces -- connected and disconnected -- that minimize the entanglement entropy expression in Eq.~(\ref{hee}). The expression of the entanglement entropy for the connected surface is found to be
\begin{eqnarray}
S^{\parallel}_{con}=\frac{\ell_{y_2} \ell_{y_3} L^3}{2 G_{(5)}} \int_{0}^{z_{*}^{\parallel}} dz \ \left(\frac{z_{*}^{\parallel}}{z} \right)^3 \frac{{e^{3 A(z)-3 A(z_{*}^{\parallel})}}{{e^{B^2z^2-B^2(z_{*}^{\parallel})^2}}}}{\sqrt{g(z)[(z_{*}^{\parallel})^6 e^{-2 B^2 (z_{*}^{\parallel})^2}e^{-6A(z_{*}^{\parallel})}-z^6 e^{-2 B^2 z^2} e^{-6A(z)}]}}\,,
\label{SEEcon}
\end{eqnarray}
where $z_{*}^{\parallel}$ is the turning point of the connected surface in the bulk and is defined by $z'(y_1)|_{z=z_{*}^{\parallel}}=0$. The strip length $\ell^{\parallel}$ in terms of $z_{*}^{\parallel}$ is given by
\begin{eqnarray}
\ell^{\parallel}=2\int_{0}^{z_{*}^{\parallel}} dz \ \frac{z^3 e^{-3 A(z)}e^{-B^2 z^2}}{\sqrt{g(z)[(z_{*}^{\parallel})^6 e^{-2 B^2 (z_{*}^{\parallel})^2}e^{-6A(z_{*}^{\parallel})}-z^6 e^{-2 B^2 z^2} e^{-6A(z)}]}}\,.
\label{lengthSEEcon}
\end{eqnarray}
The entanglement entropy expression of the disconnected surface is similarly found to be
\begin{eqnarray}
S^{\parallel}_{discon}=\frac{\ell_{y_2} \ell_{y_3} L^3}{2 G_{(5)}} \biggl[ \int_{0}^{\infty} dz \ \frac{e^{3 A(z)}e^{B^2 z^2}}{z^3\sqrt{g(z)}} \biggr]\,.
\label{SEEdiscon}
\end{eqnarray}
Note that $S^{\parallel}_{discon}$, unlike $S^{\parallel}_{con}$, does not depend on the strip length $\ell^{\parallel}$. This is an important feature that will greatly influence the properties of the entanglement measures in the confined phase. Also, note that both $S^{\parallel}_{con}$ and $S^{\parallel}_{discon}$ are UV-sensitive quantities and contain divergences. Here we adopt the minimal regularization procedure, as is generally done in the holographic literature, where these divergences are simply subtracted from the final results.

\begin{figure}[ht]
\begin{minipage}[b]{0.5\linewidth}
\centering
\includegraphics[width=2.8in,height=2.3in]{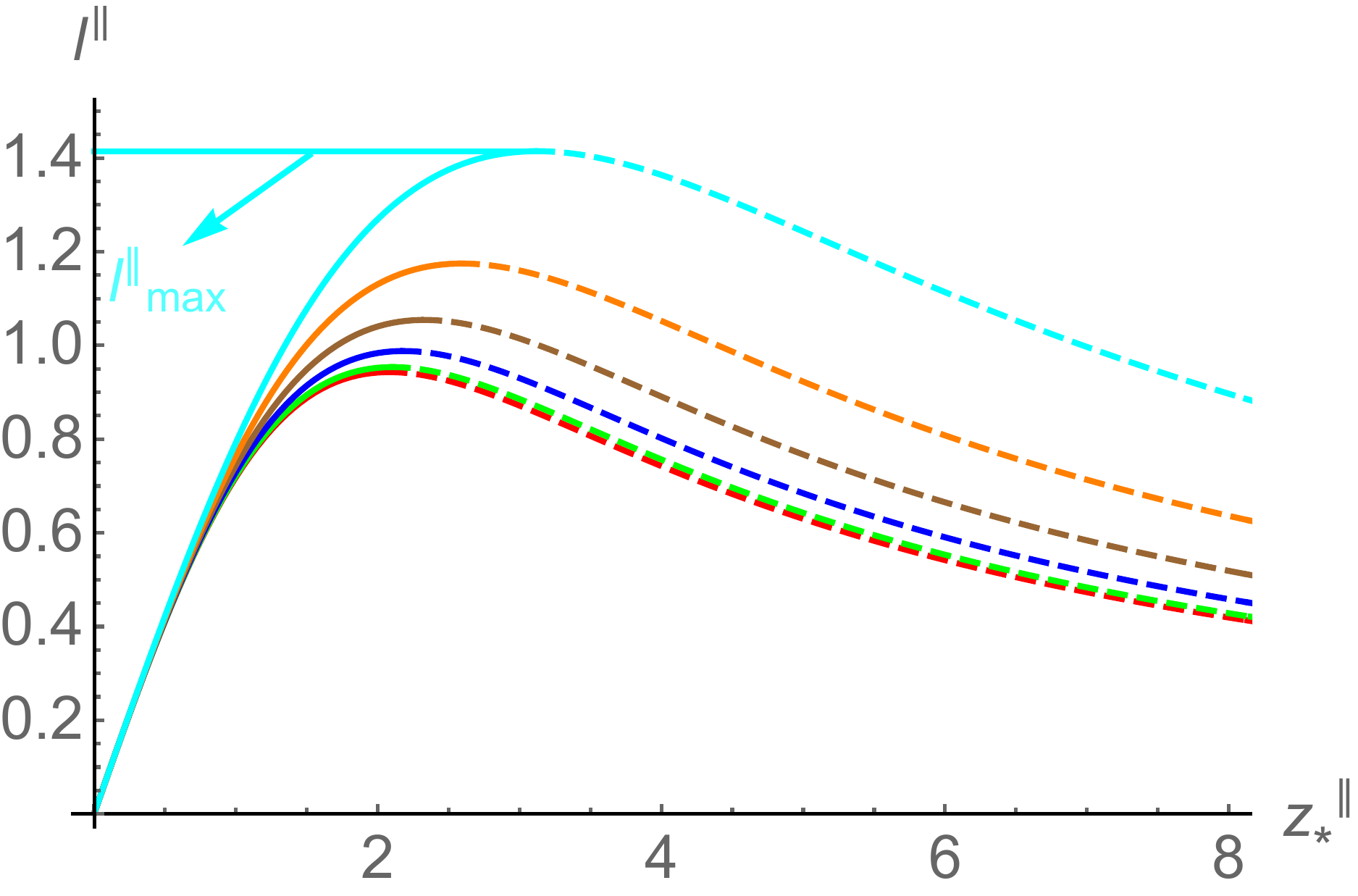}
\caption{$\ell^{\parallel}$ as a function of $z_{*}^{\parallel}$ for different values of $B$. The red, green, blue, brown, orange, and cyan
curves correspond to $B=0$, $0.1$, $0.2$, $0.3$, $0.4$, and $0.5$, respectively. In units of GeV.}
\label{lvszsTAdSparallel}
\end{minipage}
\hspace{0.4cm}
\begin{minipage}[b]{0.5\linewidth}
\centering
\includegraphics[width=2.8in,height=2.3in]{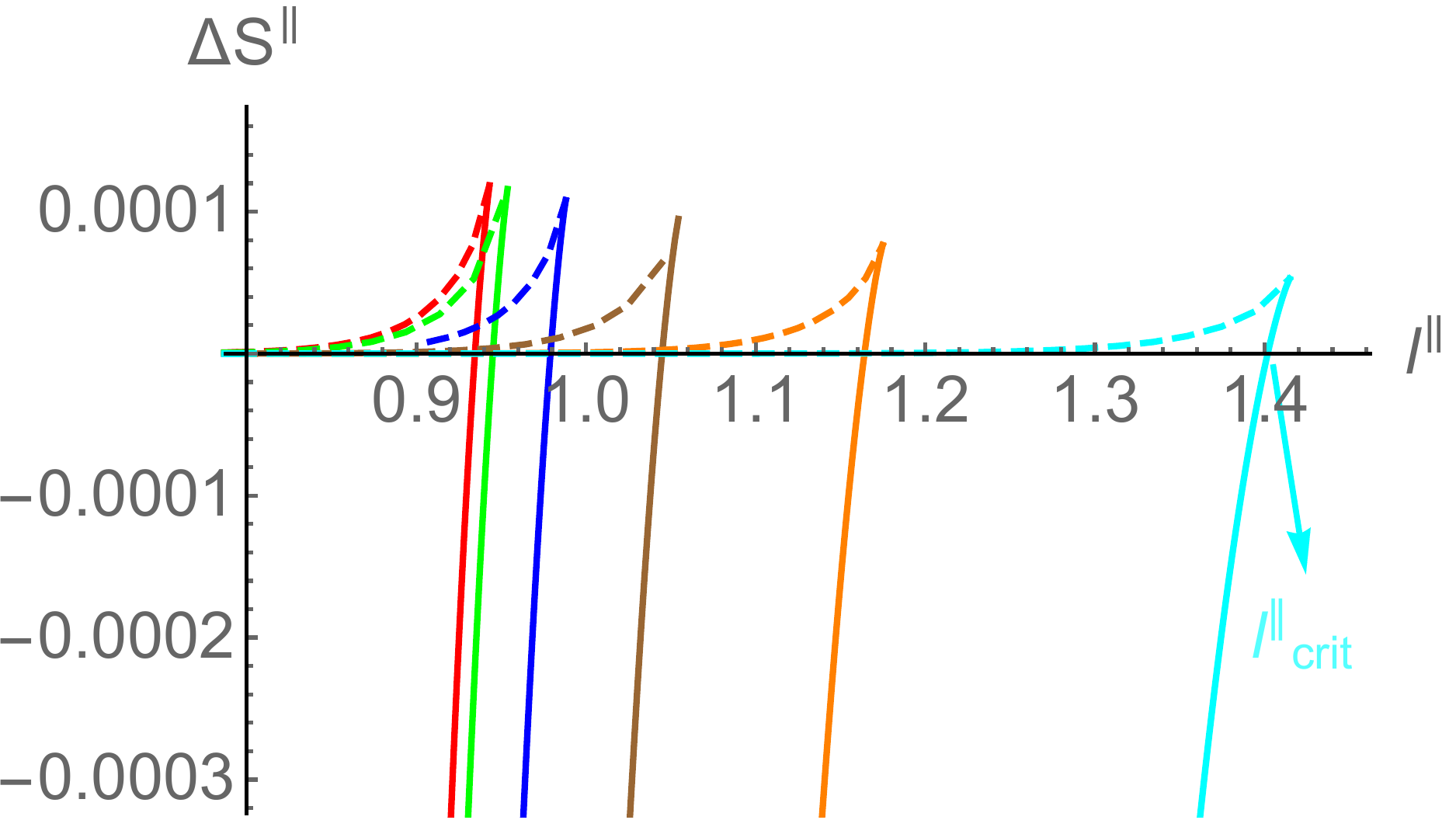}
\caption{$\Delta S^{\parallel}=S^{\parallel}_{con} - S^{\parallel}_{discon}$ as a function of $\ell^\parallel$ for different values of $B$.  The red, green, blue, brown, orange, and cyan
curves correspond to $B=0$, $0.1$, $0.2$, $0.3$, $0.4$, and $0.5$, respectively. In units of GeV.}
\label{EntropydiffTAdSparallel}
\end{minipage}
\end{figure}

Unfortunately, it is difficult to solve the above equations analytically. However, they are straightforward to solve numerically. The numerical result for the variation of the strip length $\ell^{\parallel}$ with respect to the connected surface turning point $z_{*}^{\parallel}$ for various values of $B$ is shown in Fig.~\ref{lvszsTAdSparallel}. We see that for any given value of $B$, there is a maximum length $\ell_{max}^{\parallel}$ above which no connected surface exists and only the disconnected surface exists. This $\ell_{max}^{\parallel}$ is a $B$-dependent quantity, whose magnitude not only increases but also appears at a larger $z_{*}^{\parallel}$ value as $B$ increases. This indicates that the connected entangling surface prorogates deeper into the bulk for larger $B$ values. We also see that below $\ell_{max}^{\parallel}$ there are two solutions that can minimize the connected surface area. The actual minima correspond to a solution that appears for small $z_{*}^{\parallel}$ (represented by solid lines), whereas the large $z_{*}^{\parallel}$ solution corresponds to the saddle point (represented by dashed lines).

The difference between the connected and disconnected entropies $\Delta S^{\parallel}=S^{\parallel}_{con}-S^{\parallel}_{discon}$ is shown in Fig.~\ref{EntropydiffTAdSparallel} for various values of the background magnetic field \footnote{The perfector $\ell_{y_2} \ell_{y_3} L^3/2 G_{(5)}$, appearing in Eqs.~(\ref{SEEcon}) and (\ref{SEEdiscon}), is set to one in numerical calculations.}. Again, the solution for small $z_{*}^{\parallel}$ is represented by solid lines, whereas the solution for large $z_{*}^{\parallel}$ is represented by dashed lines. It is interesting to see that $\Delta S^{\parallel}$ goes from negative to positive values as $\ell^\parallel$ increases, suggesting that for small values of $\ell^\parallel$ $S^{\parallel}_{con}$ minimizes the entanglement entropy, whereas for large values of $\ell^\parallel$ it is $S^{\parallel}_{discon}$ that minimizes the entanglement entropy. This indicates a phase transition from connected to disconnected entropy as $\ell^{\parallel}$ increases. This phase transition occurs at $\ell_{crit}^{\parallel}$, which is defined by the length at which $\Delta S^{\parallel}$ becomes zero.

\begin{figure}[ht]
\centering
\includegraphics[width=2.8in,height=2.3in]{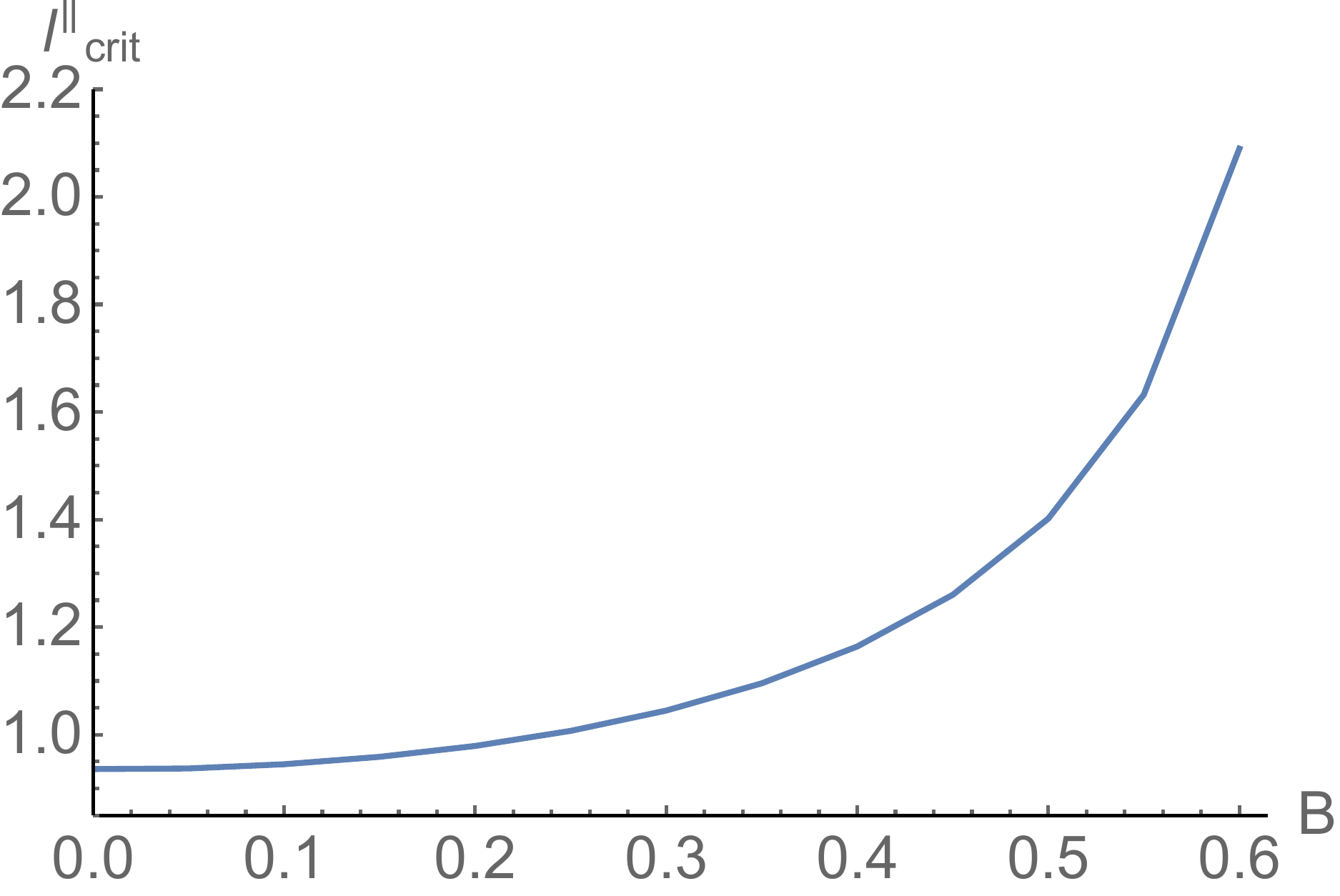}
\caption{$\ell_{crit}^{\parallel}$ as a function of $B$. In units of GeV.}
\label{BvsLcritParallel}
\end{figure}

We further find that $\ell_{crit}^{\parallel}$ depends nontrivially on the magnetic field. In particular, its magnitude increases with $B$ in the parallel direction. The overall behavior of the dependence of $\ell_{crit}^{\parallel}$ on $B$ is shown in Fig.~\ref{BvsLcritParallel}.

This type of phase transition between connected and disconnected entanglement entropies was first observed in top-down models in \cite{Klebanov:2007ws} and was suggested as a probe for confinement \footnote{In \cite{Jokela:2020wgs}, it was recently suggested that such connected and disconnected entanglement entropy phase transitions might be related to the mass gap rather than the linear confinement.}. In particular, such a geometric phase transition appears only in the confined phase, whereas no such phase transition is observed in the finite-temperature deconfined phase. Recalling the fact that for large $\ell^{\parallel}(>\ell_{crit}^{\parallel})$, the disconnected solution becomes independent of $\ell^{\parallel}$, this phase transition can be seen as follows:
\begin{eqnarray}\label{HEEphasetransition}
\frac{\partial S^{\parallel}}{\partial \ell^{\parallel}} &\propto &\frac{1}{G_{(5)}} = \mathcal{O}(N^2)\quad\text{for}\quad \ell^{\parallel} < \ell_{crit}^{\parallel}\,, \nonumber \\
&\propto& \frac{1}{G_{(5)}^{0}} = \mathcal{O}(N^0)\quad\text{for}\quad \ell^{\parallel} > \ell_{crit}^{\parallel} \,,
\end{eqnarray}
where $N$ denotes the number of colors in the dual boundary theory. This implies nonanalytic behavior at $\ell_{crit}^{\parallel}$, where the number of degrees of freedom changes from $\mathcal{O}(N^2)$ to $\mathcal{O}(N^0)$, in the entanglement entropy structure of the confined phase. This type of phase transition has been observed in other holographic confining theories as well. Here we have reconfirmed this already established result, but now in a consistent bottom-up holographic QCD model in the presence of a background magnetic field. Interestingly, a similar type of nonanalyticity in the entanglement entropy has also been observed in $SU(2)$ and $SU(3)$ gauge theories using lattice simulations \cite{Buividovich:2008kq,Buividovich:2008gq}. Therefore, it seems that nonanalyticity is a generic feature of the entanglement entropy in confining theories irrespective of whether it has a gravity dual or not. To further appreciate these results, note that our holographic estimate for the length scale at which nonanalyticity appears ($\ell_{crit}^{\parallel}\simeq0.2~fm$) is in the same ballpark as that estimated by lattice simulations ($\ell_{crit}^{\parallel}\simeq0.5~fm$). This lends further support to the notion that certain modeling of holographic theories can yield compelling predictions for real QCD-like theories. Moreover, the result that the magnitude of $\ell_{crit}^{\parallel}$ increases with the increase of the magnetic field in the parallel direction is an important prediction of our model and could be verified in independent lattice settings (as we would not have to worry about various numerical issues, like the famous sign problem, with a finite magnetic field in lattice calculations).

\subsubsection{Strip in the perpendicular direction}
We now analyze the entanglement entropy in the perpendicular case. In this case, the strip subsystem, with the domain $\{0\leq y_1 \leq \ell_{y_1}$, $-\ell^{\perp}/2\leq y_2 \leq \ell^{\perp}/2$, $0\leq y_3 \leq \ell_{y_3} \}$, is aligned perpendicular to the magnetic field. There are again connected and disconnected bulk surfaces that minimize the entanglement entropy expression. The expression of the connected surface now reduces to
\begin{eqnarray}
S^{\perp}_{con}=\frac{\ell_{y_1} \ell_{y_3} L^3}{2 G_{(5)}} \int_{0}^{z_{*}^{\perp}} dz \ \left(\frac{z_{*}^{\perp}}{z}\right)^3 \frac{{e^{3 A(z)-3 A(z_{*}^{\perp})}}{{e^{B^2z^2-B^2(z_{*}^{\perp})^2}}}e^{-B^2 z^2/2}}{\sqrt{g(z)[(z_{*}^{\perp})^6 e^{-2 B^2 (z_{*}^{\perp})^2}e^{-6A(z_{*}^{\perp})}-z^6 e^{-2 B^2 z^2} e^{-6A(z)}]}}\,.
\label{SEEconperp}
\end{eqnarray}
Similarly, the strip length $\ell^{\perp}$ in terms of the turning point $z_{*}^{\perp}$ is
\begin{eqnarray}
\ell^{\perp}=2\int_{0}^{z_{*}^{\perp}} dz \ \frac{z^3 e^{-3 A(z)}e^{-3B^2 z^2/2}}{\sqrt{g(z)[(z_{*}^{\perp})^6 e^{-2 B^2 (z_{*}^{\perp})^2}e^{-6A(z_{*}^{\perp})}-z^6 e^{-2 B^2 z^2} e^{-6A(z)}]}}\,.
\label{lengthSEEconperp}
\end{eqnarray}
The expression for the disconnected surface is again independent of the strip length $\ell^{\perp}$ and is now given by
\begin{eqnarray}
S^{\perp}_{discon}=\frac{\ell_{y_1} \ell_{y_3} L^3}{2 G_{(5)}} \biggl[ \int_{0}^{\infty} dz \ \frac{e^{3 A(z)}e^{B^2 z^2/2}}{z^3\sqrt{g(z)}} \biggr] \,.
\label{SEEdisconperp}
\end{eqnarray}

\begin{figure}[h]
\begin{minipage}[b]{0.5\linewidth}
\centering
\includegraphics[width=2.8in,height=2.3in]{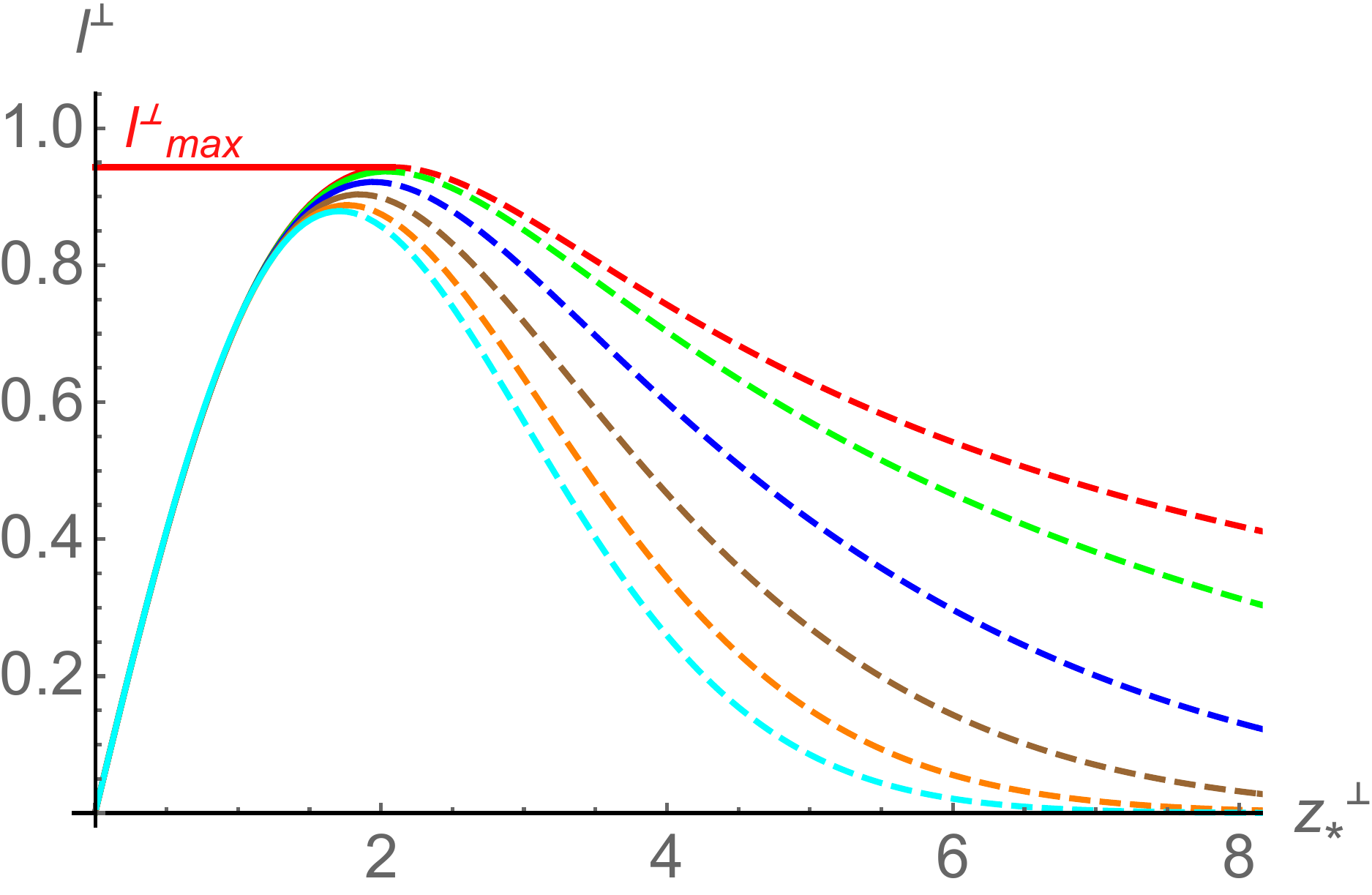}
\caption{$\ell^{\perp}$ as a function of $z_{*}^{\perp}$ for different values of $B$. The red, green, blue, brown, orange, and cyan
curves correspond to $B=0$, $0.1$, $0.2$, $0.3$, $0.4$, and $0.5$, respectively. In units of GeV.}
\label{lvszsTAdSperp}
\end{minipage}
\hspace{0.4cm}
\begin{minipage}[b]{0.5\linewidth}
\centering
\includegraphics[width=2.8in,height=2.3in]{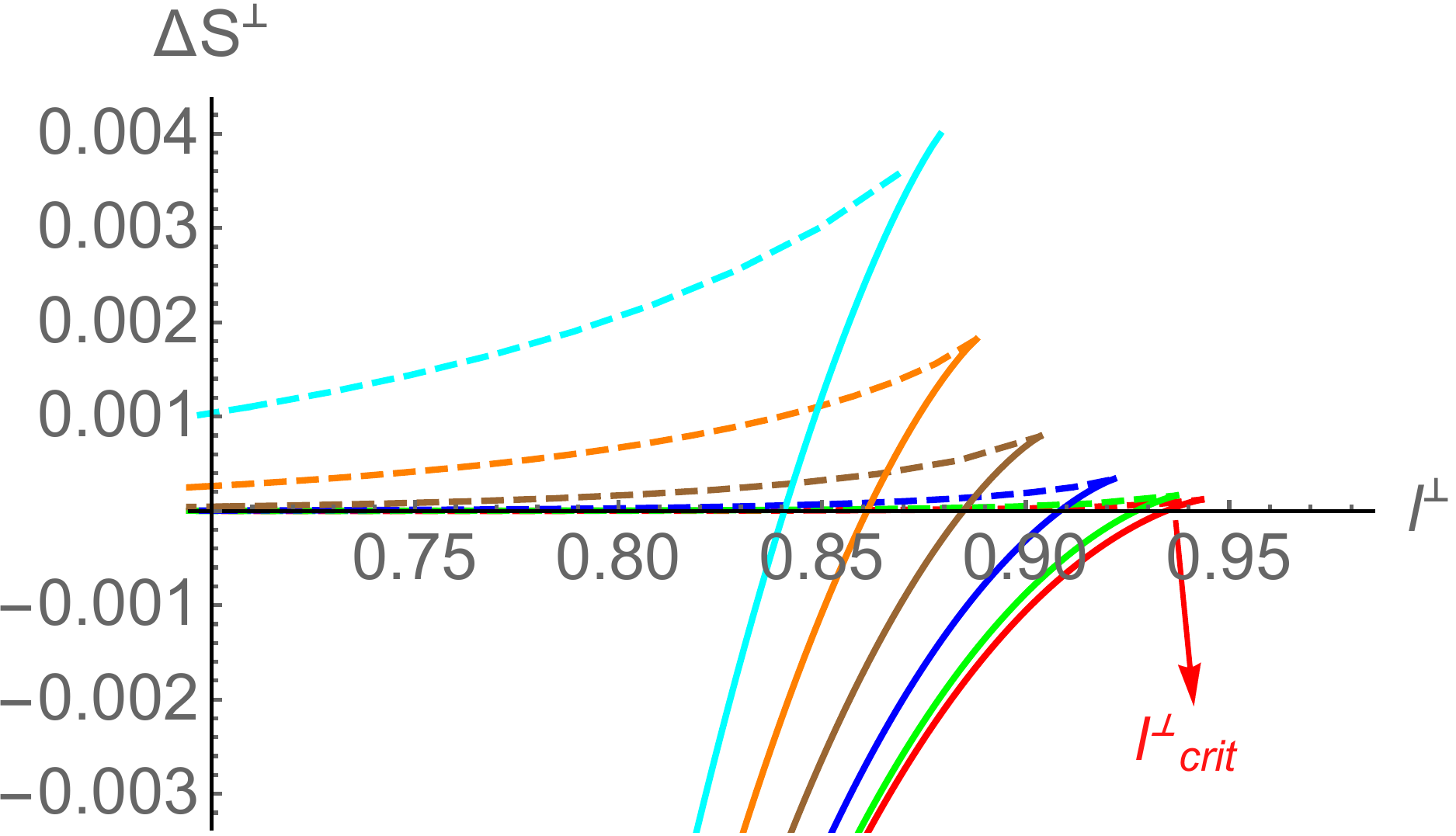}
\caption{$\Delta S^{\perp}=S^{\perp}_{con} - S^{\perp}_{discon}$ as a function of $\ell^\perp$ for different values of $B$.  The red, green, blue, brown, orange, and cyan
curves correspond to $B=0$, $0.1$, $0.2$, $0.3$, $0.4$, and $0.5$, respectively. In units of GeV.}
\label{EntropydiffTAdSperp}
\end{minipage}
\end{figure}

We can clearly see some differences in the above equations compared to the parallel case. Accordingly, some differences in the entanglement entropy result are also expected. The variation of $\ell^{\perp}$ with respect to the connected surface $z_{*}^{\perp}$ for different values of $B$ is shown in Fig.~\ref{lvszsTAdSperp}. We observe that for any given value of $B$, like in the parallel case, there is again a maximum length $\ell_{max}^{\perp}$ above which no connected solution exists and only the disconnected solution exists. However, as opposed to the parallel case, now not only the magnitude of $\ell_{max}^{\perp}$ but also the value of the turning point $z_{*}^{\perp}$ at which it appears decreases with $B$. This suggests a lesser penetration of the entangling surface into the bulk as compared to the parallel case as $B$ increases. Further, below $\ell_{max}^{\perp}$, there are again two connected solutions (shown by solid and dashed lines) which can minimize the surface area. The solid line corresponds to the actual minima and appears for small $z_{*}^{\perp}$, whereas the dashed line corresponds to the saddle point and appears for large $z_{*}^{\perp}$.

\begin{figure}[ht]
\centering
\includegraphics[width=2.8in,height=2.3in]{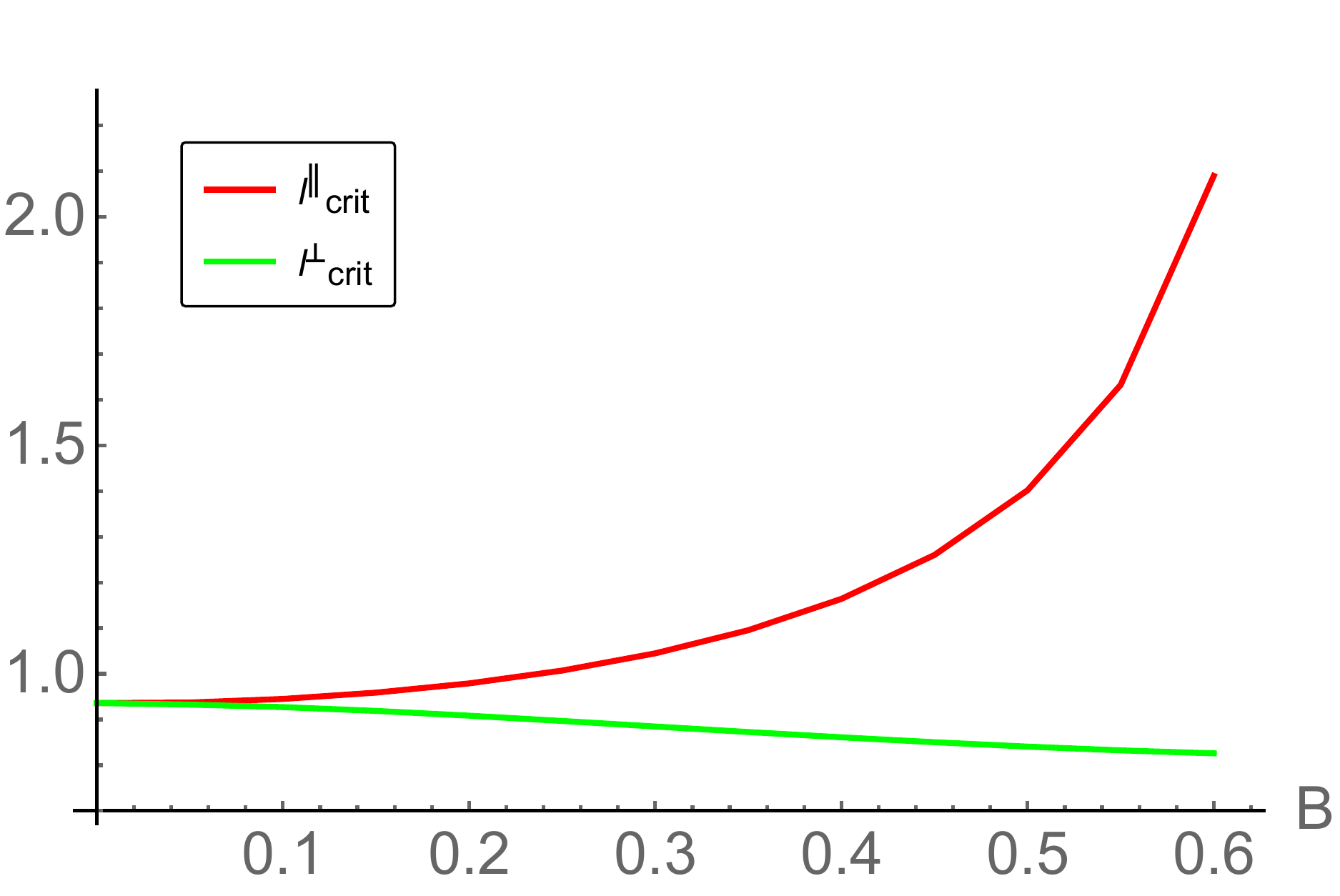}
\caption{Variation of $\ell_{crit}^{\perp}$ (green line) and $\ell_{crit}^{\parallel}$ (red line) as a function of $B$. In units of GeV.}
\label{BvsLcritPerpendicular}
\end{figure}

The difference between the connected and disconnected entropies $\Delta S^{\perp}=S^{\perp}_{con}-S^{\perp}_{discon}$ for the perpendicular case is shown in Fig.~\ref{EntropydiffTAdSperp} for various values of $B$. The connected solution with small $z_{*}^{\perp}$ (indicated by solid lines) always has a lower entanglement entropy than the large $z_{*}^{\perp}$ solution (indicated by dashed lines). Further, $\Delta S^{\perp}$ goes from negative to positive values as $\ell^{\perp}$ increases, indicating that $S^{\perp}_{con}$ ($S^{\perp}_{discon}$) minimizes the entropy for small $\ell^{\perp}$ (large $\ell^{\perp}$). This results in a phase transition from connected to disconnected surfaces, similar to the ones in the parallel case, as we increase $\ell^{\perp}$. The critical length at which this phase transition appears is now defined as $\ell_{crit}^{\perp}$, where
$\ell_{crit}^{\perp}<\ell_{max}^{\perp}$. Therefore, similar to Eq.~(\ref{HEEphasetransition}) for the parallel case, we again have a length scale at which the order of the entanglement entropy changes from $\mathcal{O}(N^2)$ to $\mathcal{O}(N^0)$. However, in contrast with the parallel case, this critical length in the perpendicular case now decreases with $B$. This is shown in Fig.~\ref{BvsLcritPerpendicular}, where the parallel-case result is also included for comparison. We find that the difference between $\ell_{crit}^{\parallel}-\ell_{crit}^{\perp}$ is small for small $B$; however, it can be appreciable for large $B$. This suggests that the nonanalyticity in the entanglement entropy appears at larger lengths in the parallel case compared to the perpendicular case for all values of $B$. Our whole analysis therefore suggests appreciable anisotropic changes in the entanglement entropy structure of the confined phase in the presence of a magnetic field.

\subsection{Holographic mutual information}
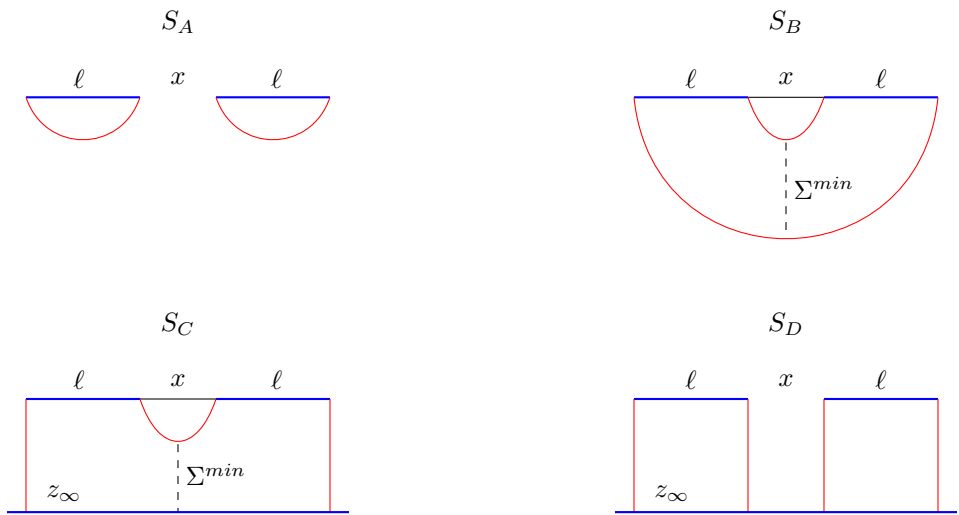
\begin{figure}[h]
\begin{center}
\begin{tikzpicture}
\draw [blue, thick] (-13,0) -- (-11.5,0);
\draw [blue, thick](-10.5,0) -- (-9,0);
\draw[red] (-13,0) .. controls (-12.75,-0.75) and (-11.75,-0.75) .. (-11.5,0);
\draw[red] (-10.5,0) .. controls (-10.25,-0.75) and (-9.25,-0.75) .. (-9,0);
\node()at (-9.7,0.25){$\ell$};
\node()at (-11,0.25){$x$};
\node()at (-11,1){$S_A$};
\node()at (-12.3,0.25){$\ell$};
\draw [blue, thick] (-5,0) -- (-3.5,0);
\draw (-3.5,0) -- (-2.5,0);
\draw [blue, thick](-2.5,0) -- (-1,0);
\draw[red] (-5,0) .. controls (-4.75,-2.5) and (-1.25,-2.5) .. (-1,0);
\draw[red] (-3.5,0) .. controls (-3.25,-0.75) and (-2.75,-0.75) .. (-2.5,0);
\draw[dashed] (-3,-0.6) -- (-3,-1.86);
\node()at (-2.5,-1.2){$\Sigma^{min}$};
\node()at (-3,1){$S_B$};
\node()at (-4.25,0.25){$\ell$};
\node()at (-3,0.25){$x$};
\node()at (-1.75,0.25){$\ell$};
\draw [blue, thick] (-13,-4) -- (-11.5,-4);
\draw (-11.5,-4) -- (-10.5,-4);
\draw [blue, thick](-10.5,-4) -- (-9,-4);
\draw[red] (-11.5,-4) .. controls (-11.25,-4.75) and (-10.75,-4.75) .. (-10.5,-4);
\draw[red] (-13,-4) -- (-13,-5.5);
\draw[red] (-9,-4) -- (-9,-5.5);
\draw[dashed] (-11,-4.6) -- (-11,-5.5);
\node()at (-10.5,-5){$\Sigma^{min}$};
\draw [blue, thick](-13.25,-5.5) -- (-8.75,-5.5);
\node()at (-12.5,-5.25){$z_{\infty}$};
\node()at (-9.7,-3.75){$\ell$};
\node()at (-11,-3.75){$x$};
\node()at (-11,-3){$S_C$};
\node()at (-12.3,-3.75){$\ell$};
\draw [blue, thick] (-5,-4) -- (-3.5,-4);
\draw [blue, thick](-2.5,-4) -- (-1,-4);
\draw[red] (-5,-4) -- (-5,-5.5);
\draw[red] (-3.5,-4) -- (-3.5,-5.5);
\draw[red] (-1,-4) -- (-1,-5.5);
\draw[red] (-2.5,-4) -- (-2.5,-5.5);
\draw [blue, thick](-5.25,-5.5) -- (-0.75,-5.5);
\node()at (-4.5,-5.25){$z_{\infty}$};
\node()at (-3,-3){$S_D$};
\node()at (-4.25,-3.75){$\ell$};
\node()at (-3,-3.75){$x$};
\node()at (-1.75,-3.75){$\ell$};
\end{tikzpicture}
\end{center}
\caption{Pictorial representation of the four different minimal surface configurations for the case of two strips of equal length $\ell$ separated by a distance $x$ in the thermal-AdS background. The dashed line represents the entanglement wedge.}
\label{ES2equalstrips}
\end{figure}

We now study the holographic mutual information with two strips in the confined phase. For simplicity, we concentrate only on equal-size strip subsystems ($\ell_1=\ell_2=\ell$),
which are separated by a distance $x$. The entanglement structure with two subsystems is much more intriguing than that with one subsystem. In particular, depending on the magnitudes of $\ell$ and $x$, there can be four possible surfaces that minimize the entropy. These four surfaces are illustrated in Fig.~\ref{ES2equalstrips}. We can now have only connected surfaces (i.e., $S_A$ and $S_B$), both connected and disconnected surface ($S_C$), or only the disconnected surface ($S_D$). The holographic entanglement entropies for these four configurations are as follows:
\begin{eqnarray}
 S_A  (\ell,x) &=& 2 S_{con} (\ell), \hspace{2.0cm} S_B(\ell,x)= S_{con} (x) + S_{con} (2\ell+x)  \,, \nonumber \\
  S_C  (\ell,x) &=& S_{con} (x) + S_{discon} , \hspace{0.7cm} S_D(\ell,x)= 2 S_{discon} \,,
 \label{eq2strips}
\end{eqnarray}
where $S_{con}$ and $S_{discon}$ are the single-interval holographic entanglement entropies for the connected and disconnected surfaces, respectively.

\subsubsection{Parallel case}
\begin{figure}[h]
\centering
\includegraphics[width=2.8in,height=2.3in]{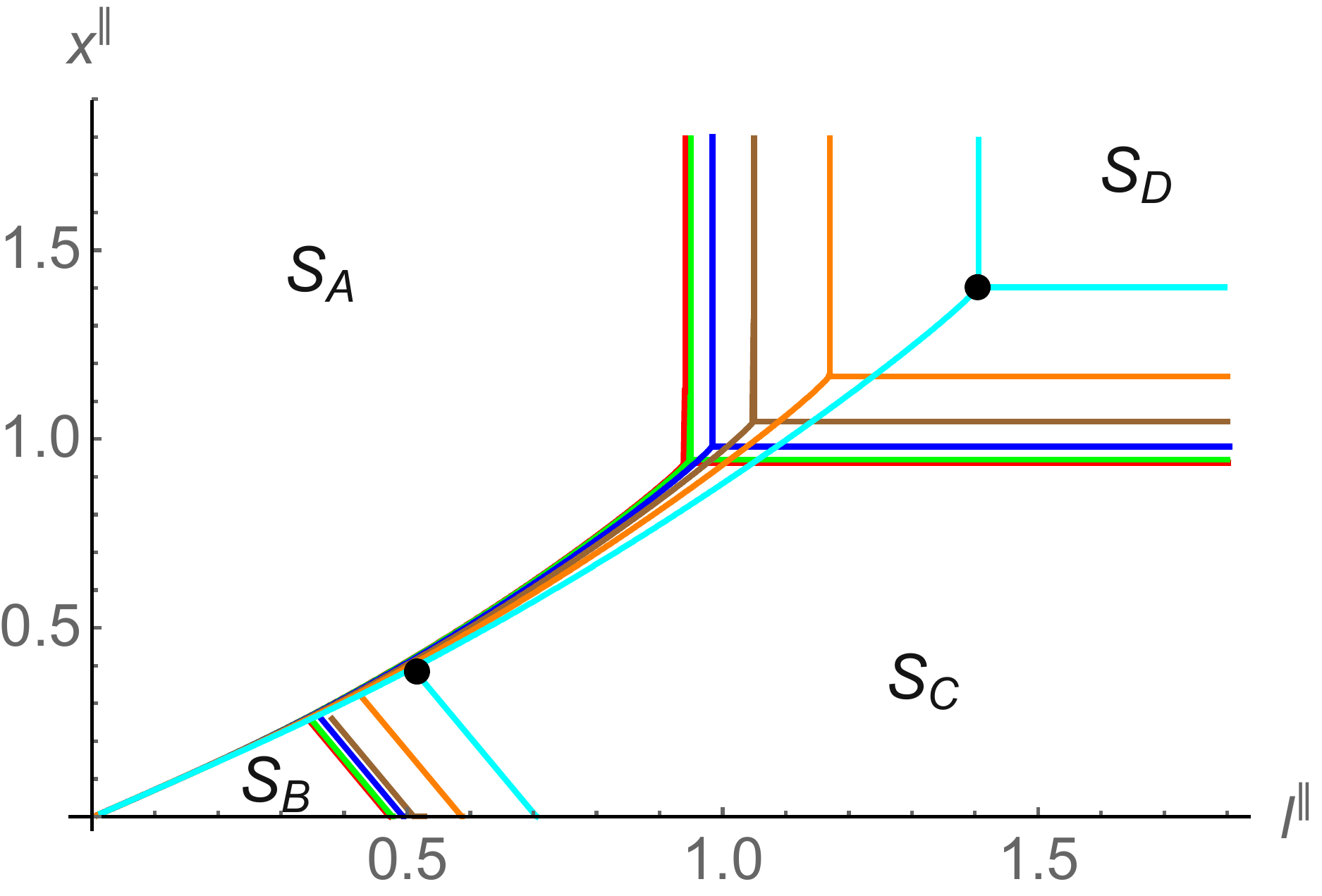}
\caption{Phase diagram of various minimal area surfaces for the case of two strips of equal length $\ell$
separated by a distance $x$ in the confining background for the parallel case. The red, green, blue, brown, orange, and cyan
curves correspond to $B=0$, $0.1$, $0.2$, $0.3$, $0.4$, and $0.5$, respectively. The two black dots indicate the two tricritical points for $B=0.5$. In units of GeV.}
\label{TAdSparallelphasediag}
\end{figure}

Let us first discuss the results when the strips are oriented in a parallel direction relative to the magnetic field. We find that there can be different phase transitions between the above-mentioned four configurations. This phase diagram can be illustrated better in the $(\ell^{\parallel},x^{\parallel})$ plane and is shown in Fig.~\ref{TAdSparallelphasediag}. We find that for small $x^{\parallel}, \ell^{\parallel} \ll \ell_{crit}^{\parallel}$, $S_A$ phase is preferred as it has the lowest entropy. As $\ell^{\parallel}$ increases, the $S_B$ phase becomes dominant. As we keep increasing $\ell^{\parallel}$ but keep $x^{\parallel} (\ll \ell_{crit}^{\parallel})$ fixed, a phase transition from $S_B$ to $S_C$ ocuurs. For $x^{\parallel}=0$, this $S_B$ to $S_C$ phase transition happens at $\ell^{\parallel}=\ell_{crit}^{\parallel}/2$, whereas for a general value of $x^{\parallel}$, it happens at $2\ell^{\parallel}+x^{\parallel}=\ell_{crit}^{\parallel}$. Further, if we take $x^{\parallel}, \ell^{\parallel} \gg \ell_{crit}^{\parallel}$, then the $S_D$ configuration becomes the dominant one. We also observe that these phase transitions depend nontrivially on $B$. For example, the $S_B/S_C$ phase transition line shifts to the right in the $\ell^{\parallel}-x^{\parallel}$ plane and appears for larger values of $x^{\parallel}$ and $\ell^{\parallel}$,  whereas the $S_A/S_C$ transition occurs for lower values of $x^{\parallel}$ when $B$ increases.

Also, there are two tricritical points in this phase diagram. For $B=0.5$, these are indicated by two black dots. The first tricritical point is recognized when the $S_A$, $S_B$, and $S_C$ phases coexist, and the second tricritical point occurs when the $S_A$, $S_C$, and $S_D$ phases coexist. The presence of these phase transitions and critical points reflects the nonanalytic nature of the entanglement entropy with multiple strips. The magnitudes of $x^{\parallel}$ and $\ell^{\parallel}$ at these tricritical points also depend nontrivially on $B$ and can be observed in Fig.~\ref{TAdSparallelphasediag}.

It is also interesting to note that the order of the entanglement entropy (from $N^2$ to $N^0$ or vice versa) may or may not change as we pass through various phase transition lines in the two-strips case. For instance, there is no change in the order if the $S_A$, $S_B$, or $S_C$ phases are involved, whereas the order can change if the $S_D$ phase is involved.

\begin{figure}[h]
\begin{minipage}[b]{0.5\linewidth}
\centering
\includegraphics[width=2.8in,height=2.3in]{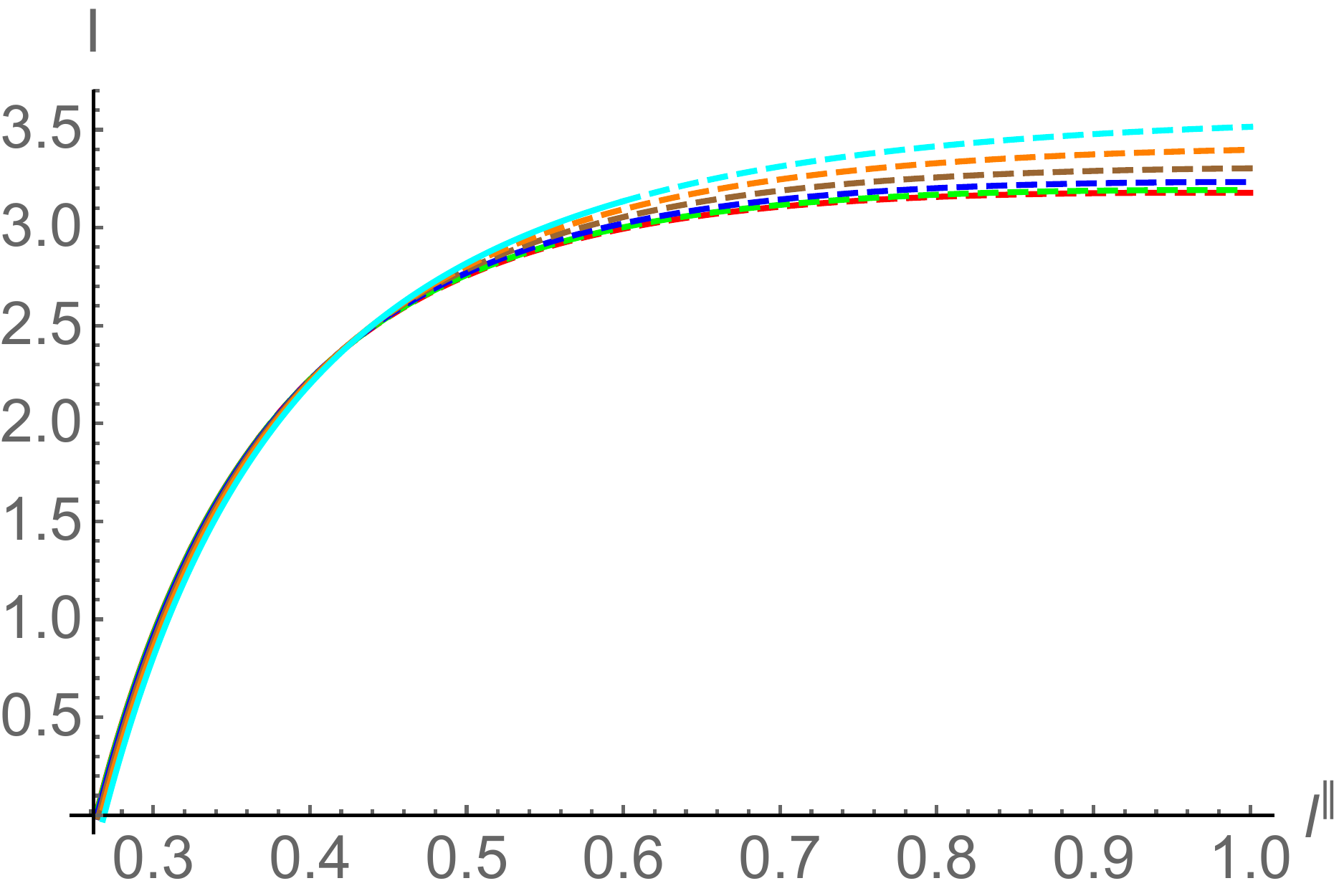}
\caption{Variation of the mutual information with $\ell^{\parallel}$ for different values of $B$. Here $x^{\parallel}=0.2$ is used, and the red, green, blue, brown, orange, and cyan
curves correspond to $B=0$, $0.1$, $0.2$, $0.3$, $0.4$, and $0.5$, respectively. In units of GeV.}
\label{TAdSParallelMIvsl}
\end{minipage}
\hspace{0.4cm}
\begin{minipage}[b]{0.5\linewidth}
\centering
\includegraphics[width=2.8in,height=2.3in]{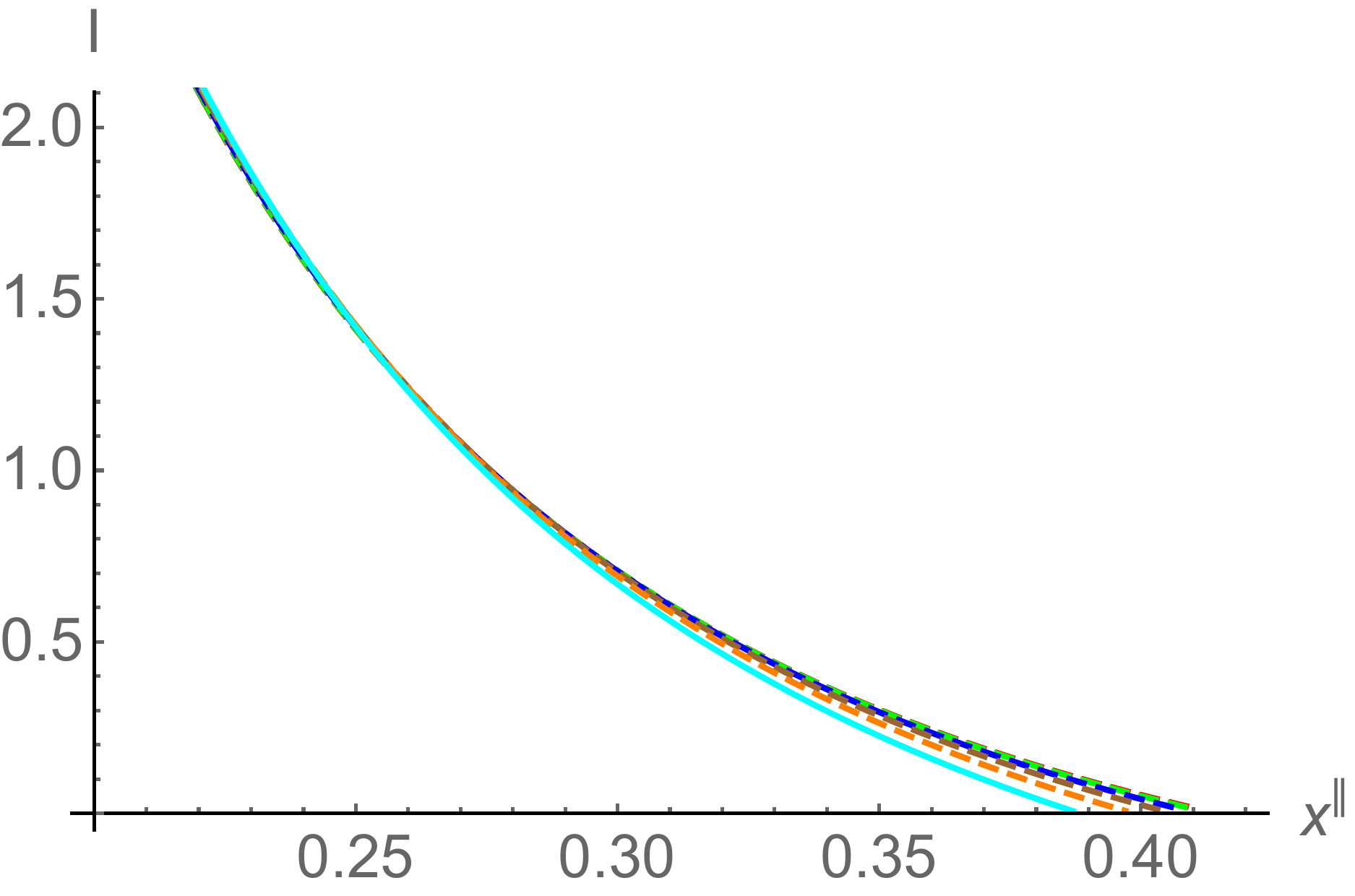}
\caption{Variation of the mutual information with $x^{\parallel}$ for different values of $B$. Here $\ell^{\parallel}=0.5$ is used, and the red, green, blue, brown, orange, and cyan
curves correspond to $B=0$, $0.1$, $0.2$, $0.3$, $0.4$, and $0.5$, respectively. In units of GeV.}
\label{TAdSParallelMIvsx}
\end{minipage}
\end{figure}

Let us now discuss the holographic mutual information, $I = S_1+S_2 - S_1 \cup S_2$, which in the above four different phases has the form
\begin{eqnarray}
 I_A  (\ell^{\parallel},x^{\parallel}) &=& S^{\parallel}_{con}(\ell^{\parallel}) + S^{\parallel}_{con}(\ell^{\parallel}) - 2 S^{\parallel}_{con}(\ell^{\parallel}) = 0 \,, \nonumber \\
 I_B (\ell^{\parallel},x^{\parallel})  &=&  S^{\parallel}_{con}(\ell^{\parallel}) + S^{\parallel}_{con}(\ell^{\parallel})-S^{\parallel}_{con} (x^{\parallel}) - S^{\parallel}_{con} (2\ell^{\parallel}+x^{\parallel}) \geq 0  \,, \nonumber \\
 I_C  (\ell^{\parallel},x^{\parallel}) &=& S^{\parallel}_{con}(\ell^{\parallel}) + S^{\parallel}_{con}(\ell^{\parallel}) - S^{\parallel}_{con} (x^{\parallel}) - S^{\parallel}_{discon} \geq 0 \,, \nonumber \\
 I_D (\ell^{\parallel},x^{\parallel})  &=& S^{\parallel}_{discon} + S^{EE}_{discon} - 2 S^{EE}_{discon} =0 \,,
 \label{mutual2strips}
\end{eqnarray}
which in turn means that
\begin{eqnarray}\label{MIorder}
\frac{\partial I_A}{\partial \ell^{\parallel}} \propto \frac{1}{G_{(5)}^0} = \mathcal{O}(N^0), \ \ \ \ \ \ \frac{\partial I_B}{\partial \ell^{\parallel}} \propto \frac{1}{G_{(5)}} = \mathcal{O}(N^2)\,,  \nonumber \\
\frac{\partial I_C}{\partial \ell^{\parallel}} \propto \frac{1}{G_{(5)}} = \mathcal{O}(N^2), \ \ \ \ \ \ \frac{\partial I_D}{\partial \ell^{\parallel}} \propto \frac{1}{G_{(5)}^{0}} = \mathcal{O}(N^0) \,.
\end{eqnarray}
Therefore, depending on the transition line, the order of the mutual information may or may not change as we go from one phase to another. For instance, going from the $S_A$ phase to the $S_B$ phase (by decreasing $x^\parallel$) causes a change in its order [from $\mathcal{O}(N^0)$ to $\mathcal{O}(N^2)$], whereas no such change occurs when we go from the $S_B$ phase to the $S_C$ phase (by increasing $\ell^\parallel$).

The variation of the mutual information with respect to strip length $\ell^{\parallel}$ and separation length $x^{\parallel}$ for different values of $B$ is shown in Figs.~(\ref{TAdSParallelMIvsl}) and (\ref{TAdSParallelMIvsx}). Here the mutual information in the $S_B$ ($S_C$) phase is represented by the solid (dashed) lines. We observe that the mutual information varies smoothly as we move from $S_B$ to $S_C$ via the $S_B/S_C$ transition line. In Fig.~\ref{TAdSParallelMIvsl} we have shown the results for a fixed $x^{\parallel}=0.2$ line, but similar results exist for other values of $x^{\parallel}$ as well. As we increase $B$ along the parallel direction, $I_B$ almost remains the same but $I_C$ increases slightly. Similarly, the mutual information smoothly goes to zero as we approach the $S_A$ (or $S_D$) phase from the $S_B$ (or $S_C$) phase. This is shown in Fig.~\ref{TAdSParallelMIvsx}.

\subsubsection{Perpendicular case}
\begin{figure}[ht]
\centering
\includegraphics[width=2.8in,height=2.3in]{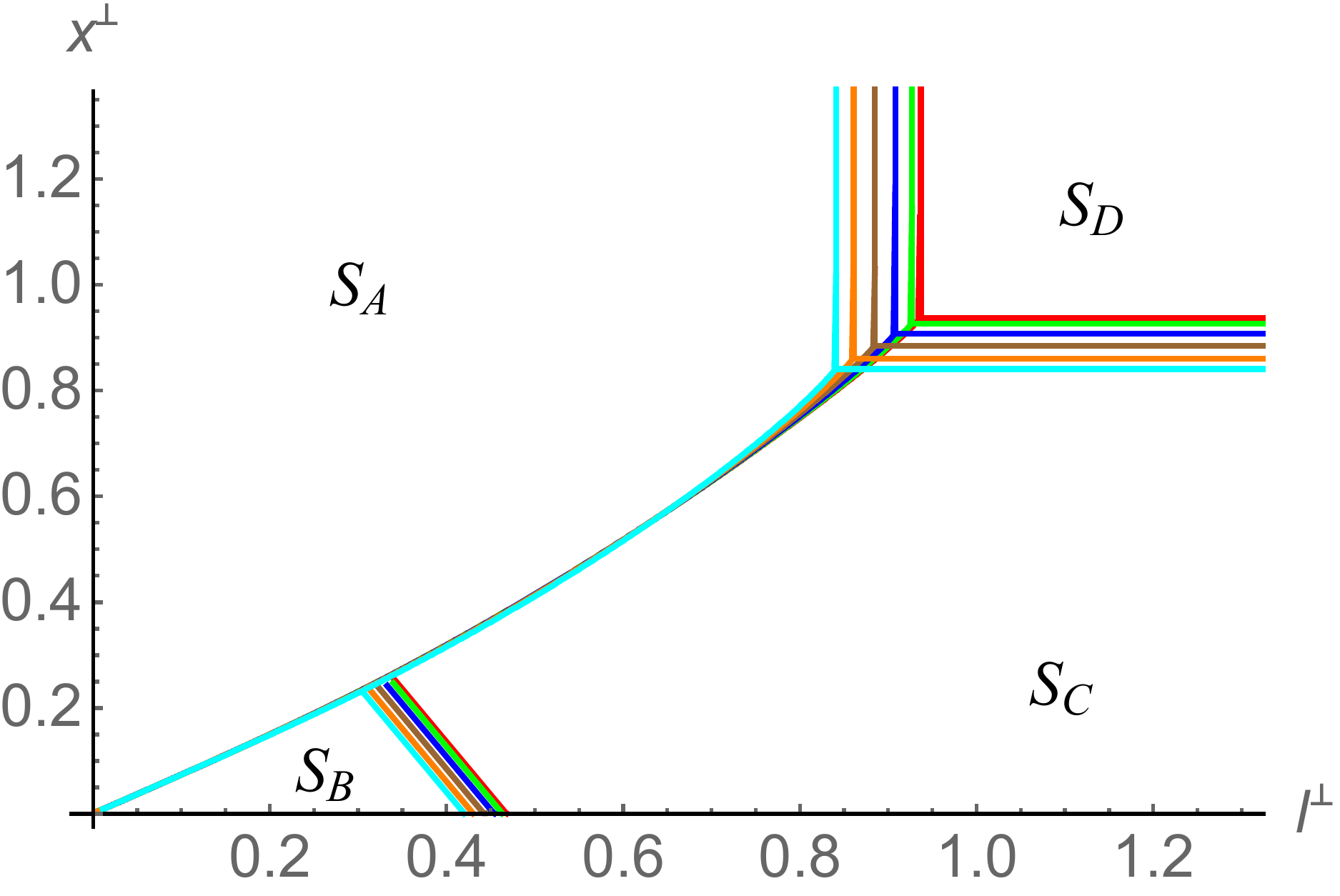}
\caption{Phase diagram of various minimal area surfaces for the case of two strips of equal length $\ell^\perp$ separated by a distance $x^\perp$ in the confining background for the perpendicular case. The red, green, blue, brown, orange, and cyan curves correspond to $B=0$, $0.1$, $0.2$, $0.3$, $0.4$, and $0.5$, respectively. In units of GeV.}
\label{TAdSperpphasediag}
\end{figure}
We now move on to discuss the two-strip phase diagram and the corresponding mutual information when the strips are oriented in the perpendicular direction. This phase diagram is shown in Fig.~\ref{TAdSperpphasediag}. We see that there are again four phases, with each one dominating different parts of the $\ell^{\perp}-x^{\perp}$ phase space, and they undergo various phase transitions as we vary $\ell^{\perp}$ and $x^{\perp}$. There are again two tricritical points, which are $B$-dependent. This is qualitatively similar to the parallel-case phase diagram. However, there are some differences as well. In particular, the values of $\{\ell^{\perp},x^{\perp}\}$ at both tri-critical points now decrease with $B$, in contrast to the parallel case where these values at the second tricritical point increase with $B$. Similarly, in contrast to the parallel case, the size of the $S_B$ phase now decreases for higher values of $B$. Moreover, the $S_A/S_C$ transition line also moves slightly upward for higher values of $B$.

\begin{figure}[h]
\begin{minipage}[b]{0.5\linewidth}
\centering
\includegraphics[width=2.8in,height=2.3in]{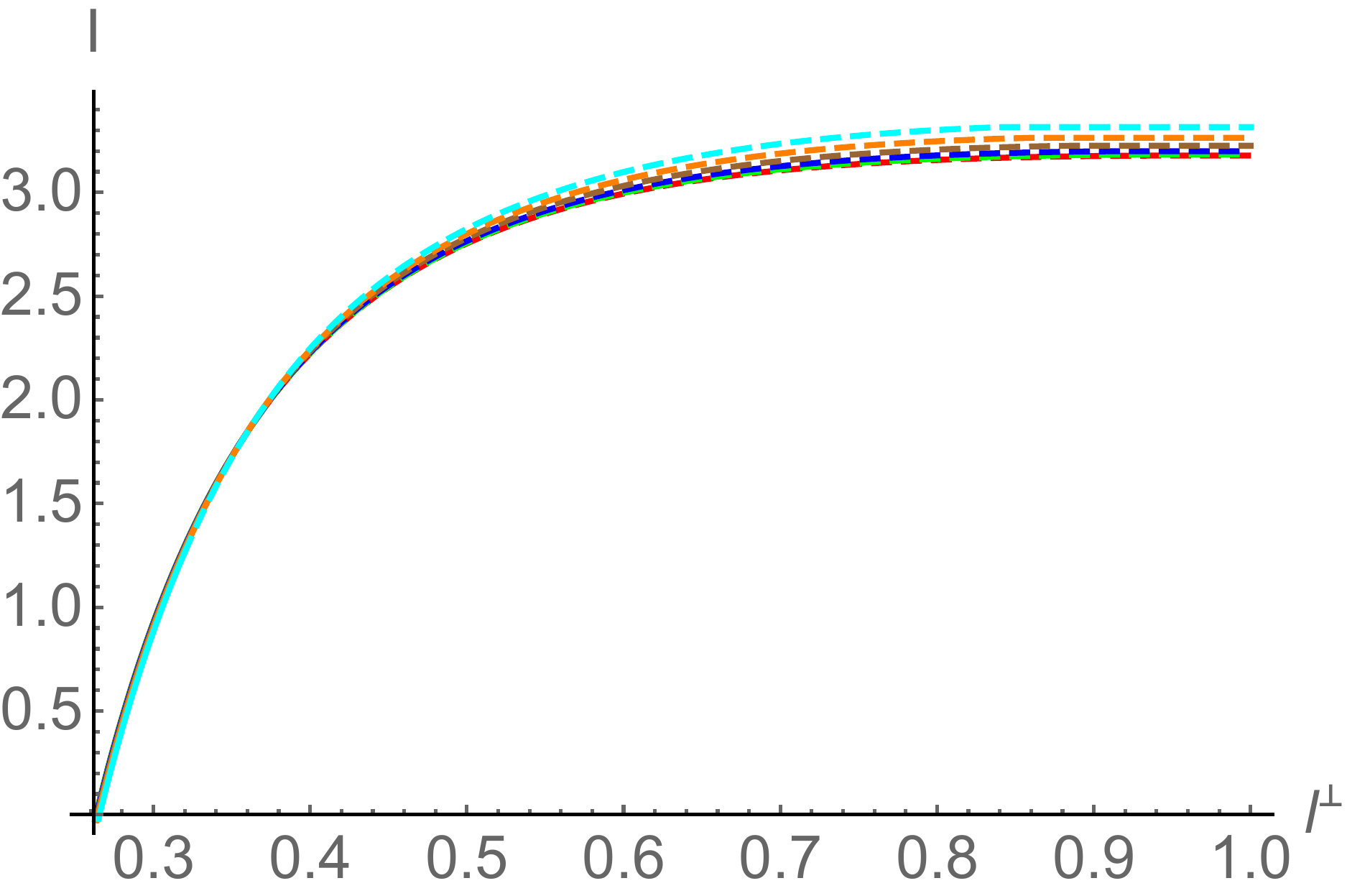}
\caption{Variation of the mutual information with $\ell^{\perp}$ for different values of $B$. Here $x^{\perp}=0.2$ is used, and the red, green, blue, brown, orange, and cyan
curves correspond to $B=0$, $0.1$, $0.2$, $0.3$, $0.4$, and $0.5$, respectively. In units of GeV.}
\label{TAdSperpMIvsl}
\end{minipage}
\hspace{0.4cm}
\begin{minipage}[b]{0.5\linewidth}
\centering
\includegraphics[width=2.8in,height=2.3in]{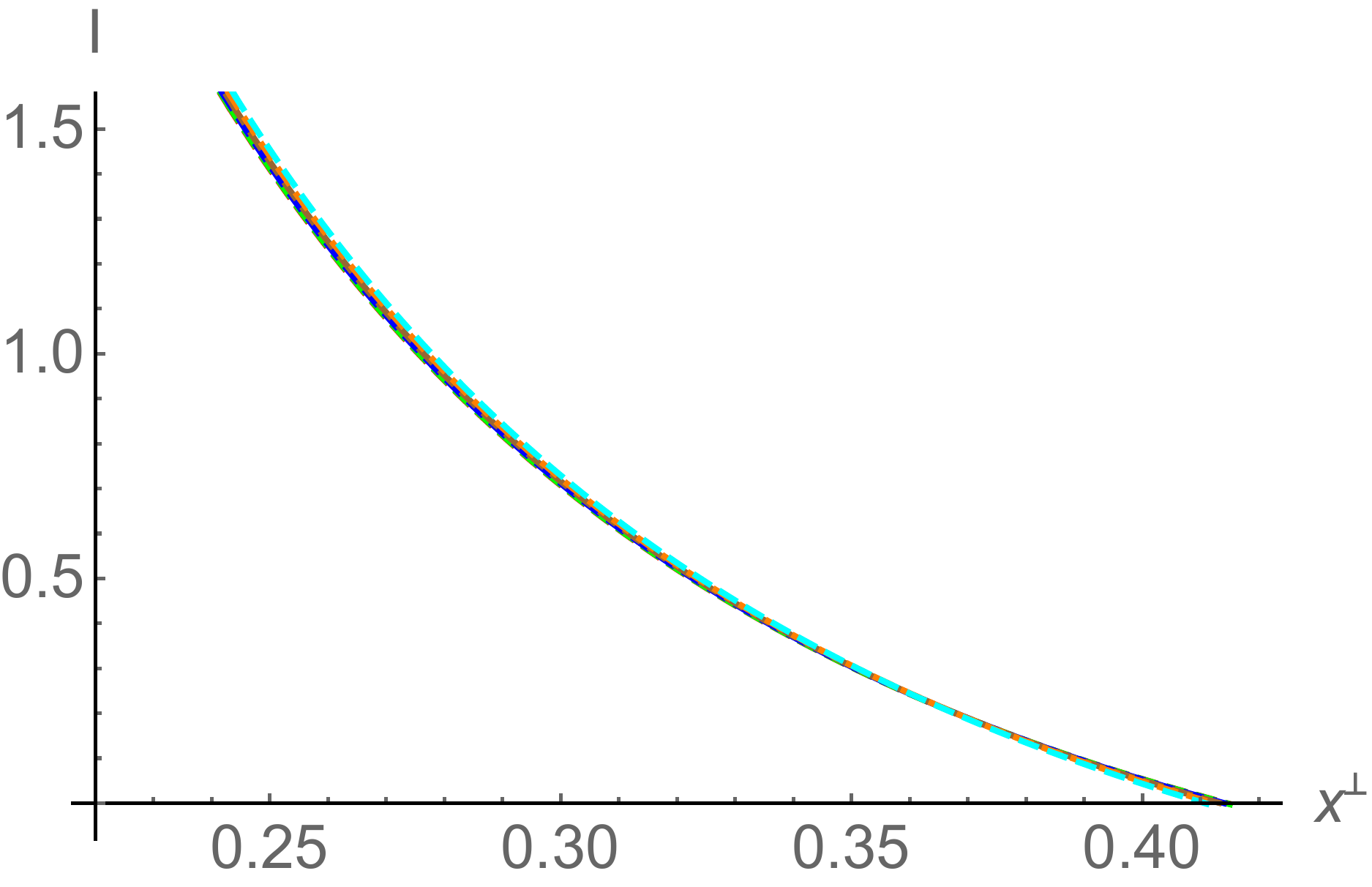}
\caption{Variation of the mutual information with $x^{\perp}$ for different values of $B$. Here $\ell^{\perp}=0.5$ is used, and the red, green, blue, brown, orange, and cyan
curves correspond to $B=0$, $0.1$, $0.2$, $0.3$, $0.4$, and $0.5$, respectively. In units of GeV.}
\label{TAdSperpMIvsx}
\end{minipage}
\end{figure}

We can similarly compute the mutual information. The structure of the mutual information is qualitatively similar to the parallel case [Eq.~(\ref{mutual2strips})]. In particular, it goes to zero in the $S_A$ and $S_D$ phases, whereas it is finite and positive in the $S_B$ and $S_C$ phases. Therefore, it is again of order $N^2$ in the $\{S_B,S_C\}$ phases and is of order $N^0$ in the $\{S_A,S_D\}$ phases. The variation of the mutual information with $\ell^{\perp}$ for different values of $B$ is shown in Fig.~\ref{TAdSperpMIvsl}, where the solid and dashed lines are used to represent the mutual information in the $S_B$ and $S_C$ phases, respectively. We find that it varies smoothly as we move from the $S_B$ phase to the $S_C$ phase (or vice versa) via the $S_B/S_C$ transition line. Further, the mutual information also varies smoothly with $x^{\perp}$ and it goes to zero as the $S_A$ or $S_D$ phase is approached. This is shown in Fig.~~\ref{TAdSperpMIvsx}. This is consistent with the physical expectation that the entanglement between the two subsystems should decrease when they are moved farther apart.

It is interesting to point out that, unlike the entanglement entropy, lattice results for the QCD mutual information are not available yet. These results from holography can have analogous correlations in real QCD, and therefore these results for the mutual information can be treated as a prediction from holography.

\subsection{Entanglement wedge cross-section}
We now discuss the entanglement wedge cross-section $E_W$ in the confining phase. The surface that divides the entanglement wedge, associated with two strip subsystems $A$ and $B$, into two parts can be identified as a vertical flat surface $\Sigma$ from the symmetry consideration. Therefore, for the strip subsystems under consideration, $E_W$ is given by the area of a constant $y_1$ (for the parallel case) or $y_2$ (for the perpendicular case) hypersurface located in the middle of the strips (see Fig.~\ref{ES2equalstrips}).

\subsubsection{Parallel case}
The entanglement wedge cross-section in this case is given by the minimum area of the constant $(y_1,t)$ hypersurface. The induced metric on this hypersurface is
\begin{eqnarray}
& &(ds^{2})_{\Sigma}^{ind}=\frac{L^2e^{2A(z)}}{z^2}\biggl[ \frac{dz^2}{g(z)} + e^{B^2 z^2} \biggl( dy_{2}^2 + dy_{3}^2 \biggr) \biggr]\,,
\label{TAdSEWindparallel}
\end{eqnarray}
from which we obtain the entanglement wedge cross-section as
\begin{eqnarray}
E_{W}^{\parallel}=\frac{\ell_{y_2} \ell_{y_3} L^3}{4 G_{(5)}} \biggl[ \int dz \ \frac{e^{3 A(z)}e^{B^2 z^2}}{z^3\sqrt{g(z)}} \biggr]
\label{TAdSparallelEW}
\end{eqnarray}
From the phase diagram, we can conclude that the entanglement wedge only exists for the $S_B$ and $S_C$ phases, whereas it is zero in the $S_A$ and $S_D$ phases. For the $S_B$ phase, it is given by
\begin{eqnarray}
& & E_{W}^{\parallel}(S_B)=\frac{\ell_{y_2} \ell_{y_3} L^3}{4 G_{(5)}} \biggl[ \int_{z_{*}^{\parallel}(x)}^{z_{*}^{\parallel}(2\ell+x)} dz \ \frac{e^{3 A(z)}e^{B^2 z^2}}{z^3\sqrt{g(z)}} \biggr] \,.
\end{eqnarray}
Interestingly, the above integral can be evaluated explicitly as
\begin{eqnarray}
& & =\frac{\ell_{y_2} \ell_{y_3} L^3}{4 G_{(5)}} \Bigg| \frac{1}{2} \left(B^2-3 a\right) Ei \left[\left(B^2-3 a\right) z^2\right]-\frac{e^{z^2
   \left(B^2-3 a\right)}}{2 z^2}    \Bigg|_{z=z_{*}^{\parallel}(x)}^{z=z_{*}^{\parallel}(2\ell+x)} \,,
\label{TAdSparallelEWSB}
\end{eqnarray}
where $Ei$ is the exponential integral function. Similarly, for the $S_C$ phase, we have
\begin{eqnarray}
& & E_{W}^{\parallel}(S_C)=\frac{\ell_{y_2} \ell_{y_3} L^3}{4 G_{(5)}} \biggl[ \int_{z_{*}^{\parallel}(x)}^{\infty} dz \ \frac{e^{3 A(z)}e^{B^2 z^2}}{z^3\sqrt{g(z)}} \biggr] \nonumber \\
& & = - \frac{\ell_{y_2} \ell_{y_3} L^3}{4 G_{(5)}} \left[ \frac{1}{2} \left(B^2-3 a\right) Ei \left[\left(B^2-3 a\right) z^2\right]-\frac{e^{z^2
   \left(B^2-3 a\right)}}{2 z^2}    \right]_{z=z_{*}^{\parallel}(x)} \,.
\label{TAdSparallelEWSC}
\end{eqnarray}
From the above results, it is clear that both $E_{W}^{\parallel}(S_B)$ and $E_{W}^{\parallel}(S_C)$ are not only positive as $z=z_{*}^{\parallel}(x) \leq z_{*}^{\parallel}(2\ell+x) \leq \infty$, but also UV finite. The analytic expressions of $E_{W}^{\parallel}(S_B)$ and $E_{W}^{\parallel}(S_C)$ further allow us to make several concrete observations about the entanglement wedge in the confining phase without resorting to any numerics. In particular, the difference between $E_{W}^{\parallel}(S_B)-E_{W}^{\parallel}(S_C)$, for the allowed range of the magnetic field
\begin{eqnarray}
& & E_{W}^{\parallel}(S_B)-E_{W}^{\parallel}(S_C)= \frac{\ell_{y_2} \ell_{y_3} L^3}{4 G_{(5)}} \Bigg| \frac{1}{2} \left(B^2-3 a\right) Ei \left[\left(B^2-3 a\right) z^2\right]-\frac{e^{z^2
   \left(B^2-3 a\right)}}{2 z^2}    \Bigg|_{z=z_{*}^{\parallel}(2\ell+x)} \,,
\label{TAdSparallelEWSBSCdiff}
\end{eqnarray}
is always negative and finite at the $S_B/S_C$ transition line (defined by $2\ell^{\parallel}+x^{\parallel}=\ell_{crit}^{\parallel}$). This indicates that, irrespective of the values of the magnetic field, the entanglement wedge cross-section will exhibit a discontinuous behavior at the $S_B/S_C$ transition line. This should be contrasted with the mutual information which behaves smoothly near this transition line. Similarly, since $z_{*}^{\parallel}(x) \neq \infty$, this implies that $E_{W}^{\parallel}(S_C)$ does not go to zero continuously as the $S_C/S_D$ transition line is approached. The same is true for $E_{W}^{\parallel}(S_B)$, as it also does not go to zero when the $S_A/S_B$ transition line is approached [since $z_{*}^{\parallel}(x) \neq z_{*}^{\parallel}(2\ell^\parallel+x^\parallel)$]. Therefore, we clearly see that, unlike the mutual information, the entanglement wedge exhibits discontinuity every time we pass through a transition line in the $\ell^\parallel-x^\parallel$ phase space.

\begin{figure}[h]
\begin{minipage}[b]{0.5\linewidth}
\centering
\includegraphics[width=2.8in,height=2.3in]{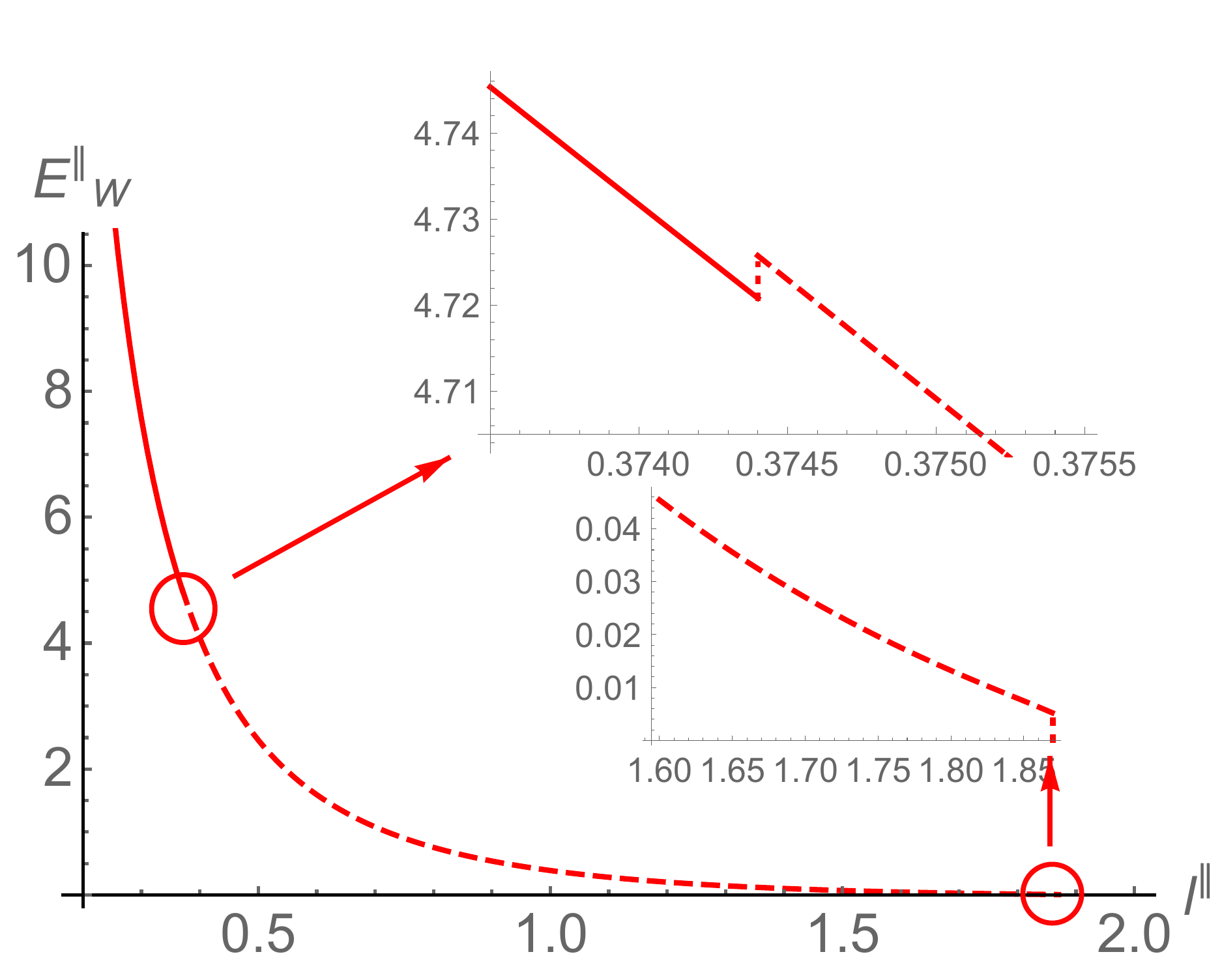}
\caption{$E_{W}^{\parallel}$ as a function of separation length $\ell^{\parallel}$ along a fixed
line $x^\parallel = 0.5 \ell^\parallel$. Here $B=0$ is used. Solid and dashed lines correspond to $E_{W}^{\parallel}$ of
the $S_B$ and $S_C$ phases, respectively. In units of GeV.}
\label{TAdSparaEWvsl2}
\end{minipage}
\hspace{0.4cm}
\begin{minipage}[b]{0.5\linewidth}
\centering
\includegraphics[width=2.8in,height=2.3in]{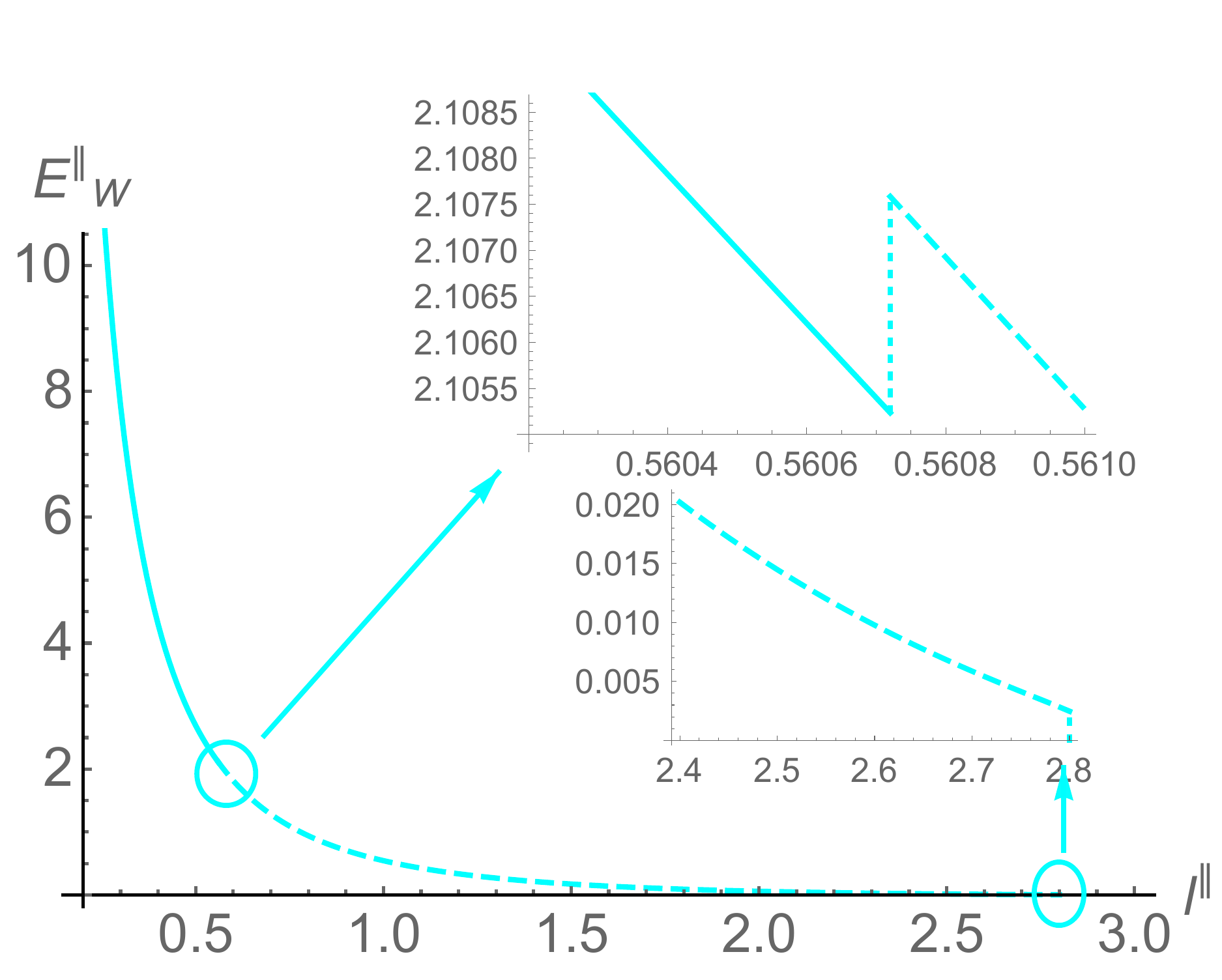}
\caption{$E_{W}^{\parallel}$ as a function of separation length $\ell^{\parallel}$ along a fixed
line $x^\parallel = 0.5 \ell^\parallel$. Here $B=0.5$ is used. Solid and dashed lines correspond to $E_{W}^{\parallel}$ of
the $S_B$ and $S_C$ phases, respectively. In units of GeV.}
\label{TAdSparaEWvsl3}
\end{minipage}
\end{figure}

Further details pertaining to the behavior of $E_{W}^{\parallel}$ are summarized in Figs.~\ref{TAdSparaEWvsl2} and \ref{TAdSparaEWvsl3}  for two different values of $B$. Here, a particular line $x^\parallel = 0.5 \ell^\parallel$ is considered so that the behavior of $E_{W}^{\parallel}$ in the $S_B$, $S_C$, and $S_D$ phases can be probed simultaneously. The solid and dashed lines are used to represent $E_{W}^{\parallel}$ of
the $S_B$ and $S_C$ phases, respectively \footnote{In Figs.~\ref{TAdSparaEWvsl2} and \ref{TAdSparaEWvsl3}, the perfector $\frac{\ell_{y_2} \ell_{y_3} L^3}{2 G_{(5)}}$ is again set to one.}. From the subplots, we clearly see that $E_W^{\parallel}$ becomes discontinuous at the $S_B/S_C$ transition line. Moreover, there is an upward jump in the magnitude of $E_{W}^{\parallel}$  when the $S_B/S_C$ transition line is approached from the $S_B$ phase, i.e., $E_{W}^{\parallel}(S_C)>E_{W}^{\parallel}(S_B)$, indicating that the area of the wedge grows at this transition line. These results are in complete agreement with our analytical analysis. Similarly, $E_{W}^{\parallel}(S_C)$ does not go to zero at the $S_C/S_D$ transition line, indicating that the entanglement wedge cross-section vanishes abruptly for large values of $\ell^\parallel$ and $x^\parallel$. The same results are true for other values of $B$ as well.

\begin{figure}[h]
\begin{minipage}[b]{0.5\linewidth}
\centering
\includegraphics[width=2.8in,height=2.3in]{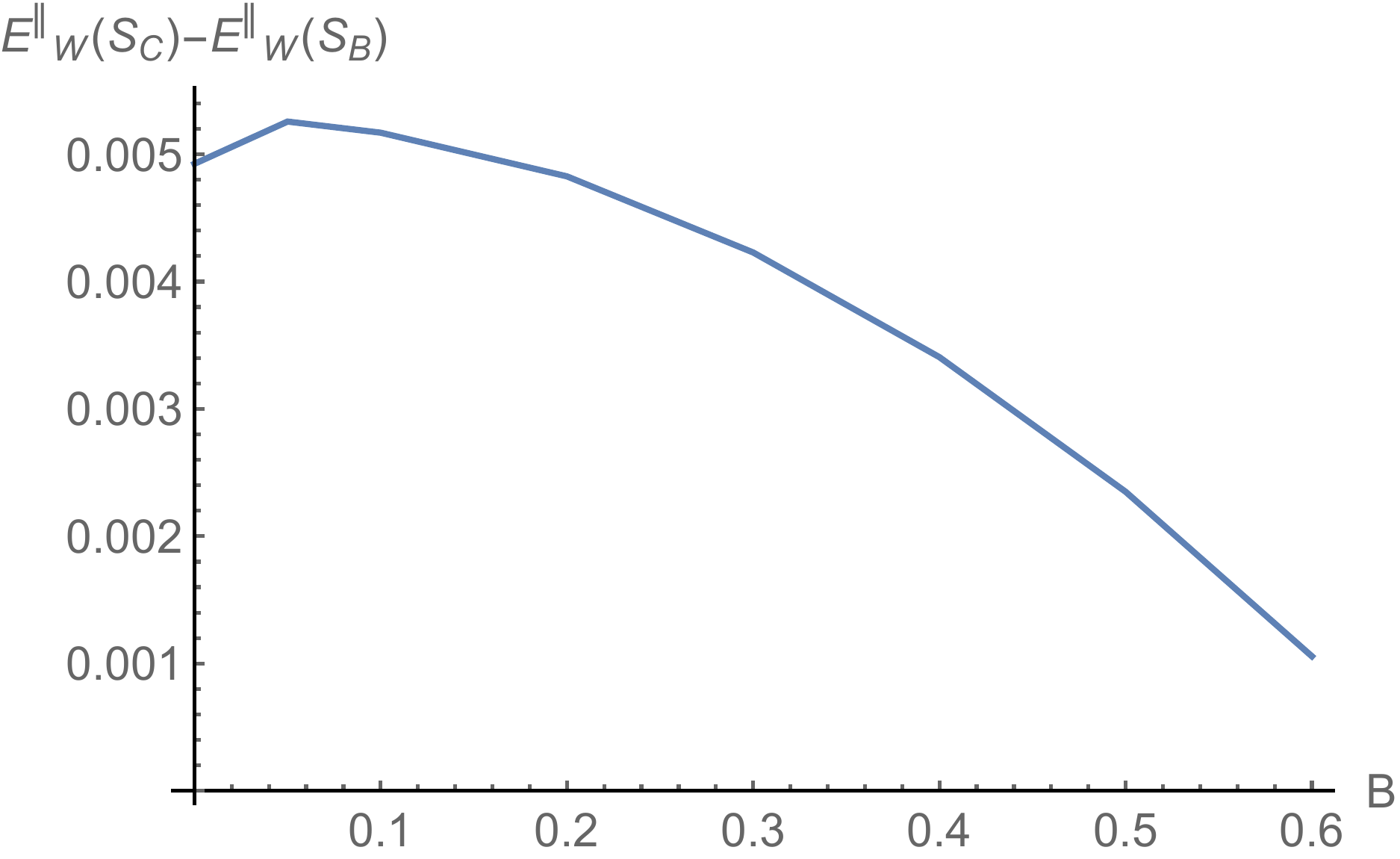}
\caption{Variation of $E_{W}^{\parallel}(S_C)-E_{W}^{\parallel}(S_B)$ with $B$ at the $S_B/S_C$ transition line along a fixed
line $x^\parallel = 0.5 \ell^\parallel$.}
\label{BvsSCSBdiffparallel}
\end{minipage}
\hspace{0.4cm}
\begin{minipage}[b]{0.5\linewidth}
\centering
\includegraphics[width=2.8in,height=2.3in]{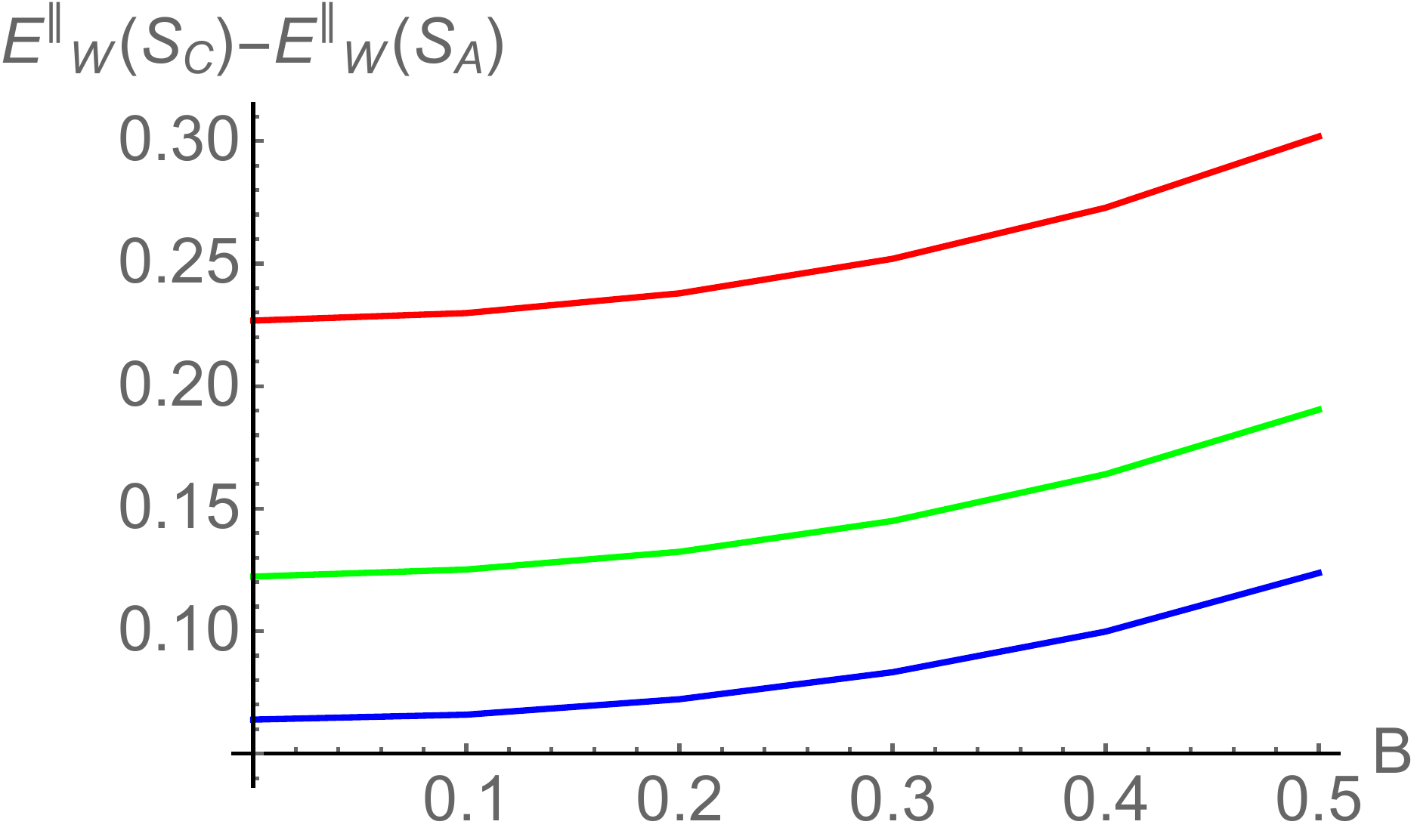}
\caption{Variation of $E_{W}^{\parallel}(S_C)-E_{W}^{\parallel}(S_A)$ with $B$ at the $S_C/S_A$ transition line. Here the red, green, and blue curves correspond to $x^\parallel = 0.5$, $0.6$, and $0.7$, respectively.}
\label{BvsSCSAvsXdiffparallel}
\end{minipage}
\end{figure}

It is also interesting to see how the area of the entanglement wedge changes at the, e.g., $S_B/S_C$ transition line for different values of $B$. This is shown in Fig.~\ref{BvsSCSBdiffparallel}. We see that the difference $E_{W}^{\parallel}(S_C)-E_{W}^{\parallel}(S_B)$ is always positive at the transition point for all values of $B$. However, we further find that this difference decreases with $B$ for relatively large $B$, suggesting a smaller discontinuity in the structure of $E_{W}^{\parallel}$ at this transition line due to $B$. Moreover, the difference $E_{W}^{\parallel}(S_C)-E_{W}^{\parallel}(S_D)$ at the $S_C/S_D$ transition line is found to be exactly similar to the behavior shown in Fig.~\ref{BvsSCSBdiffparallel}. This can again be traced back to the fact that these differences depend only on the critical values $\ell_{crit}^{\parallel}$ ($=x_{crit}^{\parallel}$) at the corresponding transition lines. On the other hand, the difference $E_{W}^{\parallel}(S_C)-E_{W}^{\parallel}(S_A)$ at the $S_A/S_C$ transition line is found to be increasing with $B$ for all values of $x^\parallel$ and $\ell^\parallel$, implying a strengthening of the wedge discontinuity at this transition line with $B$. This is shown in Fig.~\ref{BvsSCSAvsXdiffparallel}.  Overall, we find that $E_{W}^{\parallel}(S_C)$ is a monotonically decreasing function of $x^\parallel$ which abruptly vanishes at $x^\parallel=\ell_{crit}^{\parallel}$.

\begin{figure}[ht]
\centering
\includegraphics[width=2.8in,height=2.3in]{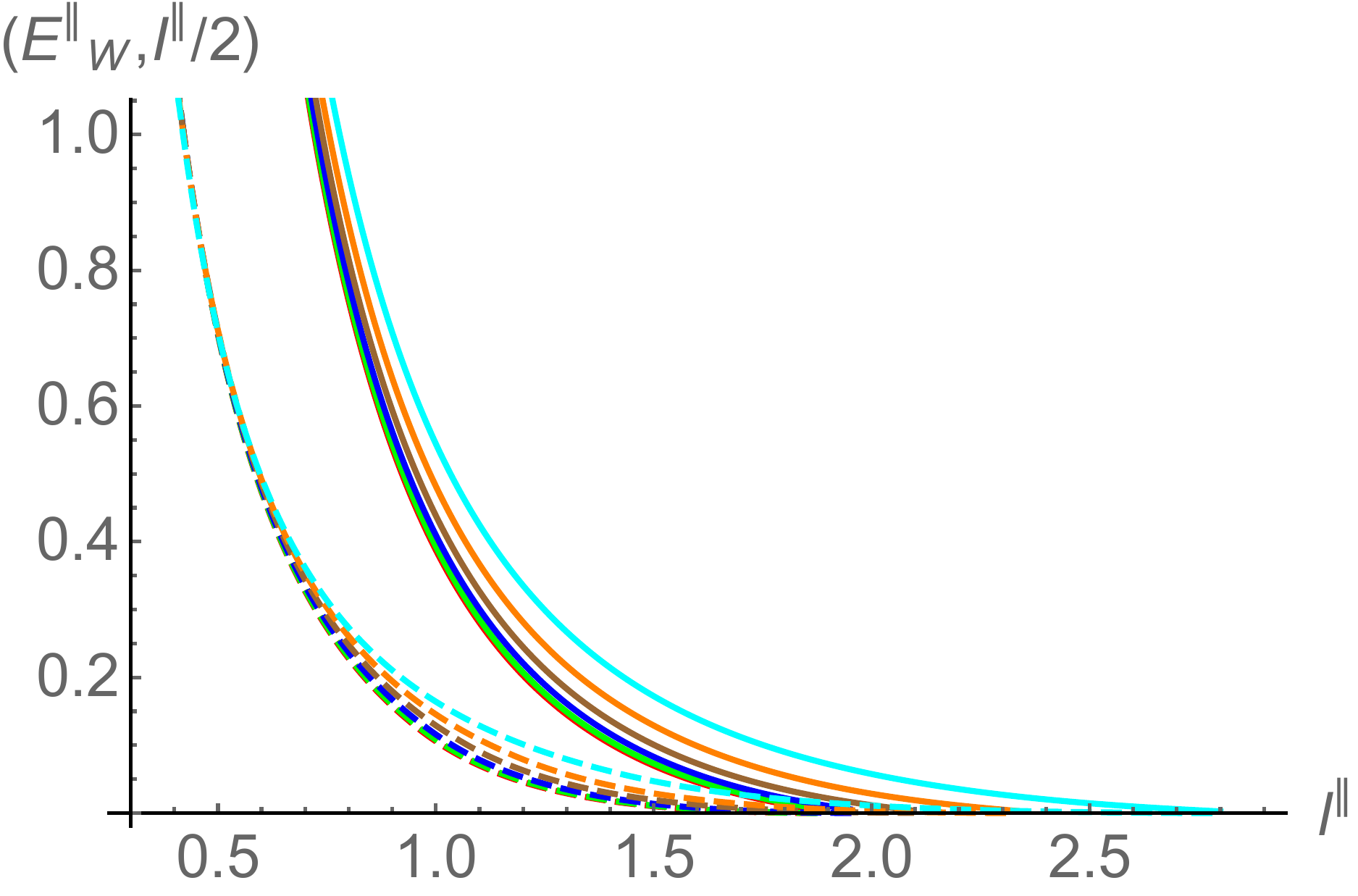}
\caption{Entanglement wedge $E_{W}^{\parallel}$ and mutual information $I^\parallel$ as functions of $\ell^\parallel$ along a fixed line $x^\parallel=0.5\ell^\parallel$. The solid curves
correspond to $E_{W}^{\parallel}$, whereas the dashed curves correspond to $I^\parallel/2$. The red, green, blue, brown, orange, and cyan
curves correspond to $B=0$, $0.1$, $0.2$, $0.3$, $0.4$, and $0.5$, respectively. }
\label{lvsEWandMIby2fordiiffBparallel}
\end{figure}

In holography, it has been suggested that the entanglement wedge always at least exceeds half the mutual information, i.e., $E_{W}^\parallel \geq I^{\parallel}/2$ \cite{Takayanagi:2017knl}. Therefore, it is interesting to check if this inequality is satisfied in the current holographic model. The comparison between the entanglement wedge and mutual information is shown in Fig.~\ref{lvsEWandMIby2fordiiffBparallel} along the line $x^\parallel=0.5\ell^\parallel$ for different values of $B$. We find that, irrespective of the phases involved, this inequality is always satisfied for all values of $B$.

\subsubsection{Perpendicular case}
The computation of the entanglement wedge cross-section $E_{W}^{\perp}$ in the perpendicular direction is completely analogous to the parallel case. In this case, it is given by the minimum area of the constant $(y_2,t)$ hypersurface. The induced metric on this hypersurface is
\begin{eqnarray}
& &(ds^{2})_{\Sigma}^{ind}=\frac{L^2 e^{2A(z)}}{z^2}\biggl[ \frac{dz^2}{g(z)} + dy_{1}^2 + e^{B^2 z^2} \biggl(dy_{3}^2 \biggr) \biggr]\,,
\label{TAdSEWindperp}
\end{eqnarray}
from which the expression of the entanglement wedge cross-section can be obtained as
\begin{eqnarray}
E_{W}^{\perp}=\frac{\ell_{y_1} \ell_{y_3} L^3}{4 G_{(5)}} \biggl[ \int dz \ \frac{e^{3 A(z)}e^{B^2 z^2/2}}{z^3\sqrt{g(z)}} \biggr]\,.
\label{TAdSperpEW}
\end{eqnarray}
The two-strip phase diagram of the perpendicular case again tells us that the nontrivial entanglement wedge can exist only in the $S_B$ and $S_C$ phases. For the $S_B$ phase, we have
\begin{eqnarray}
& & E_{W}^{\perp}(S_B)=\frac{\ell_{y_1} \ell_{y_3} L^3}{4 G_{(5)}} \biggl[ \int_{z_{*}^{\perp}(x)}^{z_{*}^{\perp}(2\ell+x)} dz \ \frac{e^{3 A(z)}e^{B^2 z^2/2}}{z^3\sqrt{g(z)}} \biggr] \nonumber\\
& & = \frac{\ell_{y_1} \ell_{y_3} L^3}{4 G_{(5)}} \Bigg| \frac{1}{4} \left(B^2-6 a\right) Ei\left[\frac{1}{2} \left(B^2-6 a\right)
   z^2\right]-\frac{e^{\frac{1}{2} z^2 \left(B^2-6 a\right)}}{2 z^2} \Bigg|_{z=z_{*}^{\perp}(x)}^{z=z_{*}^{\perp}(2\ell^\perp+x^\perp)} \,,
\label{TAdSperpEWSB}
\end{eqnarray}
whereas for $S_C$ it is given by
\begin{eqnarray}
& & E_{W}^{\perp}(S_C)=\frac{\ell_{y_1} \ell_{y_3} L^3}{4 G_{(5)}} \biggl[ \int_{z_{*}^{\perp}(x)}^{\infty} dz \ \frac{e^{3 A(z)}e^{B^2 z^2/2}}{z^3\sqrt{g(z)}} \biggr] \nonumber\\
& & = - \frac{\ell_{y_1} \ell_{y_3} L^3}{4 G_{(5)}} \Bigg| \frac{1}{4} \left(B^2-6 a\right) Ei\left[\frac{1}{2} \left(B^2-6 a\right)
   z^2\right]-\frac{e^{\frac{1}{2} z^2 \left(B^2-6 a\right)}}{2 z^2} \Bigg|_{z=z_{*}^{\perp}(x)} \,.
\label{TAdSperpEWSC}
\end{eqnarray}

\begin{figure}[h]
\begin{minipage}[b]{0.5\linewidth}
\centering
\includegraphics[width=2.8in,height=2.3in]{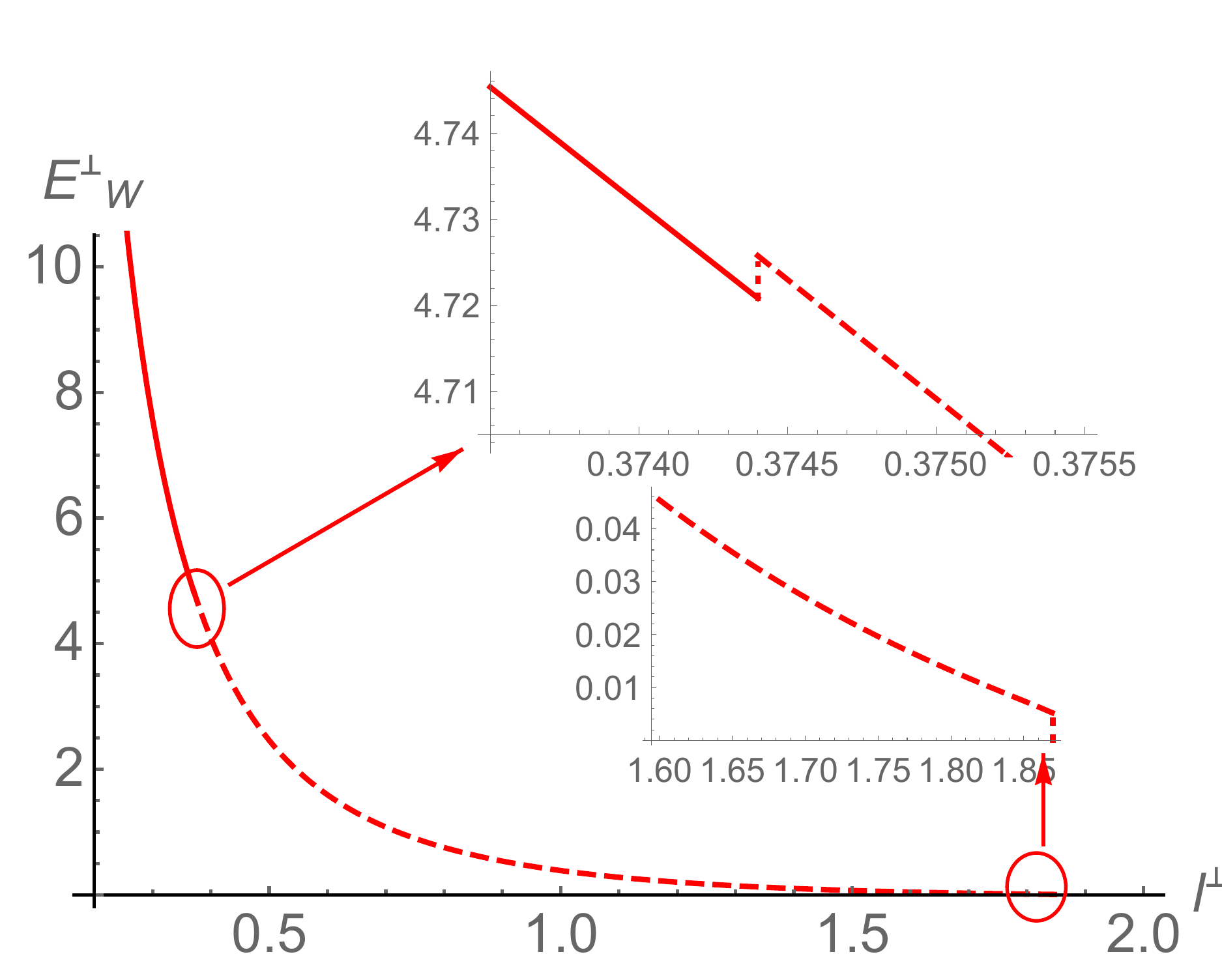}
\caption{$E_{W}^{\perp}$ as a function of separation length $\ell^{\perp}$ along a fixed
line $x^\perp = 0.5 \ell^\perp$. Here $B=0$ is used. Solid and dashed lines correspond to $E_{W}^{\perp}$ of
the $S_B$ and $S_C$ phases, respectively. In units of GeV.}
\label{TAdSperpEWvsl2}
\end{minipage}
\hspace{0.4cm}
\begin{minipage}[b]{0.5\linewidth}
\centering
\includegraphics[width=2.8in,height=2.3in]{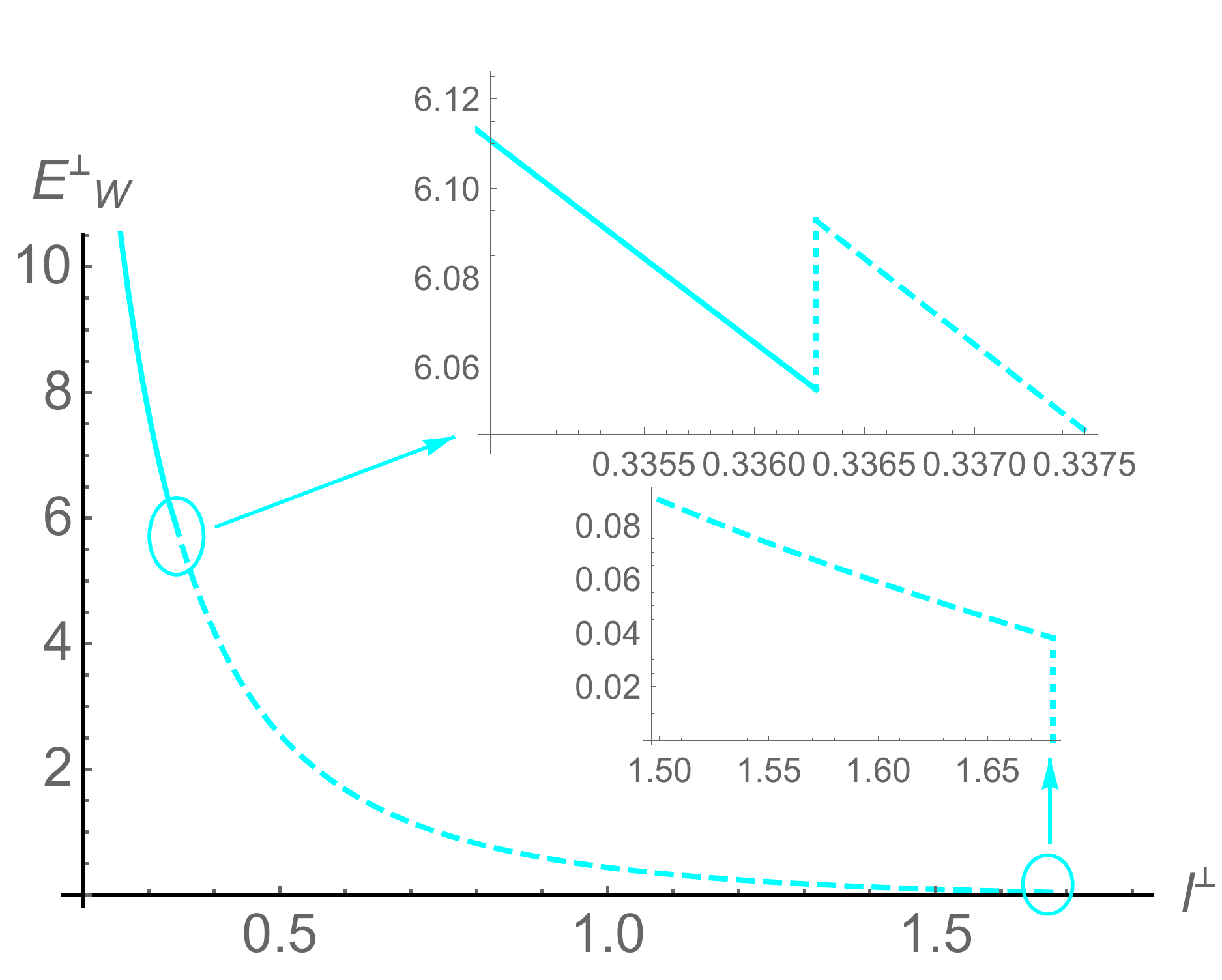}
\caption{$E_{W}^{\perp}$ as a function of separation length $\ell^{\perp}$ along a fixed
line $x^\perp = 0.5 \ell^\perp$. Here $B=0.5$ is used. Solid and dashed lines correspond to $E_{W}^{\perp}$ of
the $S_B$ and $S_C$ phases, respectively. In units of GeV.}
\label{TAdSperpEWvsl3}
\end{minipage}
\end{figure}
The variation of $E_{W}^{\perp}$ with $\ell^{\perp}$ along the line $x^\perp=0.5\ell^\perp$ for two different values of $B$ is shown in Figs.~\ref{TAdSperpEWvsl2} and \ref{TAdSperpEWvsl3}.
We find that the behavior of $E_{W}^{\perp}$ is qualitatively similar to the parallel case. In particular, $E_{W}^{\perp}$ again behaves discontinuously at the $S_B/S_C$ transition line. This can be seen
mathematically from Eqs.~(\ref{TAdSperpEWSB}) and (\ref{TAdSperpEWSC}), where the condition  $z_{*}^{\perp}(2\ell^\perp+x^\perp)\neq\infty$ ensures that $E_{W}^{\perp}(S_B)$ and $E_{W}^{\perp}(S_C)$ do not attain the same value at the $S_B/S_C$ transition line. Moreover, the entanglement wedge does not vanish smoothly as the $S_A$ (or $S_D$) phase is approached from the $S_B$ (or $S_C$) phase. This result can again be traced back to the fact that $z_{*}^{\perp}(x^\perp) \neq z_{*}^{\perp}(2\ell^\perp+x^\perp)\neq  \infty$. Accordingly, we find that the entanglement wedge is a monotonic function of $x^\perp$, which vanishes discontinuously at $x^\perp=\ell_{crit}^{\perp}$.  Therefore, like in the parallel case, the entanglement wedge exhibits discontinuity each time a phase transition between different phases occurs in the perpendicular case as well.

\begin{figure}[h]
\begin{minipage}[b]{0.5\linewidth}
\centering
\includegraphics[width=2.8in,height=2.3in]{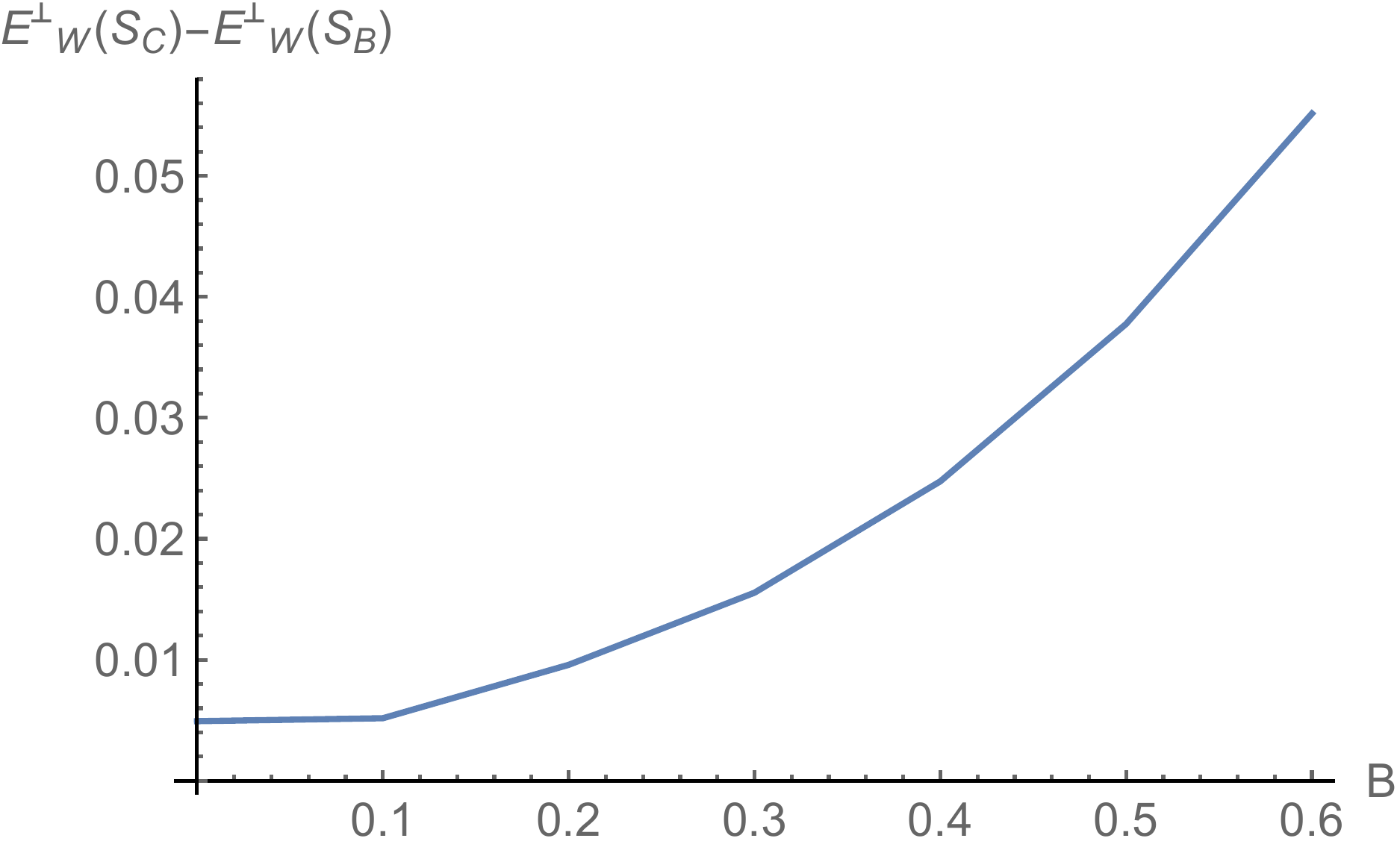}
\caption{Variation of $E_{W}^{\perp}(S_C)-E_{W}^{\perp}(S_B)$ with $B$ at the $S_B/S_C$ transition line along a fixed
line $x^\perp = 0.5 \ell^\perp$.}
\label{BvsSCSBdiffperpdicular}
\end{minipage}
\hspace{0.4cm}
\begin{minipage}[b]{0.5\linewidth}
\centering
\includegraphics[width=2.8in,height=2.3in]{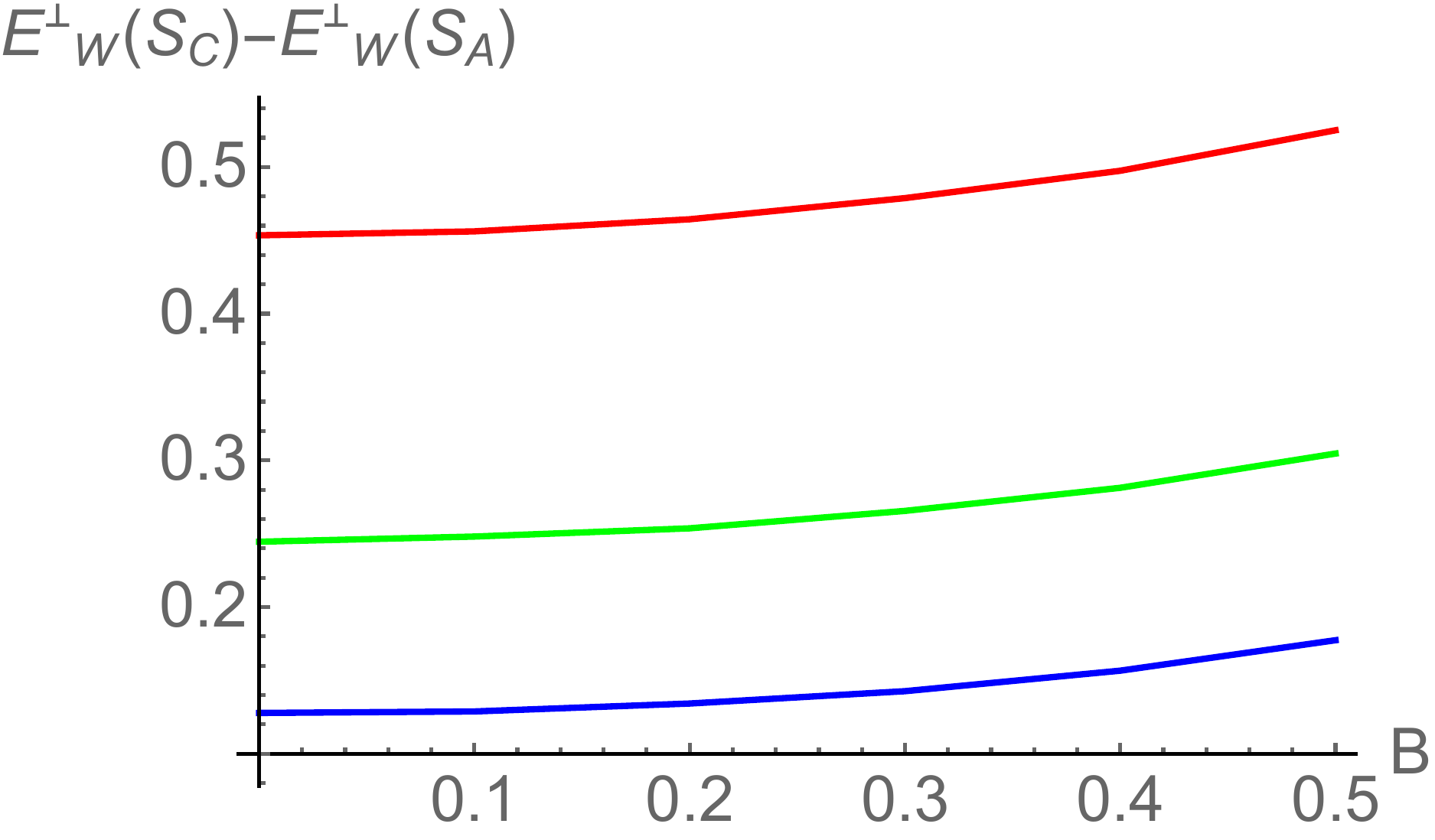}
\caption{Variation of $E_{W}^{\perp}(S_C)-E_{W}^{\perp}(S_A)$ with $B$ at the $S_C/S_A$ transition line. Here the red, green, and blue curves correspond to $x^\perp = 0.5$, $0.6$, and $0.7$, respectively.}
\label{BvsSCSAvsXdiffperpendicular}
\end{minipage}
\end{figure}

We can further analyze how much the area of the entanglement wedge changes at the transition point in the perpendicular case. At the $S_B/S_C$ transition line, this is shown in Fig.~\ref{BvsSCSBdiffperpdicular}. This can be compared with Fig.~\ref{BvsSCSBdiffparallel} of the parallel case.  We find that the difference between $E_{W}^{\perp}(S_C)-E_{W}^{\perp}(S_B)$ is always positive [since $z_{*}^{\perp}(x^\perp) < z_{*}^{\perp}(2\ell^\perp+x^\perp)$], suggesting an increment in the area of the entanglement wedge at the transition point. This result is similar to the parallel case. However, in contrast to the parallel case, the difference $E_{W}^{\perp}(S_C)-E_{W}^{\perp}(S_B)$ increases with $B$. This points to a larger discontinuity in the entanglement wedge cross-section at the $S_B/S_C$ transition point with $B$ in the perpendicular direction. Similarly, the difference $E_{W}^{\perp}(S_C)-E_{W}^{\perp}(S_A)$ at the $S_A/S_C$ transition line is found to be an increasing function of $B$ for all values of $x^\perp$ and $\ell^\perp$. This behavior is quite similar to the parallel case, though the magnitude of the difference is slightly higher now. This is shown in Fig.~\ref{BvsSCSAvsXdiffperpendicular}.

\begin{figure}[ht]
\centering
\includegraphics[width=2.8in,height=2.3in]{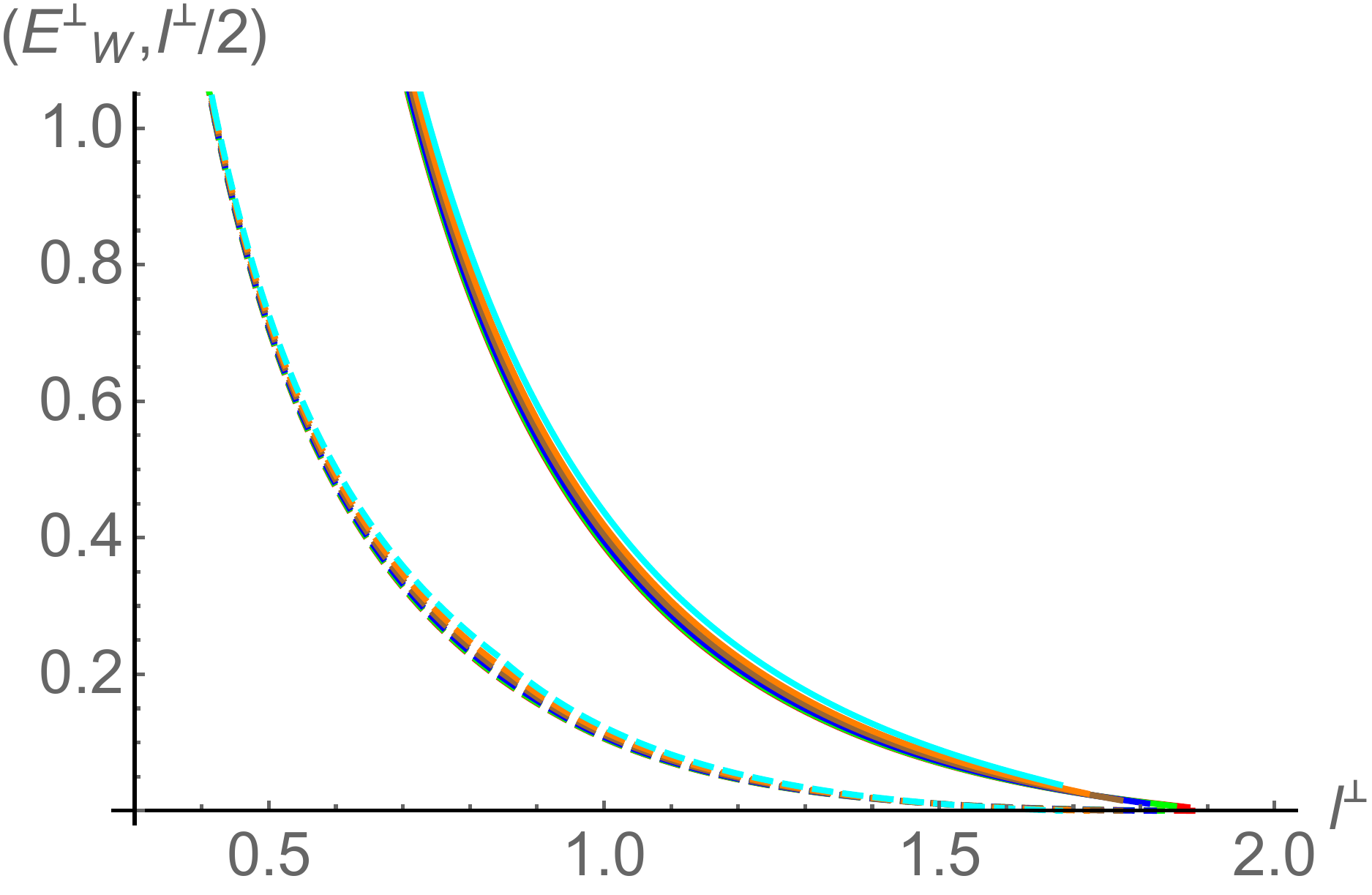}
\caption{Entanglement wedge $E_{W}^{\perp}$ and mutual information $I^\perp$ as functions of $\ell^\perp$ along a fixed line $x^\perp=0.5\ell^\perp$. The solid curves
correspond to $E_{W}^{\perp}$, whereas the dashed curves correspond to $I^\perp/2$. The red, green, blue, brown, orange, and cyan
curves correspond to $B=0$, $0.1$, $0.2$, $0.3$, $0.4$, and $0.5$, respectively. }
\label{lvsEWandMIby2fordiiffBperp}
\end{figure}

We further test the inequality $E_{W}^{\perp}\geq I^\perp/2$ in the perpendicular case. The results are shown in Fig.~\ref{lvsEWandMIby2fordiiffBperp}. We find that this inequality is again satisfied everywhere in the $\ell^\perp-x^\perp$ plane for all values of $B$. The inequality saturates only at the critical points, at which $I^\perp/2$ continuously goes to zero, whereas $E_{W}^{\perp}$ exhibits a sharp drop to zero.

From the above analysis, we see that the entanglement wedge not only exhibits nontrivial features each time a phase transition between different phases occurs but also is sensitive to the orientation of the magnetic field. This is an important result considering that the entanglement wedge has been suggested as the holographic dual of many mixed-state entanglement measures. Therefore, our whole analysis suggests that nontrivial and anisotropic features are expected in these measures in the presence of a magnetic field, especially in the confined phase.

\subsection{Holographic entanglement negativity}
We now study the holographic entanglement negativity in the confined phase. We begin with the single-interval case. This is given in Eq.~(\ref{HEEsc1}). In the limit $B\rightarrow A^c\rightarrow\infty$, the disconnected entropy dominates for both the parallel and perpendicular cases (see Fig.~\ref{entanglenegativitypicture} for more details). So for both cases, we have
\begin{eqnarray}\label{HEETAdS}
& & \mathcal{E} = \lim_{B\rightarrow A^c}\frac{3}{4}\left[2S(A)+S(B_1)+S(B_2)-S(A\cup B_1)-S(A\cup B_2)\right]\,, \nonumber \\
& & \mathcal{E} = \frac{3}{2} S(A) \,.
\end{eqnarray}
This is an interesting new result implying that the holographic entanglement negativity is just $3/2$ times the entanglement entropy in the single-interval case. This suggests that the entanglement negativity is also discontinuous at the critical lengths $\ell_{crit}^{\parallel}$ ($\ell_{crit}^{\perp}$) for the parallel (perpendicular) case. Therefore, the entanglement negativity also undergoes an order change from
$\mathcal{O}(N^2)$ to $\mathcal{O}(N^0)$ at these critical lengths. For instance, for the parallel case, we have
\begin{equation}
\begin{split}
\frac{\partial\mathcal{E}^\parallel}{\partial \ell^\parallel} = \mathcal{O}(N^2)\ \mathrm{for}\ \ell^\parallel<\ell_{crit}^\parallel \,, \\
\frac{\partial\mathcal{E}^\parallel}{\partial \ell^\parallel} = \mathcal{O}(N^0)\ \mathrm{for}\ \ell^\parallel>\ell_{crit}^\parallel\,,
\end{split}
\end{equation}
with similar results for the perpendicular case. The discontinuous aspect of $\mathcal{E}$
in the confined phase is an interesting new result and a prediction from holography (strictly speaking, a prediction from the entanglement negativity proposal of \cite{Chaturvedi:2016rft,Chaturvedi:2016rcn}) and should be amenable for independent testing. Here, we further find that this discontinuous behavior of $\mathcal{E}$ in the confined phase persists in the presence of a magnetic field as well. Moreover, the direction and $B$ dependence of the critical lengths associated with the negativity remain the same as that illustrated in Fig.~\ref{BvsLcritPerpendicular}, implying that the magnetic field induces orientation-dependent features in this particular entanglement measure as well.

We now proceed to discuss the holographic entanglement negativity when we have two disjoint intervals \cite{Malvimat:2018txq,Basak:2020bot}. In comparison to \cite{Malvimat:2018txq,Basak:2020bot}, in our case we have $l_s = x\ ,\ l_1=l_2=\ell$. Hence,
$\mathcal{E}$ is expressed as
\begin{equation}
\mathcal{E}=\frac{3}{4}\left[S(\ell+x)+S(\ell+x)-S(2\ell+x)-S(x)\right] \,,
\label{ENtwostripTAdS}
\end{equation}
wherein $S$ denotes the holographic entanglement entropy for a single interval. If $x>\ell_{crit}$, then $\mathcal{E}=0$ in Eq.~(\ref{ENtwostripTAdS}) as all terms are now dominated by the disconnected entropy $S_{discon}$.  This implies that, just like the mutual information and entanglement wedge, $\mathcal{E}$ is zero in the $S_D$ phase as well. This is true for both the parallel and perpendicular cases. However, as we will see shortly, the entanglement negativity does not vanish in the $S_A$ phase.

\subsubsection{Negativity for two strips in the parallel direction}
\begin{figure}[h]
\begin{minipage}[b]{0.5\linewidth}
\centering
\includegraphics[width=2.8in,height=2.3in]{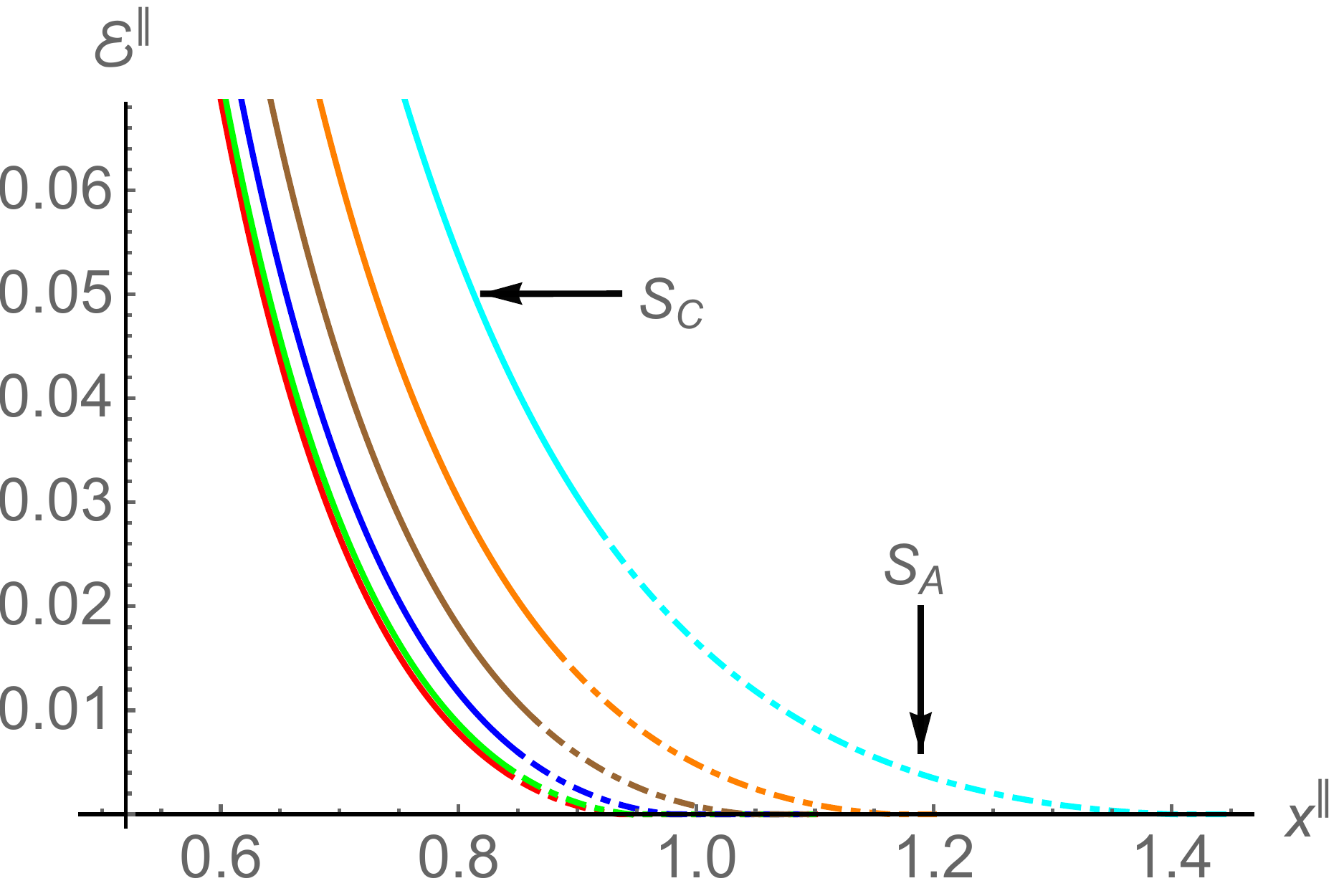}
\caption{$\mathcal{E}^\parallel$ as a function of length $x^{\parallel}$  for different values of $B$. Here $\ell^{\parallel}=0.8$ is used. The dot-dashed and solid lines correspond to $\mathcal{E}^\parallel$ of
the $S_A$ and $S_C$ phases, respectively. The red, green, blue, brown, orange, and cyan
curves correspond to $B=0$, $0.1$, $0.2$, $0.3$, $0.4$, and $0.5$, respectively. In units of GeV.}
\label{TAdSparallelENvsx}
\end{minipage}
\hspace{0.4cm}
\begin{minipage}[b]{0.5\linewidth}
\centering
\includegraphics[width=2.8in,height=2.3in]{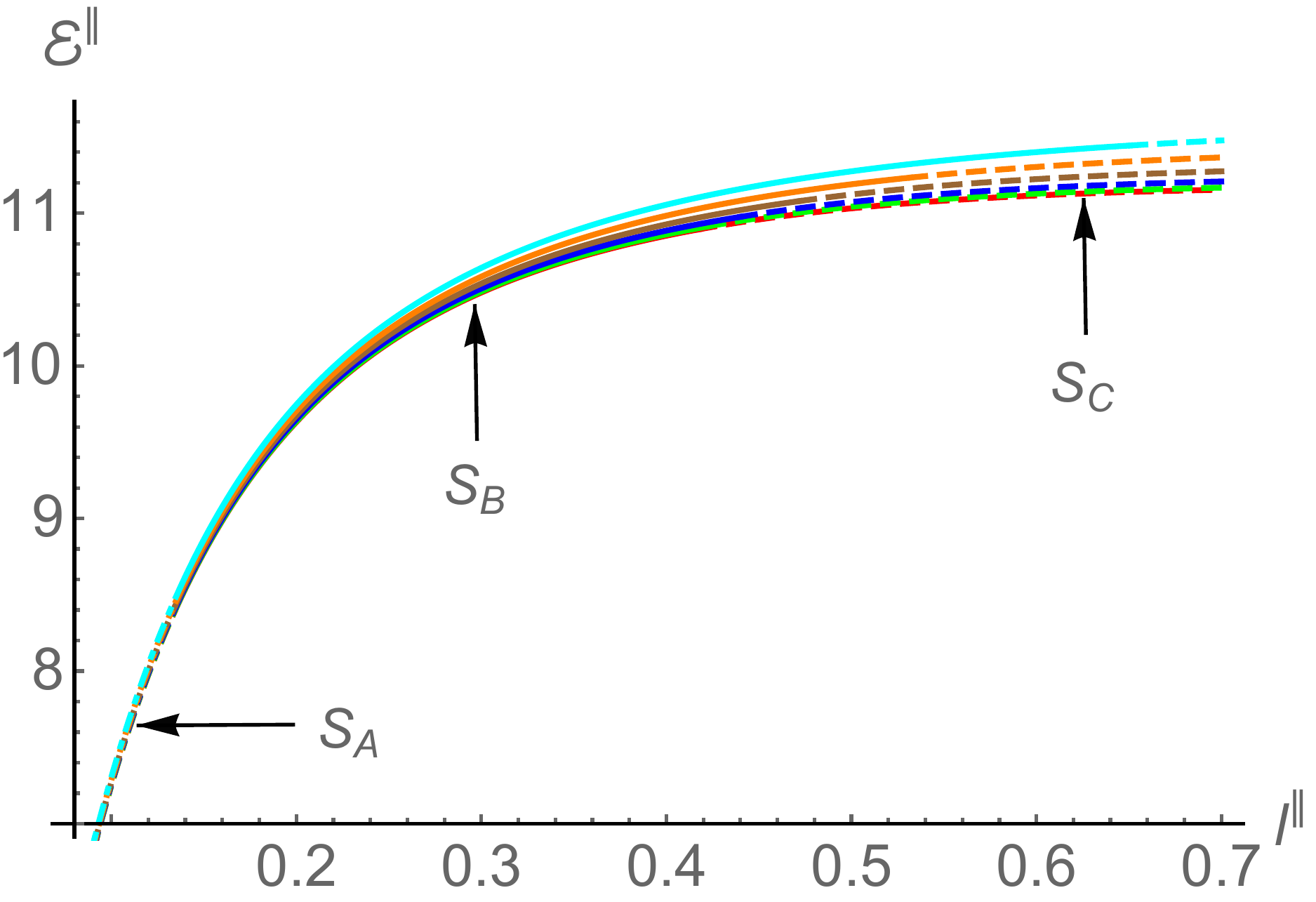}
\caption{$\mathcal{E}^\parallel$ as a function of length $\ell^{\parallel}$  for different values of $B$. Here $x^{\parallel}=0.1$ is used. The dot-dashed, solid, and dashed lines correspond to $\mathcal{E}^\parallel$ of
the $S_A$, $S_B$, and $S_C$ phases, respectively. The red, green, blue, brown, orange, and cyan
curves correspond to $B=0$, $0.1$, $0.2$, $0.3$, $0.4$, and $0.5$, respectively. In units of GeV.}
\label{TAdSparallelENvsl}
\end{minipage}
\end{figure}

The variation of $\mathcal{E}^\parallel$ with $x^\parallel$ for two strips is shown in Fig.~\ref{TAdSparallelENvsx}. Here $\ell^\parallel=0.8$ is used for illustration, but similar results exist for other values of $\ell^\parallel$ as well. We find that $\mathcal{E}^\parallel$ varies monotonically with $x^\parallel$ and smoothly approaches zero at $x^\parallel=\ell_{crit}^\parallel$. In particular, as is expected, the negativity decreases as the two subsystems are taken further and further apart, and eventually vanishes. An interesting result to note is that $\mathcal{E}^\parallel$ is finite in some parts of the $S_A$ phase. This is in sharp contrast to the behavior of the mutual information and entanglement wedge, which was zero everywhere in the $S_A$ phase. Only when $x^\parallel \geq \ell_{crit}^\parallel$ the negativity goes to zero in the $S_A$ phase.

We further find that $\mathcal{E}^\parallel$ also varies monotonically with $\ell^\parallel$. This is shown in Fig.~\ref{TAdSparallelENvsl}. Here we use a fixed $x^\parallel=0.1$ line such that all three phases can be simultaneously probed. We observe that as we increase $B$, the value of $\mathcal{E}^\parallel$ increases for all three phases $\{S_A,S_B,S_C\}$. We find that the negativity first increases as the size of the subsystems increases and then saturates to a $B$-dependent constant value. This $B$-dependent constant value, in particular, increases as $B$ increases. Moreover,  our analysis further suggests that, unlike the entanglement wedge, the entanglement negativity behaves smoothly across various phase transition lines and there is no discontinuity in its structure.

\subsubsection{Negativity for two strips in the perpendicular direction}
\begin{figure}[h]
\begin{minipage}[b]{0.5\linewidth}
\centering
\includegraphics[width=2.8in,height=2.3in]{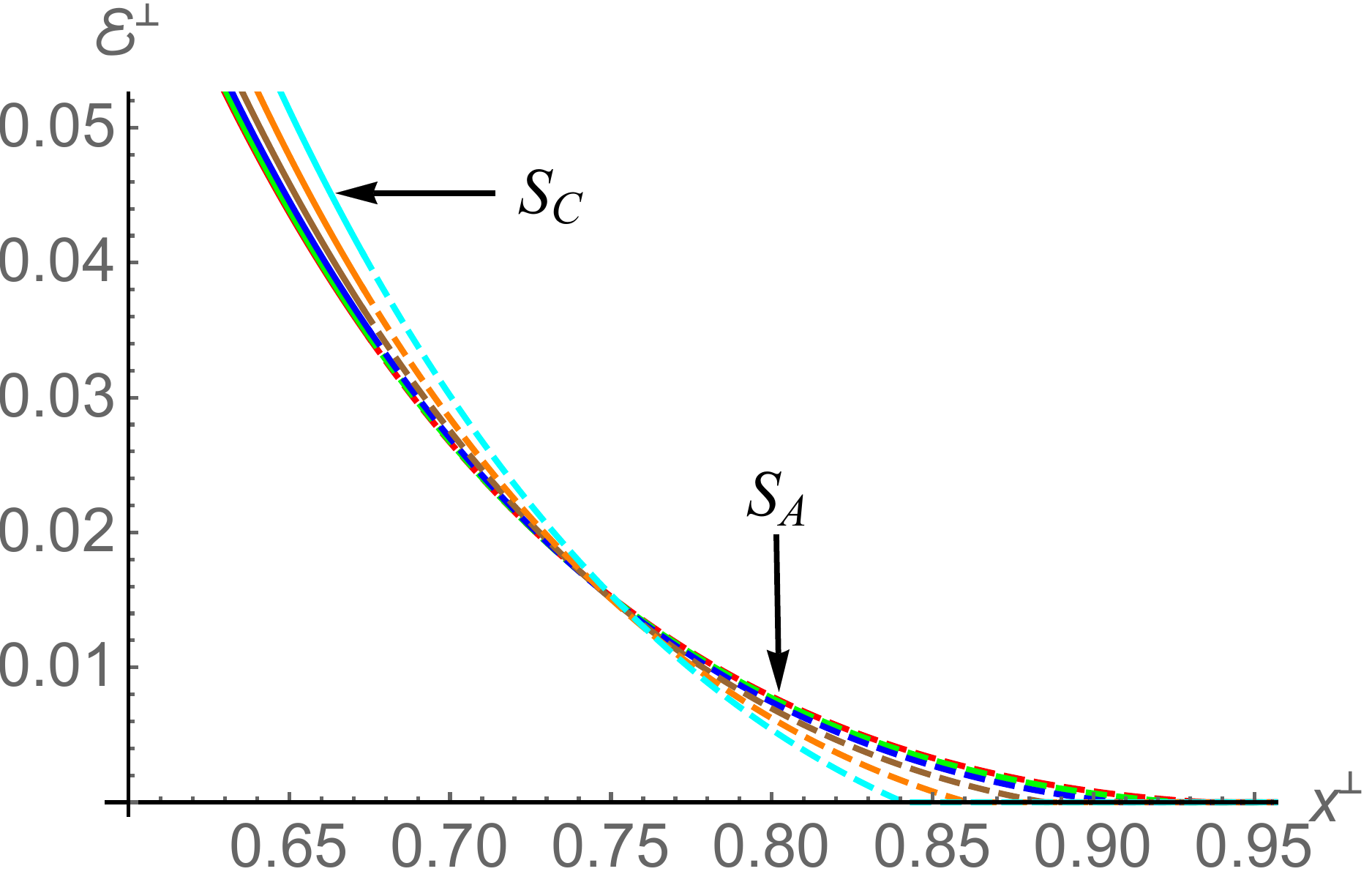}
\caption{$\mathcal{E}^\perp$ as a function of length $x^{\perp}$  for different values of $B$. Here $\ell^{\perp}=0.7$ is used. The dot-dashed and solid lines correspond to $\mathcal{E}^\perp$ of
the $S_A$ and $S_C$ phases, respectively. The red, green, blue, brown, orange, and cyan
curves correspond to $B=0$, $0.1$, $0.2$, $0.3$, $0.4$, and $0.5$, respectively. In units of GeV.}
\label{TAdSperpENvsx}
\end{minipage}
\hspace{0.4cm}
\begin{minipage}[b]{0.5\linewidth}
\centering
\includegraphics[width=2.8in,height=2.3in]{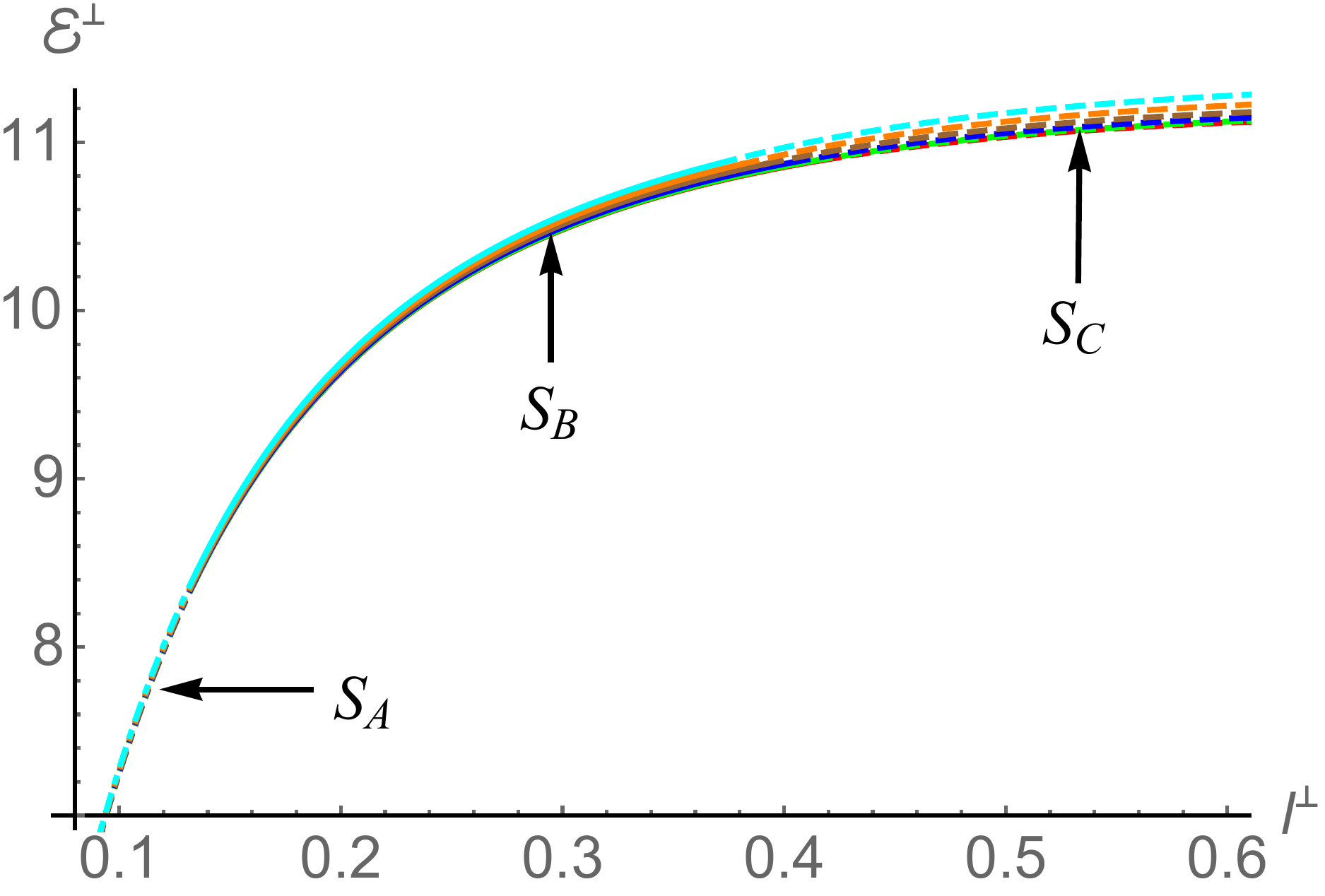}
\caption{$\mathcal{E}^\perp$ as a function of length $\ell^{\perp}$  for different values of $B$. Here $x^{\perp}=0.1$ is used. The dot-dashed, solid, and dashed lines correspond to $\mathcal{E}^\perp$ of
the $S_A$, $S_B$, and $S_C$ phases, respectively. The red, green, blue, brown, orange, and cyan
curves correspond to $B=0$, $0.1$, $0.2$, $0.3$, $0.4$, and $0.5$, respectively. In units of GeV.}
\label{TAdSperpENvsl}
\end{minipage}
\end{figure}
The negativity results for two strips in the perpendicular direction are shown in Figs.~\ref{TAdSperpENvsx} and \ref{TAdSperpENvsl}. The results are again qualitatively similar to the parallel case. The negativity again decreases monotonically with separation size and only goes to zero at the critical separation length $x^\perp=\ell_{crit}^{\perp}$, and thus it is again nonzero in some parts of the $S_A$ phase. We further observe that as we increase $B$ along the perpendicular direction, the value of $\mathcal{E}^\perp$ initially increases and then decreases. In particular, the negativity always increases with $B$ in the $S_C$ phase; however, in the $S_A$ phase it increases with $B$ for small $x^\perp$, whereas it decreases with $B$ near $x_{crit}^\perp$. This behavior is different from the parallel case wherein only the increment in negativity was observed. Similarly, we observe that $\mathcal{E}^\perp$ first monotonically increases with $\ell^\perp$ and then saturates to a $B$-dependent constant value. This $B$-dependent constant value, like in the parallel case, increases with $B$. Importantly, $\mathcal{E}^\perp$ is again continuous across various phase transitions.

We end this section by making a few observations about the entanglement negativity. As mentioned in the last section, there are two different holographic proposals for the entanglement negativity. In the first proposal \cite{Kudler-Flam:2018qjo,Kusuki:2019zsp}, the negativity is proportional to the entanglement wedge (neglecting the quantum correction term). Since the entanglement wedge is zero in the $S_A$ and $S_D$ phases, this suggests that the negativity, if computed using the proposal of \cite{Kudler-Flam:2018qjo,Kusuki:2019zsp}, would also be zero in these phases. However, as discussed above, the second proposal of \cite{Chaturvedi:2016rft,Chaturvedi:2016rcn}, gives a nonzero negativity in some parts of the $S_A$ phase. Therefore, as far as the negativity for two strips in the confined phase is concerned, these two proposals seem to provide inequivalent results. It should be mentioned that both of these proposals have been tested for conformal field theories and have independently reproduced exact known results for the negativity. Therefore, our results provide the first counterexample where the disparity between these two proposals is observed. Also, as we will see shortly, a similar feature is present for all values of the magnetic field and temperature in the deconfined phase, suggesting that the proposal of \cite{Chaturvedi:2016rft,Chaturvedi:2016rcn} points to some kind of universality in the structure of the entanglement negativity.

\section{Deconfining phase}
\label{adsblackholephase}
Having thoroughly discussed the various holographic entanglement measures in the confined phase, we now proceed to discuss them in the finite-temperature deconfined phase. This corresponds to having a black hole on the dual gravity side. Apart from the magnetic field, we also have another parameter, i.e., temperature, in the theory. There is again an option of aligning the strip subsystems parallel or perpendicular to the magnetic field.

\subsection{Holographic entanglement entropy}
\begin{figure}[h]
\begin{minipage}[b]{0.5\linewidth}
\centering
\includegraphics[width=2.8in,height=2.3in]{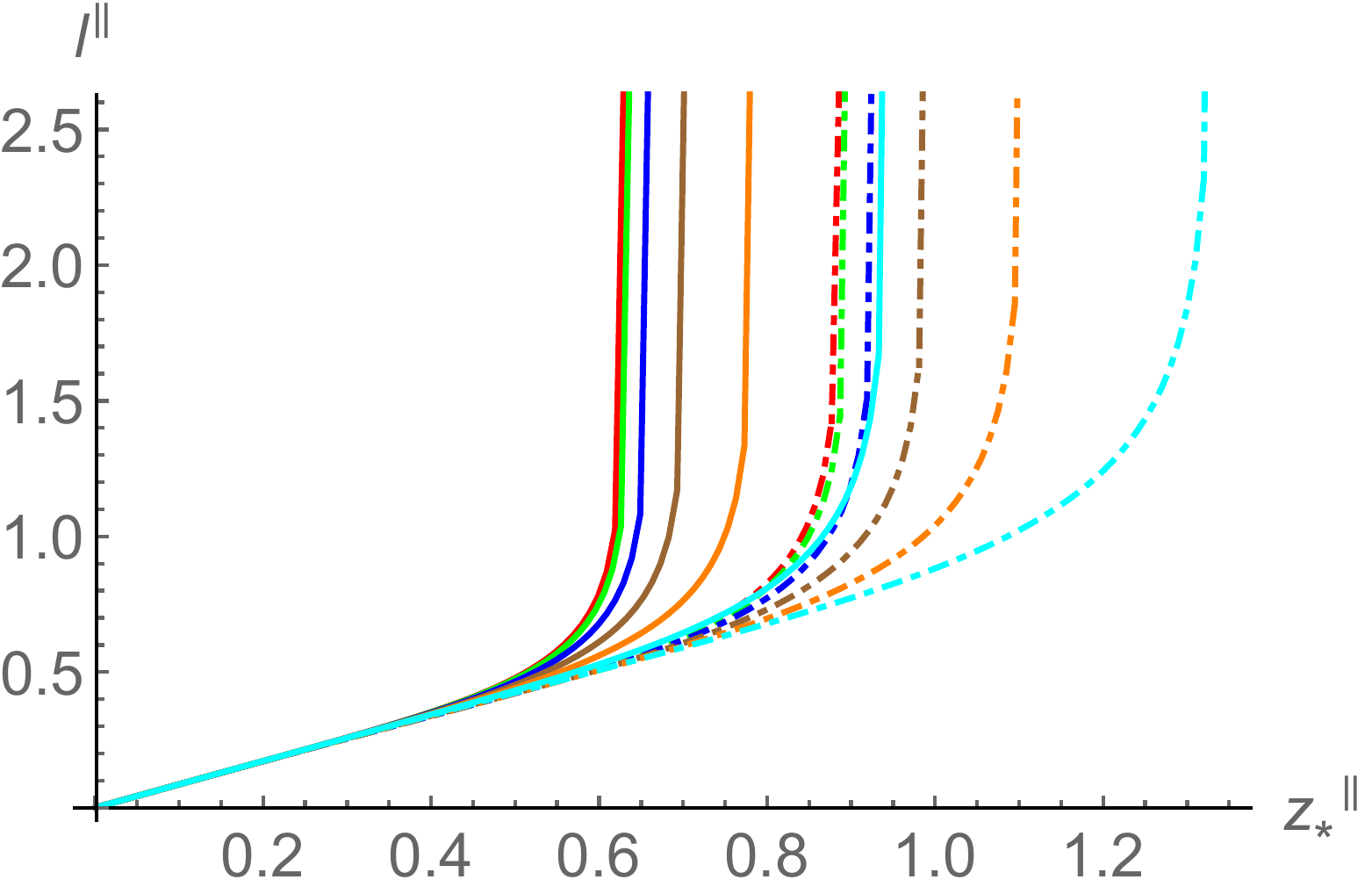}
\caption{$\ell^{\parallel}$ as a function of $z_{*}^{\parallel}$ for different values of magnetic field and temperature. The red, green, blue, brown, orange, and cyan
curves correspond to $B=0$, $0.1$, $0.2$, $0.3$, $0.4$ and $0.5$, respectively. Dot-dashed and solid lines correspond to $T/T_{crit}=1.5$ and $2.0$, respectively. In units of GeV.}
\label{BHparalvsz}
\end{minipage}
\hspace{0.4cm}
\begin{minipage}[b]{0.5\linewidth}
\centering
\includegraphics[width=2.8in,height=2.3in]{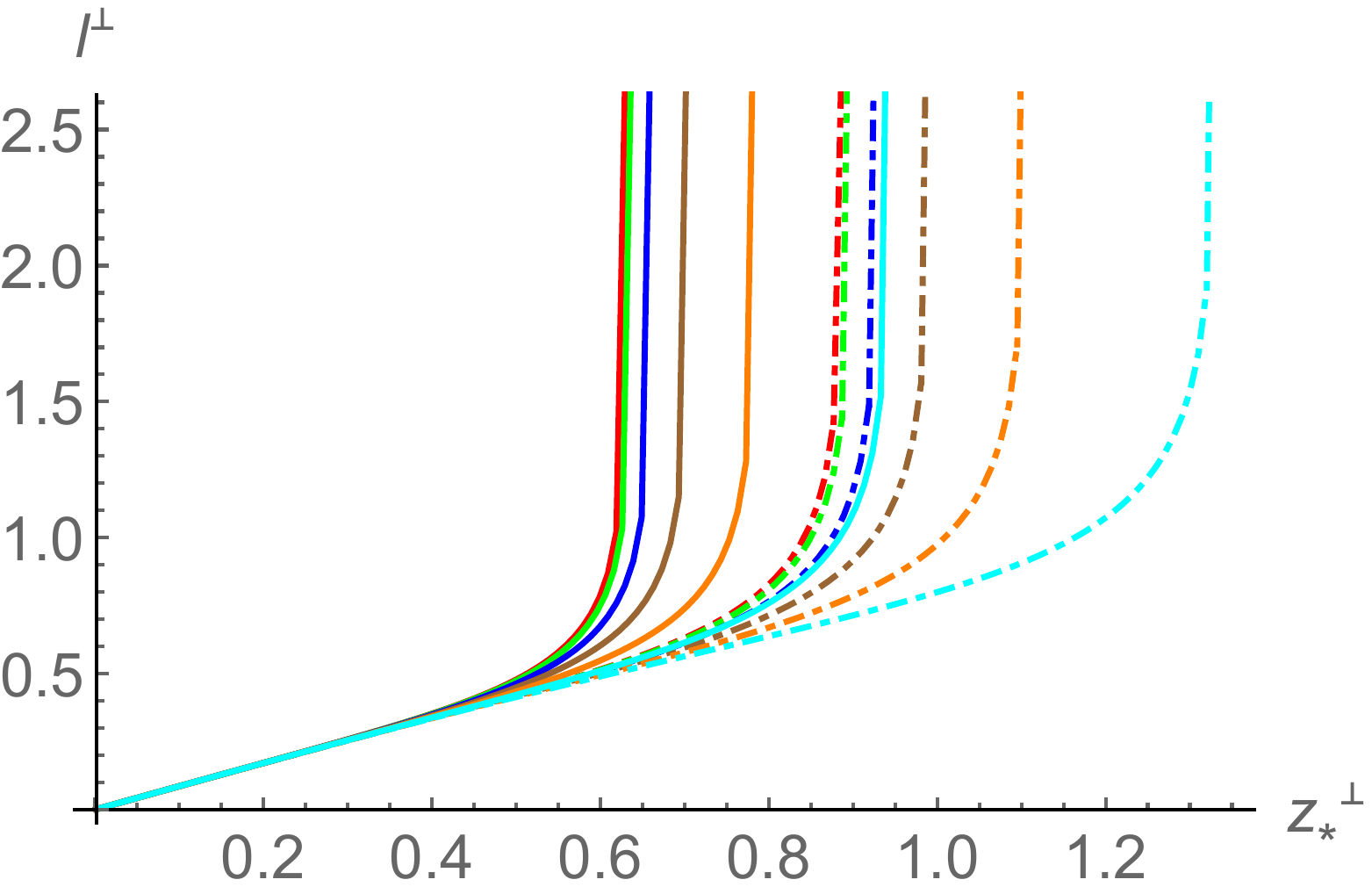}
\caption{$\ell^{\perp}$ as a function of $z_{*}^{\perp}$ for different values of magnetic field and temperature. The red, green, blue, brown, orange, and cyan curves correspond to $B=0$, $0.1$, $0.2$, $0.3$, $0.4$ and $0.5$, respectively. Dot-dashed and solid lines correspond to $T/T_{crit}=1.5$ and $2.0$, respectively. In units of GeV.}
\label{BHperplvsz}
\end{minipage}
\end{figure}
\begin{figure}[h]
\begin{minipage}[b]{0.5\linewidth}
\centering
\includegraphics[width=2.8in,height=2.3in]{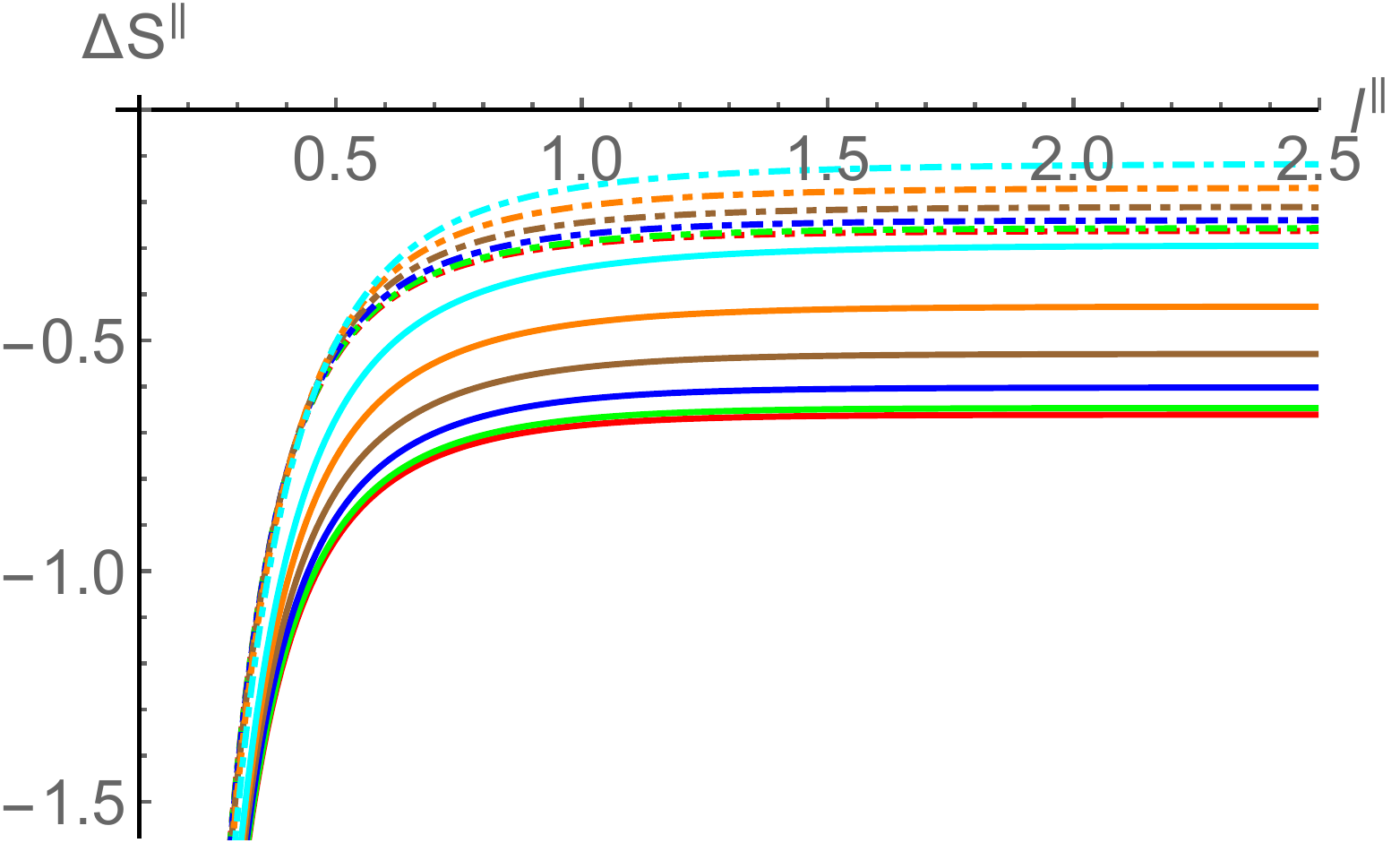}
\caption{$\Delta S^{\parallel}=S^{\parallel}_{con} - S^{\parallel}_{discon}$ as a function of $\ell^{\parallel}$ for different values of magnetic field and temperature. The red, green, blue, brown, orange, and cyan
curves correspond to $B=0$, $0.1$, $0.2$, $0.3$, $0.4$ and $0.5$, respectively. Dot-dashed and solid lines correspond to $T/T_{crit}=1.5$ and $2.0$, respectively. In units of GeV.}
\label{BHparaSvsl}
\end{minipage}
\hspace{0.4cm}
\begin{minipage}[b]{0.5\linewidth}
\centering
\includegraphics[width=2.8in,height=2.3in]{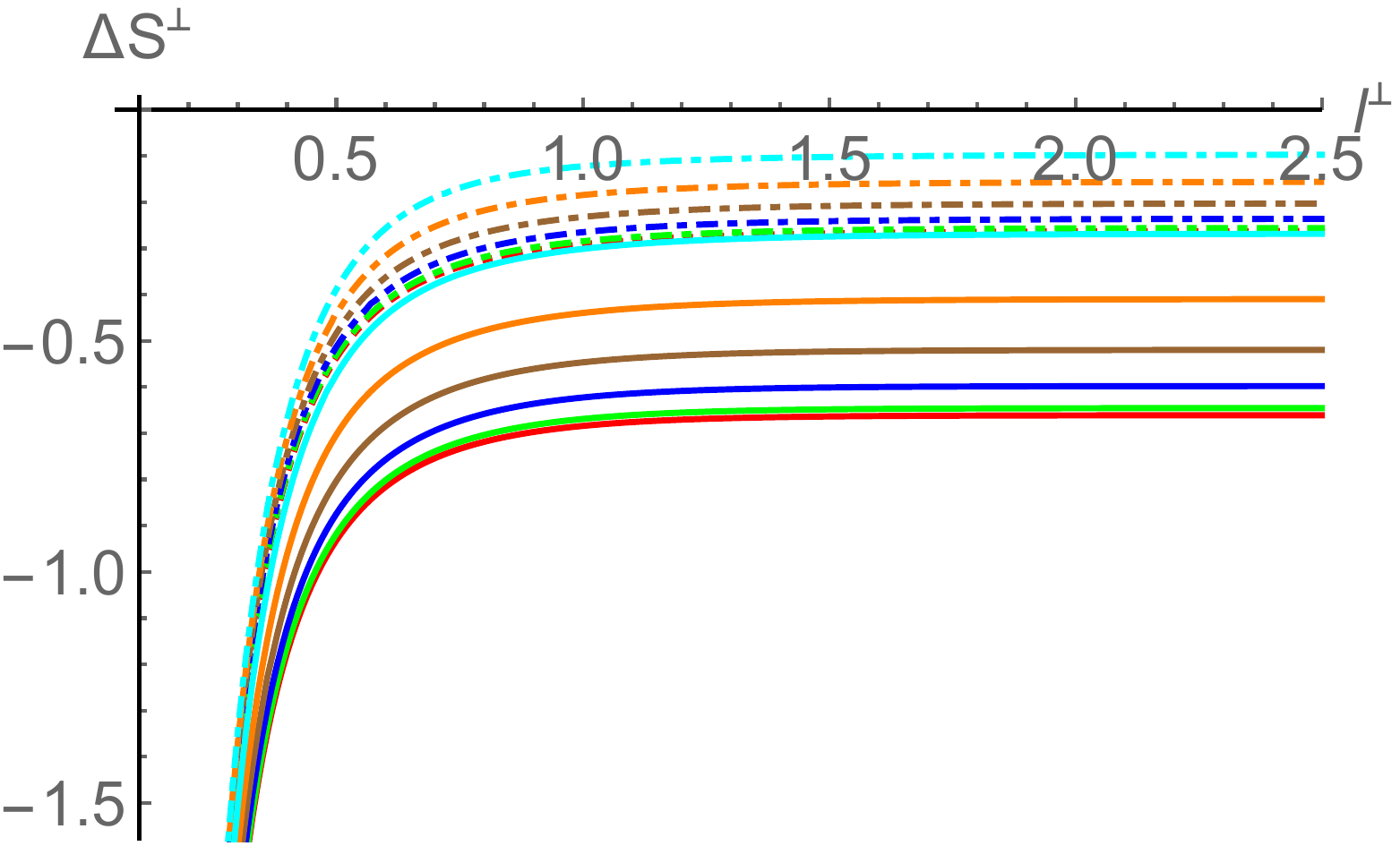}
\caption{$\Delta S^{\perp}=S^{\perp}_{con} - S^{\perp}_{discon}$ as a function of $\ell^{\perp}$ for different values of magnetic field and temperature. The red, green, blue, brown, orange, and cyan curves correspond to $B=0$, $0.1$, $0.2$, $0.3$, $0.4$ and $0.5$, respectively. Dot-dashed and solid lines correspond to $T/T_{crit}=1.5$ and $2.0$, respectively. In units of GeV.}
\label{BHperpSvsl}
\end{minipage}
\end{figure}
We start by studying the entanglement entropy for a single interval where the boundary subsystem can be aligned parallel or perpendicular to the magnetic field in a fashion similar to the thermal-AdS case. In the AdS black hole background, we again have two types of solutions for the entanglement entropy: connected and disconnected \cite{Dudal:2018ztm}. The disconnected entropy, however, turns out to be always higher than the connected entropy. The expressions of the connected entropy and strip length are the same as in the thermal-AdS case, except that $g(z)$ is now given by Eq.~(\ref{gsol}). So, for the parallel direction, we have Eqs.~(\ref{SEEcon}) and ({\ref{lengthSEEcon}}) for the connected entanglement entropy and strip length, whereas analogous equations for the perpendicular direction are given in Eqs.~(\ref{SEEconperp}) and ({\ref{lengthSEEconperp}}).

The entanglement entropy of the disconnected surface, however, will get an additional contribution. In the parallel direction we have
\begin{eqnarray}
S^{\parallel}_{discon}=\frac{\ell_{y_2} \ell_{y_3} L^3}{2 G_{(5)}} \biggl[ \int_{0}^{z_h} dz \ \frac{e^{3 A(z)}e^{B^2 z^2}}{z^3\sqrt{g(z)}} + \frac{e^{3 A(z_h)B^2 z_{h}^2}}{ 2z_{h}^3} \ell^\parallel \biggr] \,,
\label{SEEdisconBH}
\end{eqnarray}
and for the perpendicular direction we have
\begin{eqnarray}
S^{\perp}_{discon}=\frac{\ell_{y_1} \ell_{y_3} L^3}{2 G_{(5)}} \biggl[ \int_{0}^{z_h} dz \ \frac{e^{3 A(z)}e^{B^2 z^2/2}}{z^3\sqrt{g(z)}} + \frac{e^{3 A(z_h)+B^2 z_{h}^2}}{ 2z_{h}^3} \ell^\perp \biggr]\,,
\label{SEEdisconBHperp}
\end{eqnarray}
where the last term in both the parallel and perpendicular cases comes from the surface along the horizon at $z=z_h$.

We now proceed to discuss the numerical results for the entanglement entropy in the deconfined phase. The variation of the strip length with respect to the turning point of the connected surface at two different temperatures $T=1.5~T_{crit}$ and $2.0~T_{crit}$ for different values of $B$ is shown in Fig.~\ref{BHparalvsz} for the parallel case and in Fig.~\ref{BHperplvsz} for the perpendicular case.
We observe that for both orientations there exist certain common features. To begin with, unlike in the confined phase, there is no $\ell_{max}^{\parallel}$ or $\ell_{max}^{\perp}$ and the connected solution exists for the entire strip length. Second, as we increase the strip length, the connected surface's turning point moves closer to the horizon. Last, as we increase $B$, for a given value of the strip length,
the value of the turning point increases. These observations imply that, irrespective of the orientation of the strip, the strip goes deeper into the bulk by increasing $B$. This result is in contrast to the confining-phase results wherein the orientation of the magnetic field does induce anisotropy.

The corresponding entanglement entropy behavior is shown in Figs.~\ref{BHparaSvsl} and \ref{BHperpSvsl} for the parallel and perpendicular cases, respectively. We again see common features for both orientations. First, we see that there is no $\ell^{\parallel}_{crit}$ (or $\ell^{\perp}_{crit}$) for the parallel (or perpendicular) case and therefore no phase transition is observed from a connected to a disconnected surface on increasing the strip length in both cases. Next, we see that for both orientations, the difference in the entropy is always less than zero, implying that the connected entropy is always less than the disconnected entropy. Further,
in the limit $\ell^{\parallel} \rightarrow \infty$, we have
\begin{eqnarray}
S^{\parallel}_{con} = S^{\parallel}_{discon} = S_{BH} = \frac{V_3 e^{3 A(z_h)+B^2 z_{h}^{2}}}{4 G_{(5)} z_{h}^3 }\,.
\end{eqnarray}
Similarly, in the limit $\ell^{\perp} \rightarrow \infty$, we have
\begin{eqnarray}
S^{\perp}_{con} = S^{\perp}_{discon} = S_{BH} = \frac{V_3 e^{3 A(z_h)+B^2 z_{h}^{2}}}{4 G_{(5)} z_{h}^3 }\,,
\end{eqnarray}
where the $S_{BH}$ represents the Bekenstein-Hawking entropy of the AdS black hole. This reproduces the expected result that the entanglement entropy reduces to the thermal entropy when the size of the subsystem goes to infinity.  So, effectively, the entanglement entropy in the deconfined phase is always of order $N^2$,
\begin{eqnarray}
\frac{\partial S^{\parallel}}{\partial \ell^{\parallel}} \propto \frac{1}{G_{(5)}} = \mathcal{O}(N^2)\,,~~~\frac{\partial S^{\perp}}{\partial \ell^{\perp}} \propto \frac{1}{G_{(5)}} = \mathcal{O}(N^2)\,.
\end{eqnarray}
Essentially, the behavior of the entanglement entropy remains qualitatively the same for both the parallel and perpendicular cases in the deconfined phase.

\subsection{Two-strip phase diagram and mutual information}
\begin{figure}[h]
\begin{minipage}[b]{0.5\linewidth}
\centering
\includegraphics[width=2.8in,height=2.3in]{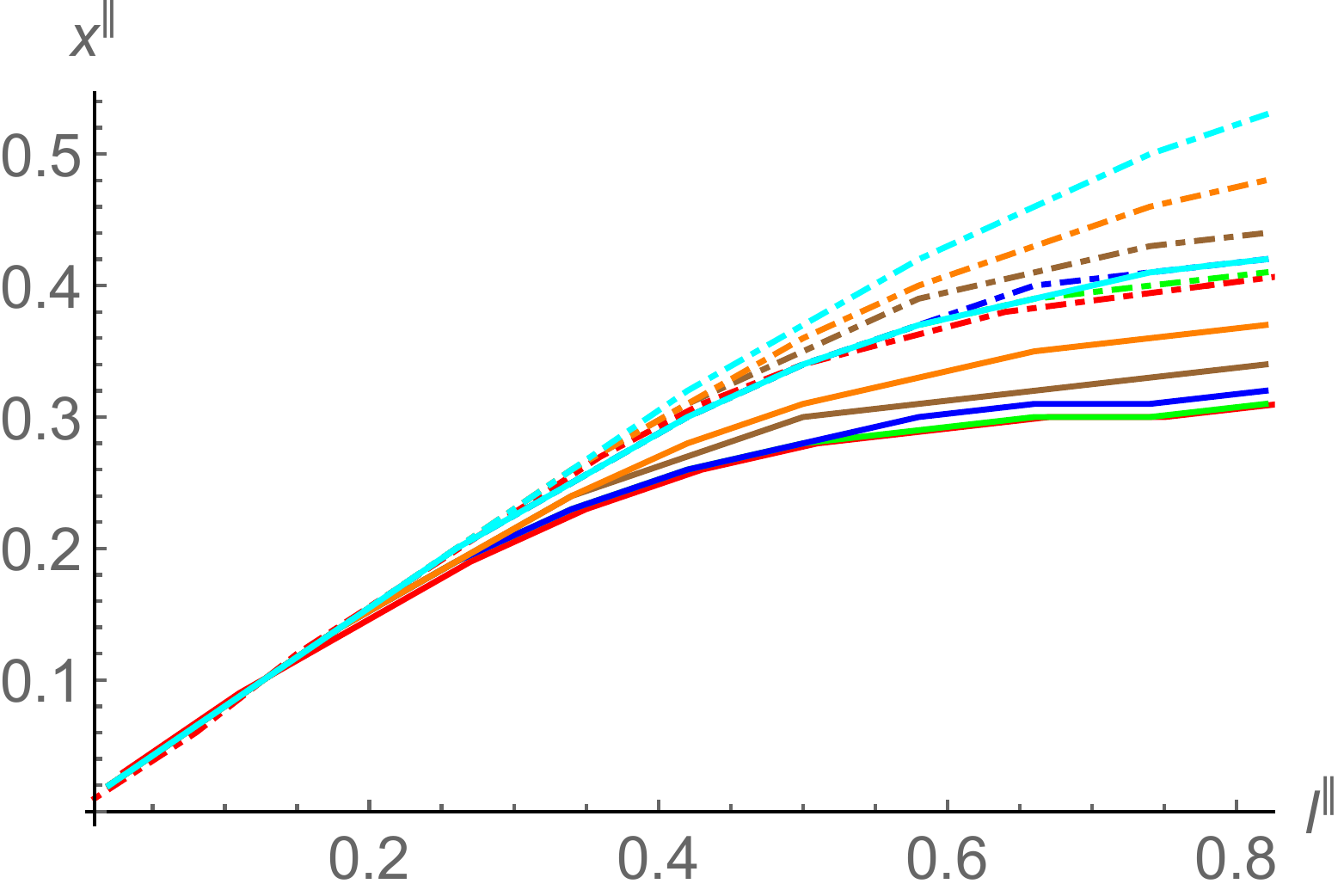}
\caption{Two-strip phase diagram in the deconfining background for the parallel case for different values of magnetic field and temperature. The red, green, blue, brown, orange, and cyan
curves correspond to $B=0$, $0.1$, $0.2$, $0.3$, $0.4$ and $0.5$, respectively. Dot-dashed and solid lines correspond to $T/T_{crit}=1.5$ and $2.0$, respectively In units of GeV.}
\label{BHparaphasediag}
\end{minipage}
\hspace{0.4cm}
\begin{minipage}[b]{0.5\linewidth}
\centering
\includegraphics[width=2.8in,height=2.3in]{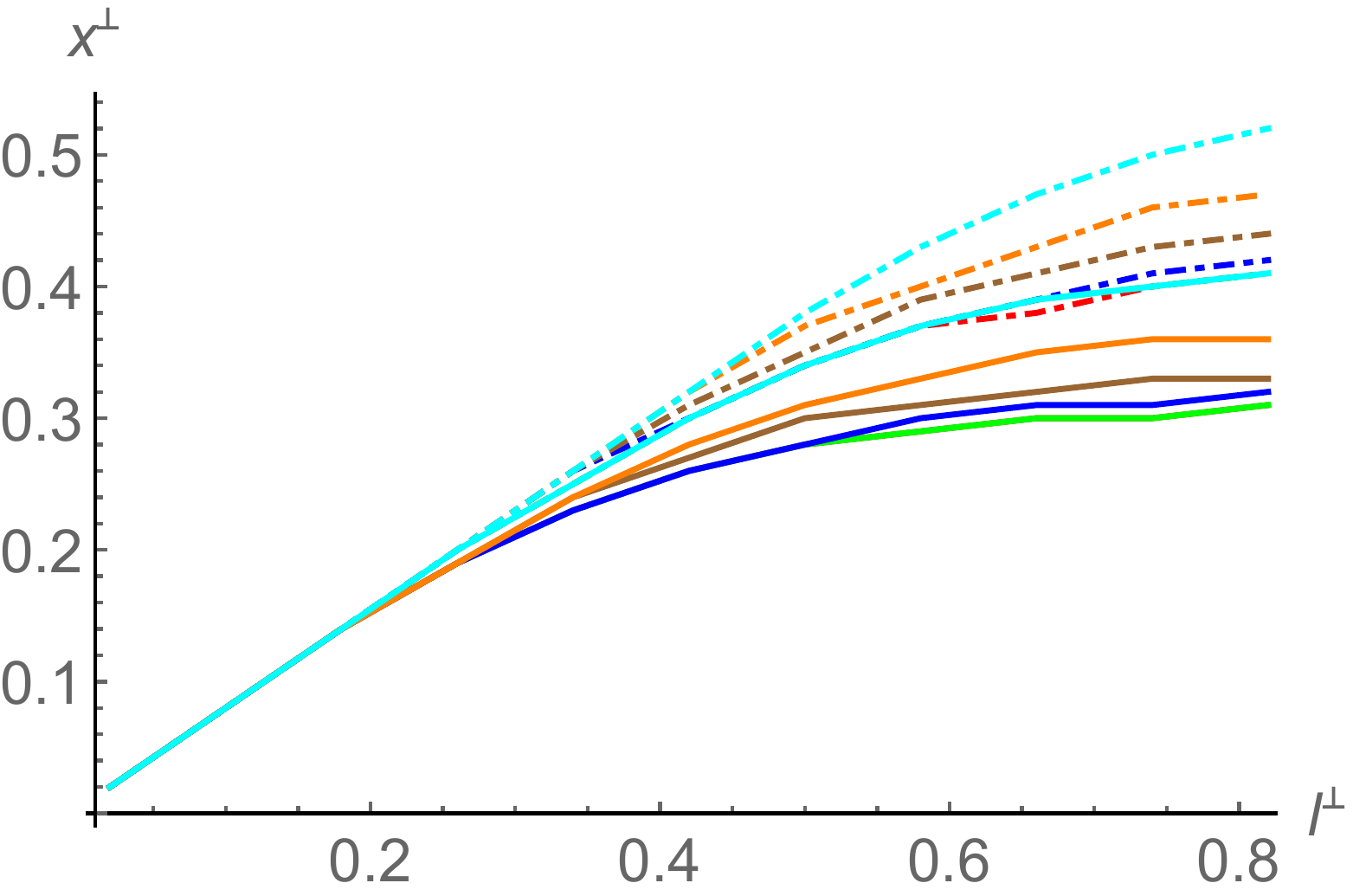}
\caption{Two-strip phase diagram in the deconfining background for the perpendicular case for different values of magnetic field and temperature. The red, green, blue, brown, orange, and cyan
curves correspond to $B=0$, $0.1$, $0.2$, $0.3$, $0.4$ and $0.5$, respectively. Dot-dashed and solid lines correspond to $T/T_{crit}=1.5$ and $2.0$, respectively. In units of GeV.}
\label{BHperpphasediag}
\end{minipage}
\end{figure}
\begin{figure}[h]
\begin{minipage}[b]{0.5\linewidth}
\centering
\includegraphics[width=2.8in,height=2.3in]{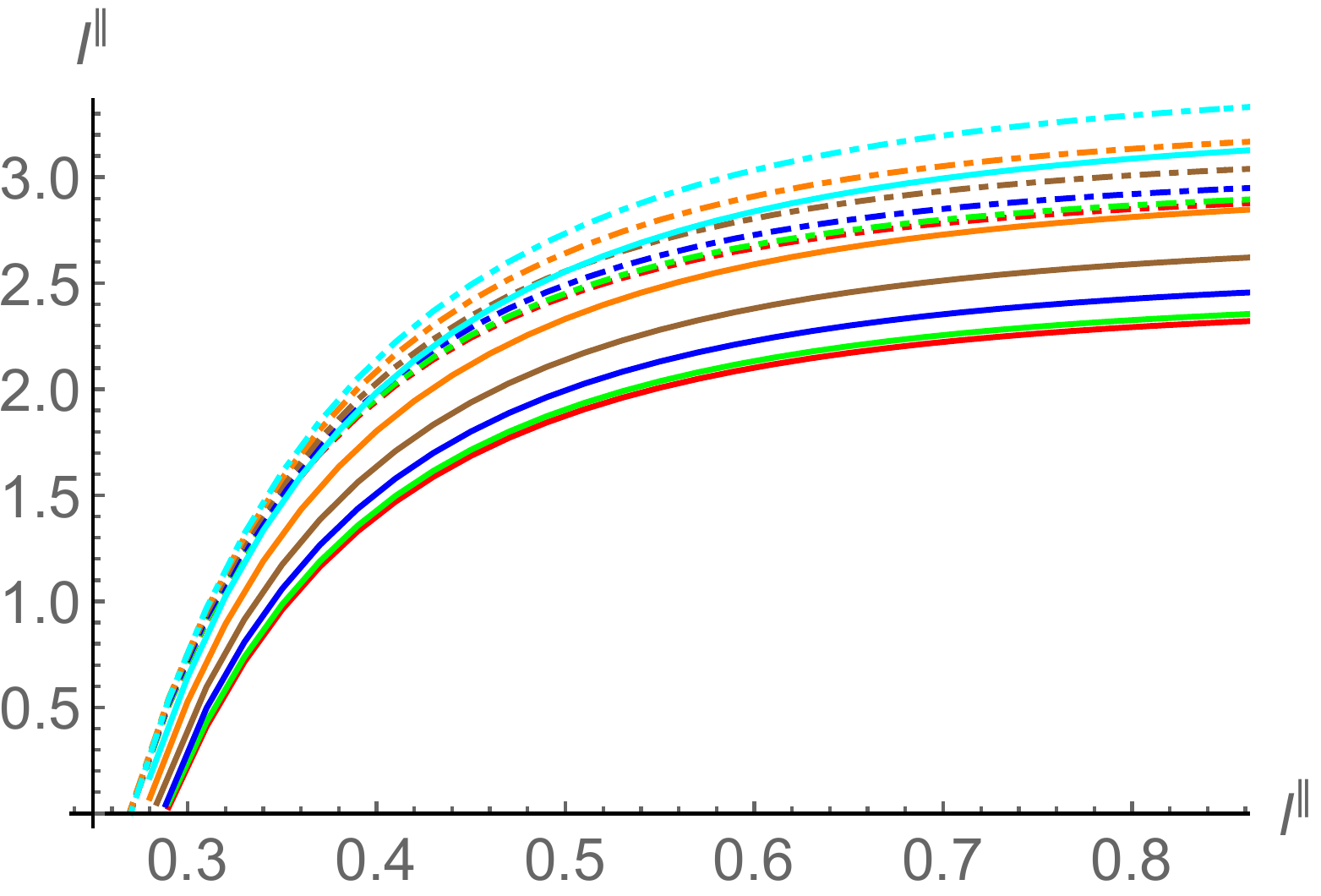}
\caption{$I^\parallel$ as a function of $\ell^{\parallel}$ for different values of magnetic field and temperature. Here $x^\parallel=0.2$ is used.  The red, green, blue, brown, orange, and cyan
curves correspond to $B=0$, $0.1$, $0.2$, $0.3$, $0.4$ and $0.5$, respectively. Dot-dashed and solid lines correspond to $T/T_{crit}=1.5$ and $2.0$, respectively. In units of GeV.}
\label{BHparaMIvsl}
\end{minipage}
\hspace{0.4cm}
\begin{minipage}[b]{0.5\linewidth}
\centering
\includegraphics[width=2.8in,height=2.3in]{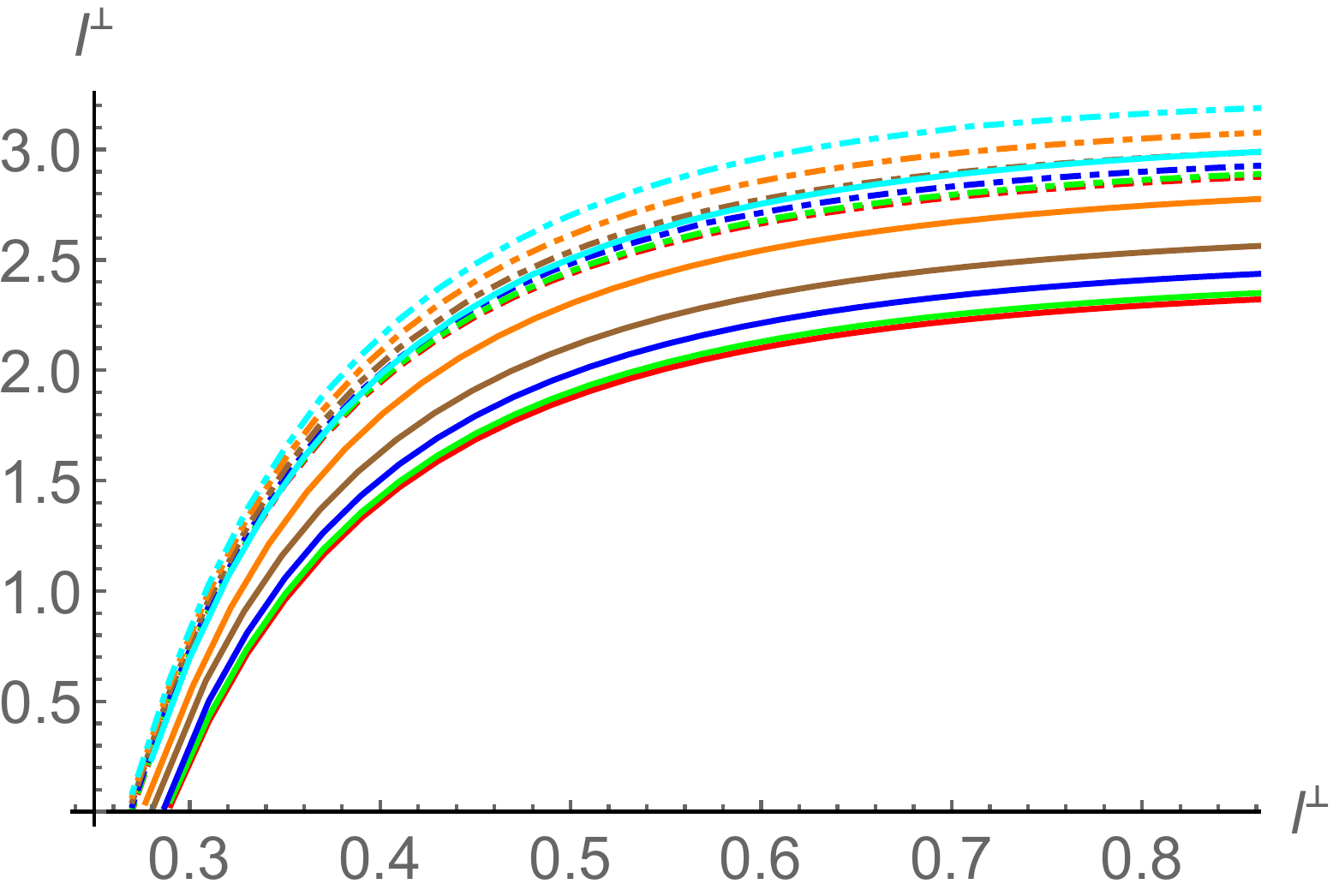}
\caption{$I^\perp$ as a function of $\ell^{\perp}$ for different values of magnetic field and temperature. Here $x^\perp=0.2$ is used. The red, green, blue, brown, orange, and cyan curves correspond to $B=0$, $0.1$, $0.2$, $0.3$, $0.4$ and $0.5$, respectively. Dot-dashed and solid lines correspond to $T/T_{crit}=1.5$ and $2.0$, respectively. In units of GeV.}
\label{BHperpMIvsl}
\end{minipage}
\end{figure}
Since there is no phase transition between connected and disconnected entanglement entropies, the corresponding two-strip phase diagram in the deconfined phase is much simpler. Here, we only have two phases $S_A$ and $S_B$ as the connected surface dominates for any given strip length. Therefore, in the phase diagram, as shown for the parallel orientation in Fig.~\ref{BHparaphasediag} and for the perpendicular orientation in Fig.~\ref{BHperpphasediag}, we can only see the phase transition between the $S_A$ and $S_B$ phases. We observe that the $S_A$ phase is preferred when $x^{\parallel}$ (or $x^{\perp}$) is large, while the $S_B$ phase is preferred for large $\ell^{\parallel}$ (or $\ell^{\perp}$). We further observe that on increasing $B$, the parameter space of the $S_B$ phase increases, suggesting its preference over the $S_A$ phase for larger magnetic field values. This is true for both the parallel and perpendicular cases. We again see that, although the magnetic field does introduce substantial changes in the phase diagram, these changes are qualitatively similar for the parallel and perpendicular cases, suggesting limited orientational effects of $B$ in the deconfined phase. Similarly, for a fixed $B$, the phase space of $S_B$ is found to increase with temperature.

The mutual information in the $S_A$ and $S_B$ phases displays similar features as in the confined phase. This is shown in Figs.~\ref{BHparaMIvsl} and \ref{BHperpMIvsl} for parallel and perpendicular cases, respectively. The mutual information is zero in the $S_A$ phase, whereas it is a monotonically increasing function of strip length in the $S_B$ phase. Moreover, the behavior of the mutual information as a function of separation length is similar to the ones shown in Fig.~\ref{TAdSParallelMIvsx}, and therefore we do not present it here for brevity. In particular, it is a monotonically decreasing function of the separation length and it goes to zero in a smooth fashion as we pass from the $S_B$ phase to the $S_A$ phase. Therefore, an order change in the mutual information appears during the $S_A/S_B$ phase transition as $I_A \propto \mathcal{O}(N^0)$ and $I_B \propto \mathcal{O}(N^2)$.  This behavior is again true for both the parallel and perpendicular orientations.

\subsection{Entanglement wedge cross-section}
We now move on to discuss the entanglement wedge cross-section $E_W$ in the deconfining phase. Guided by the symmetry of the configuration (see Fig.~\ref{Entanglementwedgecrosspic}), the area of the vertical surface $\Sigma_{AB}^{min}$ gives the entanglement wedge cross-section. In the case of an AdS black hole, $E_W$ exists only for the $S_B$ phase and its expression is
similar to the thermal-AdS case, except that $g(z)$ is now given by Eq.~(\ref{gsol}). Therefore for the parallel orientation, we have
\begin{eqnarray}
E_{W}^{\parallel}(S_B)=\frac{\ell_{y_2} \ell_{y_3} L^3}{4 G_{(5)}} \biggl[ \int_{z_{*}^{\parallel}(x)}^{z_{*}^{\parallel}(2\ell+x)} dz \ \frac{e^{3 A(z)}e^{B^2 z^2}}{z^3\sqrt{g(z)}} \biggr]\,.
\label{BHparaEWSBint}
\end{eqnarray}
Similarly, for the perpendicular orientation we have
\begin{eqnarray}
E_{W}^{\perp}(S_B)=\frac{\ell_{y_1} \ell_{y_3} L^3}{4 G_{(5)}} \biggl[ \int_{z_{*}^{\perp}(x)}^{z_{*}^{\perp}(2\ell+x)} dz \ \frac{e^{3 A(z)}e^{B^2 z^2/2}}{z^3\sqrt{g(z)}} \biggr]\,.
\label{BHperpEWSBint}
\end{eqnarray}
Since no wedge exists between two subsystems in the $S_A$ phase, accordingly the entanglement wedge cross-section is zero in this phase.

\begin{figure}[h]
\begin{minipage}[b]{0.5\linewidth}
\centering
\includegraphics[width=2.8in,height=2.3in]{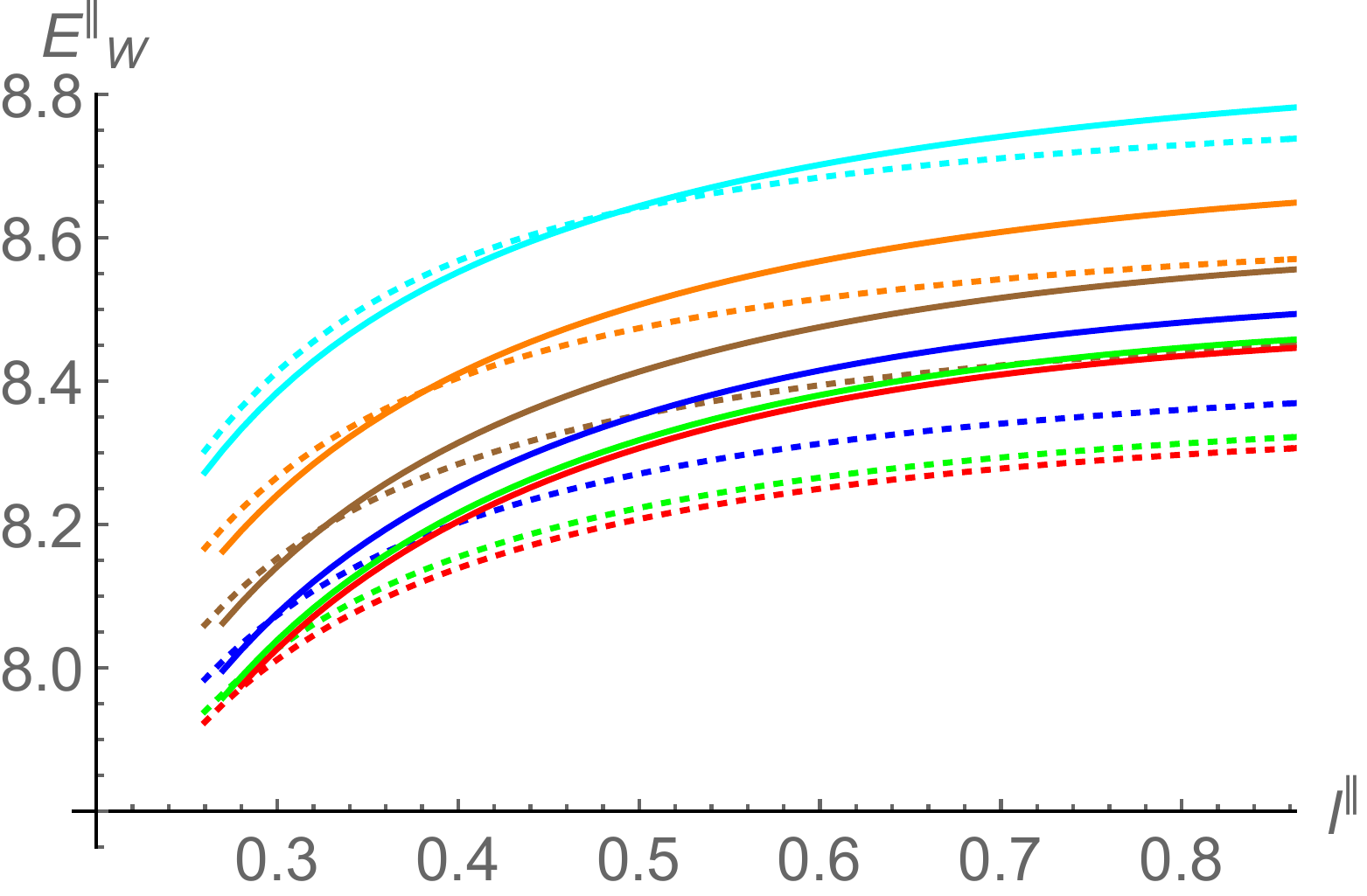}
\caption{$E_{W}^{\parallel}$ as a function of $\ell^{\parallel}$ for different values of magnetic field and temperature. Here $x^\parallel=0.2$ is used. The red, green, blue, brown, orange, and cyan
curves correspond to $B=0$, $0.1$, $0.2$, $0.3$, $0.4$ and $0.5$, respectively. Dotted and solid lines correspond to $T/T_{crit}=1.5$ and $2.0$, respectively. In units of GeV.}
\label{BHparaEWvsl}
\end{minipage}
\hspace{0.4cm}
\begin{minipage}[b]{0.5\linewidth}
\centering
\includegraphics[width=2.8in,height=2.3in]{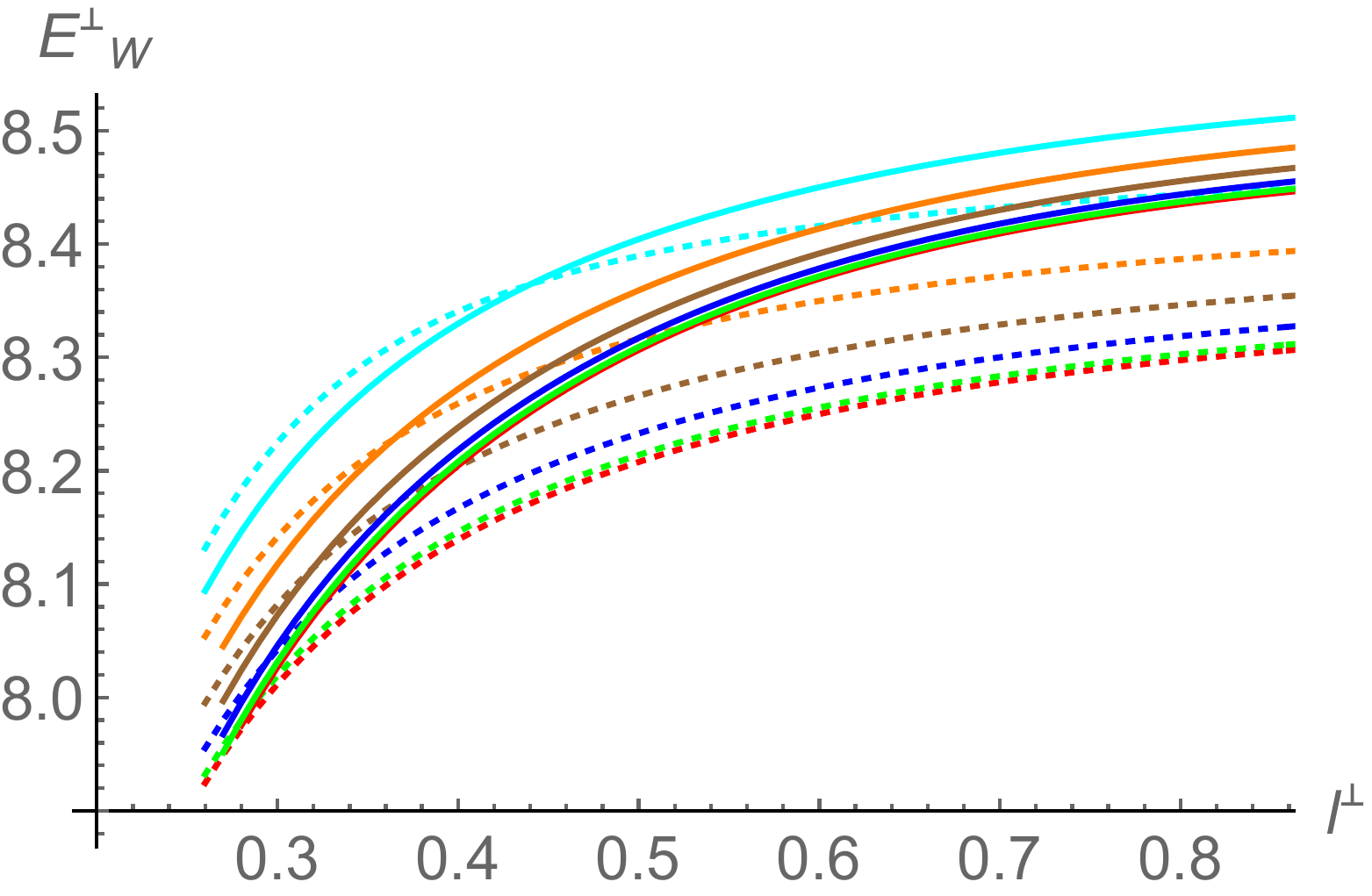}
\caption{$E_{W}^{\perp}$ as a function of $\ell^{\perp}$ for different values of magnetic field and temperature. Here $x^\perp=0.2$ is used. The red, green, blue, brown, orange, and cyan
curves correspond to $B=0$, $0.1$, $0.2$, $0.3$, $0.4$ and $0.5$, respectively. Dotted and solid lines correspond to $T/T_{crit}=1.5$ and $2.0$, respectively. In units of GeV.}
\label{BHperpEWvsl}
\end{minipage}
\end{figure}

The behavior of $E_W$ as a function of strip length for different values of magnetic field and temperature is shown in Figs.~\ref{BHparaEWvsl} and \ref{BHperpEWvsl} for the parallel and perpendicular cases, respectively. Here we choose a fixed separation length $x^{\parallel} (\text{or}~x^{\perp})=0.2$ for illustration purposes, but similar results exist for other values of $x^{\parallel} (\text{or}~x^{\perp})$ as well. The nature of $E_W$ is again qualitatively similar in both orientations.  In particular, the magnitude of $E_W$ increases with $B$ in both cases. However, the increment is slightly higher in the parallel case compared to the perpendicular case. $E_W$ again turns out to be a monotonic function of strip length in the $S_B$ phase, which vanishes discontinuously in the $S_A$ phase. This is true for all temperatures and magnetic fields. Further, in the presence of $B$, the thermal profile of $E_W$ exhibits an interesting feature, i.e., $E_W$ decreases with temperature for small trip lengths, whereas it increases with temperature for large strip lengths. This novel feature appears only in the presence of $B$ and is true for both the parallel and perpendicular cases.

\begin{figure}[h]
\begin{minipage}[b]{0.5\linewidth}
\centering
\includegraphics[width=2.8in,height=2.3in]{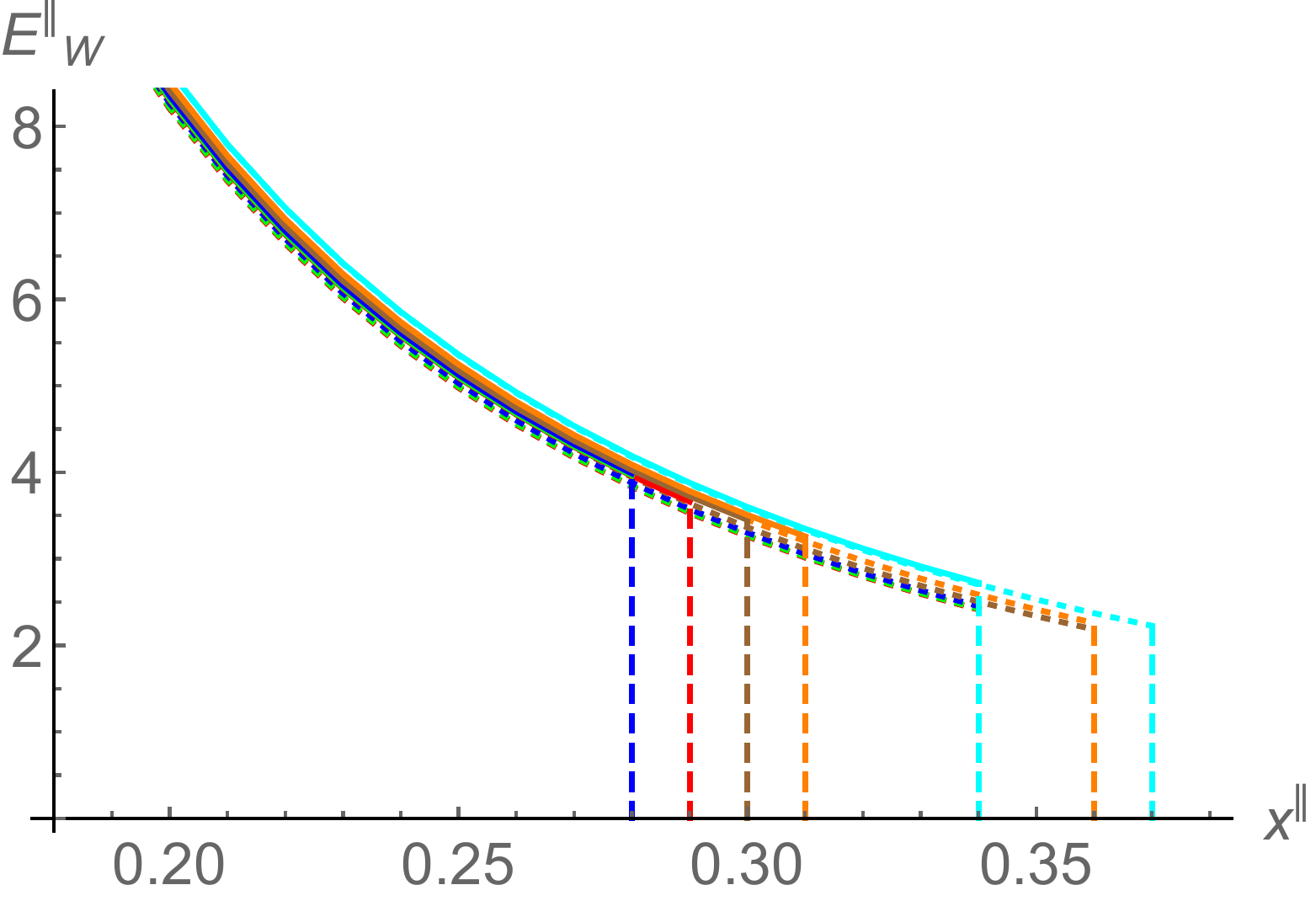}
\caption{$E_{W}^{\parallel}$ as a function of $x^{\parallel}$ for different values of magnetic field and temperature. Here $\ell^\parallel=0.5$ is used. The red, green, blue, brown, orange, and cyan
curves correspond to $B=0$, $0.1$, $0.2$, $0.3$, $0.4$ and $0.5$, respectively. Dotted and solid lines correspond to $T/T_{crit}=1.5$ and $2.0$, respectively. In units of GeV.}
\label{BHparaEWvsx}
\end{minipage}
\hspace{0.4cm}
\begin{minipage}[b]{0.5\linewidth}
\centering
\includegraphics[width=2.8in,height=2.3in]{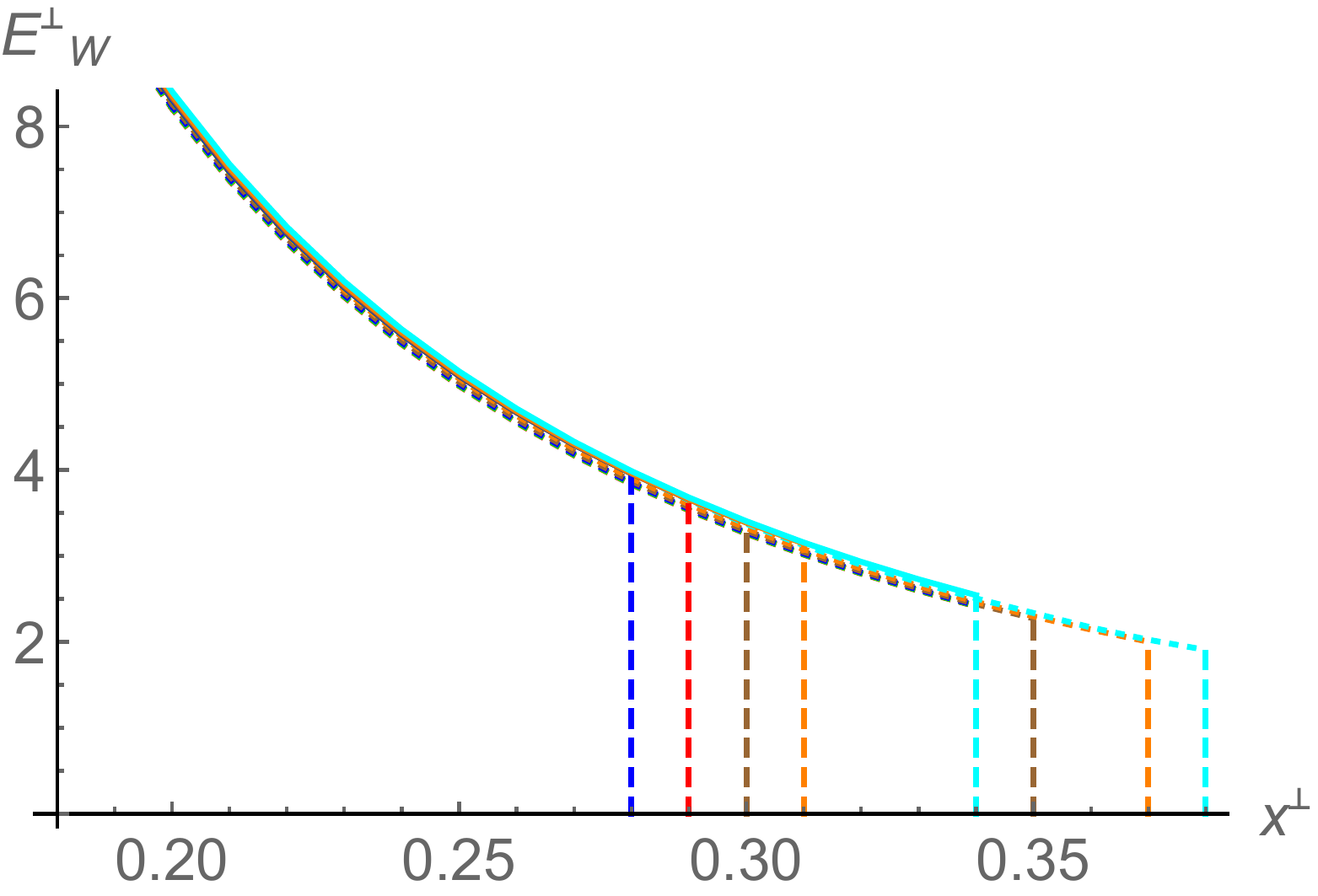}
\caption{$E_{W}^{\perp}$ as a function of $x^{\perp}$ for different values of magnetic field and temperature. Here $\ell^\perp=0.5$ is used. The red, green, blue, brown, orange, and cyan
curves correspond to $B=0$, $0.1$, $0.2$, $0.3$, $0.4$ and $0.5$, respectively. Dotted and solid lines correspond to $T/T_{crit}=1.5$ and $2.0$, respectively. In units of GeV.}
\label{BHperpEWvsx}
\end{minipage}
\end{figure}

Similarly, $E_W$ is also a monotonic function of the separation length. This is shown in Figs.~\ref{BHparaEWvsx} and \ref{BHperpEWvsx} for the parallel and perpendicular cases, respectively. For both cases, like in the confined phase, $E_W$ decreases with the separation length in the $S_B$ phase and discontinuously becomes zero as we enter the $S_A$ phase. We find that this discontinuous behavior of $E_W$ at the $S_A/S_B$ transition line is true for all values of magnetic field and temperature.

We further test (although not explicitly shown here) the inequality $E_{W} \geq I/2$ in the deconfined phase and find that this inequality is again satisfied everywhere in the $\ell$-$x$ plane of the parallel and perpendicular orientations for all values of magnetic field and temperature. The inequality saturates only at the $S_A/S_B$ transition line, at which $I/2$ continuously goes to zero, whereas $E_{W}$ exhibits a sharp drop to zero.

\subsection{Holographic entanglement negativity}
In this subsection, we talk about the holographic entanglement negativity in the deconfined phase. Beginning with the single-interval case, wherein the holographic negativity is given by Eq.~(\ref{HEEsc1}), we have in the limit $B\rightarrow A^c\rightarrow\infty$
\begin{eqnarray}\label{HEEBH}
& & \mathcal{E} = \lim_{B\rightarrow A^c}\frac{3}{4}\left[2S(A)+S(B_1)+S(B_2)-S(A\cup B_1)-S(A\cup B_2)\right]\,, \nonumber \\
& & \mathcal{E} = \frac{3}{2} S(A)
\end{eqnarray}
as, apart from $S_A$, the rest of the four terms represent the same quantity in the limit $B\rightarrow A^c\rightarrow\infty$, i.e., the black hole entropy, and therefore cancel each other. Accordingly, in the single-interval case, we have $\mathcal{E} = \frac{3}{2} S_A$ irrespective of the orientation of the magnetic field. This is the same result that we got in the confined phase as well. Therefore, for a single-interval case, the negativity in the confined and deconfined phases is always $3/2$ times the entanglement entropy. Accordingly,
\begin{eqnarray}
\frac{\partial\mathcal{E}}{\partial \ell} \propto \frac{1}{G_N} = \mathcal{O}(N^2)\,.
\end{eqnarray}
The negativity is always of order $\mathcal{O}(N^2)$ in the deconfined phase for both parallel and perpendicular magnetic fields. This is different from the confined phase, where the negativity undergoes an order change at some critical strip length.

\begin{figure}[h]
\begin{minipage}[b]{0.5\linewidth}
\centering
\includegraphics[width=2.8in,height=2.3in]{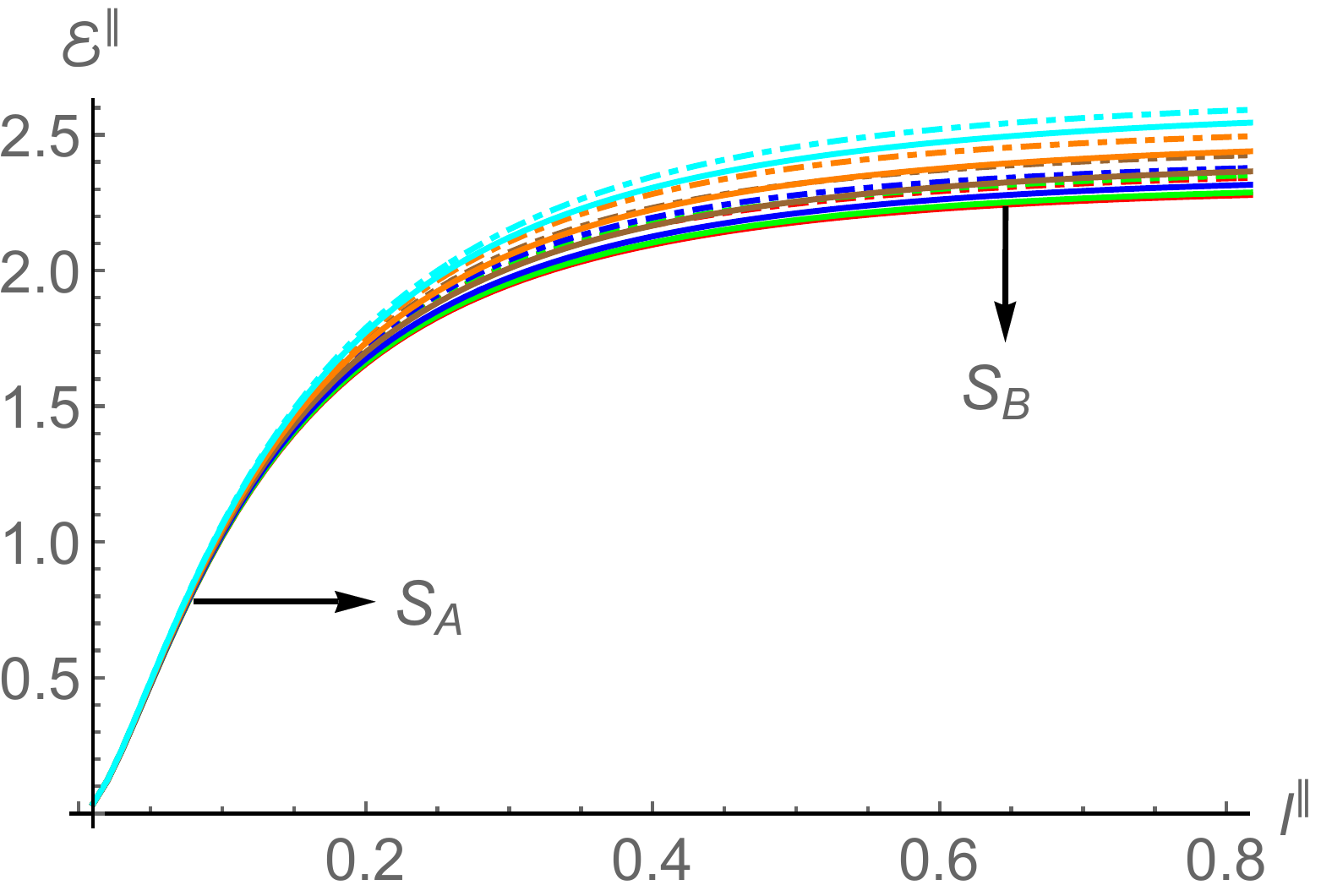}
\caption{$\mathcal{E}^{\parallel}$ as a function of $\ell^{\parallel}$ for different values of magnetic field and temperature. Here $x^\parallel=0.2$ is used. The red, green, blue, brown, orange, and cyan
curves correspond to $B=0$, $0.1$, $0.2$, $0.3$, $0.4$ and $0.5$, respectively. Dot-dashed and solid lines correspond to $T/T_{crit}=1.5$ and $2.0$, respectively. In units of GeV.}
\label{BHparaENvsl}
\end{minipage}
\hspace{0.4cm}
\begin{minipage}[b]{0.5\linewidth}
\centering
\includegraphics[width=2.8in,height=2.3in]{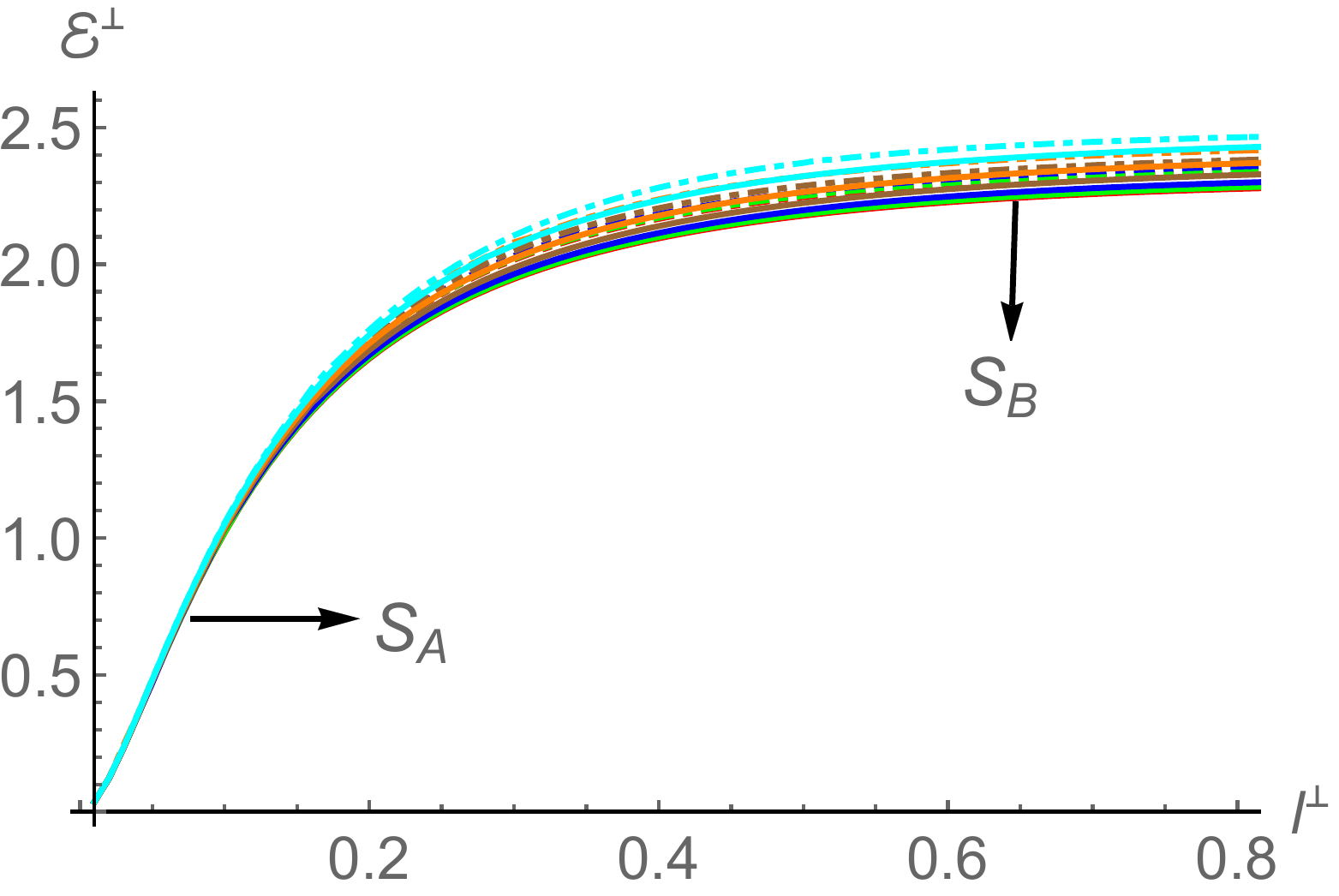}
\caption{$\mathcal{E}^{\perp}$ as a function of $\ell^{\perp}$ for different values of magnetic field and temperature. Here $x^\perp=0.2$ is used. The red, green, blue, brown, orange, and cyan
curves correspond to $B=0$, $0.1$, $0.2$, $0.3$, $0.4$ and $0.5$, respectively. Dot-dashed and solid lines correspond to $T/T_{crit}=1.5$ and $2.0$, respectively. In units of GeV.}
\label{BHperpENvsl}
\end{minipage}
\end{figure}

Moving on to the two-disjoint-interval case, we have the entanglement negativity as \cite{Malvimat:2018txq,Basak:2020bot}
\begin{equation}
\mathcal{E}=\frac{3}{4}\left[S(\ell+x)+S(\ell+x)-S(2\ell+x)-S(x)\right] \,,
\label{ENtwostripBH}
\end{equation}
where $S$ denotes the holographic entanglement entropy for a single interval. Notice that, as is expected, when $x\rightarrow\infty$, i.e., for large separations, the negativity goes to zero as all terms in the above equation represent the black hole entropy. Interestingly, like in the confined case, there can be some region in the parameter space of the $S_A$ phase where the negativity is nonzero. This once again has to be contrasted with the mutual information and entanglement wedge of the deconfined phase where these quantities were zero everywhere in the $S_A$ phase. Indeed, as shown in Figs.~\ref{BHparaENvsl} and \ref{BHperpENvsl} for the parallel and perpendicular cases, respectively, the negativity is nonzero in the $S_A$ phase as well. The nonzero negativity for large separations in the deconfined phase is again an important prediction (again, strictly speaking, a prediction of the negativity proposal of \cite{Malvimat:2018txq,Basak:2020bot}). Moreover, the negativity turns out to be a monotonic function of both strip length and separation length; in particular, it decreases for higher separation lengths, whereas it increases for higher strip lengths.  We further find that for a fixed strip length and separation length the negativity increases slightly with higher magnetic fields, whereas thermal effects try to decrease it. These results are again qualitatively similar for both the parallel and perpendicular cases.

We end this section by mentioning that our investigation suggests that the orientation-dependent effects of the magnetic field in the high-temperature deconfined phase are rather limited compared to the low-temperature confined phase. Though some nontrivial changes do arise in various entanglement measures between parallel and transverse magnetic fields in the deconfined phase, these changes are not as substantial as in the confined phase. For example, the magnetic field produced distinct effects in the entanglement phase diagram of the confined phase in the parallel and transverse directions, whereas the phase diagram is quite similar for both orientations in the deconfined phase. In the deconfined phase, the anisotropic effects might be suppressed by the large thermal effects. Indeed, if we do a large-temperature expansion of the entanglement entropy and strip length, the effect of the magnetic field appears in a similar fashion for both orientations.

\section{Discussion and conclusion}
\label{conclusion}
In this work, we performed a comprehensive analysis of the effects of a background magnetic field on various pure and mixed entanglement measures in the holographic confined/deconfined phases dual to a bottom-up phenomenological Einstein-Maxwell-dilaton gravity model. The magnetic field is expected to play an important role in QCD-related physics and here we analysed in detail how this magnetic field alters the structure of the entanglement entropy, mutual information, entanglement wedge cross-section, and entanglement negativity, in the confined/deconfined phases of QCD.

We first reestablished the known results of the entanglement entropy of a single strip in the confining phase, but now in the presence of a magnetic field. In particular, a phase transition from connected to disconnected entanglement entropy is observed at some critical strip length in the confined phase, at which the order of the entanglement entropy changes from $\mathcal{O}(N^2)$ to $\mathcal{O}(N^0)$. Interestingly, this critical length is found to increase/decrease for parallel/perpendicular magnetic fields, thereby providing anisotropic imprints of the magnetic field on the entanglement structure. We then analysed
the two-equal-strip entanglement phase diagram in the parameter space of strip length $\ell$ and separation length $x$ and found four distinct phases $\{S_A,S_B,S_C,S_D\}$. These four phases exchange dominance
as $x$ and $\ell$ are varied, leading to an interesting phase diagram. This two-strip phase diagram is again greatly modified in the presence of a magnetic field, while further exhibiting anisotropic features. The mutual information turned out to be nonzero only in the $S_B$ and $S_C$ phases and is always a monotonic function of $x$ and $\ell$. Similarly, the entanglement wedge cross-section $E_W$ was found to be nonzero only in the $S_B$ and $S_C$ phases. Interestingly, unlike the mutual information, $E_W$ vanishes discontinuously for large values of $x$ and $\ell$ and exhibits nonanalytic behavior across various transition lines. In particular, going from the $S_B$ phase to the $S_C$ phase, $E_W$ increases at the $S_B/S_C$ transition line. Interestingly, this increment in the area of the entanglement wedge at the $S_B/S_C$ transition line is found to decrease/increase for magnetic fields in parallel/perpendicular directions, yielding yet another anisotropic feature in the entanglement structure. Moreover, we tested the inequality concerning
the mutual information and $E_W$ and found that the latter always exceeds half of the former everywhere in the $\ell$-$x$ parameter space for all values of $B$. Similarly, we analysed the behavior of the entanglement negativity with one and two intervals using the holographic proposal suggested in \cite{Chaturvedi:2016rft,Chaturvedi:2016rcn} and found many interesting features in the confined phase. For a single-strip subsystem, the negativity turned out to be just $3/2$ times the entanglement entropy, implying that it also undergoes an order change, from $\mathcal{O}(N^2)$ to $\mathcal{O}(N^0)$, as the strip length is varied. This suggests that it can also be used, like the entanglement entropy, to probe confinement. The corresponding critical length is further found to increase/decrease for the parallel/perpendicular magnetic fields.  Moreover, for two strips, the negativity behaves smoothly across various phase transition lines and no discontinuity in its structure is realised. However, unlike the mutual information and entanglement wedge, the negativity can be nonzero in some parts of the $S_A$ phase, an interesting feature that may not be observed in the holographic negativity proposal of \cite{Kudler-Flam:2018qjo,Kusuki:2019zsp}. In addition, the negativity was found to display anisotropic features in parallel and perpendicular directions.

We then analysed the entanglement structure of the deconfined phase. We found that there is no connected/disconnected transition and the entanglement entropy is always given by the connected surface. Accordingly, the two-strip phase diagram is much simpler in the deconfined phase. In particular, there are only $S_A$ and $S_B$ phases, with mutual information and entanglement wedge nonzero only in the $S_B$ phase, whereas the entanglement negativity can be nonzero in both the $S_A$ and $S_B$ phases. We further found that the parameter space of the $S_B$ phase increases for both orientations of the magnetic field, suggesting a larger phase space for the nontrivial entanglement wedge in the presence of a magnetic field. Similarly, the entanglement negativity of a single strip was again found to be proportional to the entanglement entropy,  whereas for two-strips it was found to be a monotonic function of $x$ and $\ell$ for all values of magnetic field and temperature.  Our analysis suggests that, although the magnetic field introduces substantial changes in the entanglement measures, these changes remain qualitatively similar in both the parallel and perpendicular cases, suggesting a limited anisotropic effect of the magnetic field in the deconfined phase as compared to the confined phase.

We end this discussion by mentioning a few directions to extend our work. The next step in our research setup would be to include the chemical potential, as it also plays an important role in QCD physics, and to simultaneously discuss the effects of magnetic field and chemical potential on the entanglement structure of confined/deconfined phases. In the simplistic situation, this can be done in the current holographic setup as well by adding another gauge field on the gravity side. Similarly, it would also be interesting to compute $E_W$ and $\mathcal{E}$ after a global quantum quench and analyse the thermalization process via these measures, as this might also provide important information about the QGP formation in QCD. We hope to come back to these issues in the near future.

\section*{Acknowledgments}
We would like to thank D. Choudhuri for careful reading of the paper and pointing out the necessary corrections. The work of P.~J.~has been supported by an appointment to the JRG Program at the APCTP through the Science and Technology Promotion Fund and Lottery Fund of the Korean Government. P.~J. is also supported by the Korean Local Governments -- Gyeong\-sang\-buk-do Province and Pohang City -- and by the National Research Foundation of Korea (NRF) funded by the Korean government (MSIT) (grant number 2021R1A2C1010834).
The work of S.~S.~J. is supported by Grant No. 09/983(0045)/2019-EMR-I from CSIR-HRDG, India. The work of S.~M.~is supported by the Department of Science and Technology, Government of India under the Grant Agreement number IFA17-PH207 (INSPIRE Faculty Award).


\begin{thebibliography}{300}
\bibitem{Maldacena:1997re}
 J.~M.~Maldacena,
  ``The Large N limit of superconformal field theories and supergravity,''
  Int.\ J.\ Theor.\ Phys.\  {\bf 38}, 1113 (1999)
  [Adv.\ Theor.\ Math.\ Phys.\  {\bf 2}, 231 (1998)]
  [hep-th/9711200].

 \bibitem{Gubser:1998bc}
  S.~S.~Gubser, I.~R.~Klebanov and A.~M.~Polyakov,
  ``Gauge theory correlators from noncritical string theory,''
  Phys.\ Lett.\ B {\bf 428}, 105 (1998)
  [hep-th/9802109].

  \bibitem{Witten:1998qj}
  E.~Witten,
  ``Anti-de Sitter space and holography,''
  Adv.\ Theor.\ Math.\ Phys.\  {\bf 2}, 253 (1998)
  [hep-th/9802150].

  \bibitem{Klebanov:2007ws}
I.~R.~Klebanov, D.~Kutasov and A.~Murugan,
``Entanglement as a probe of confinement,''
Nucl. Phys. B \textbf{796} (2008), 274-293
[arXiv:0709.2140 [hep-th]].

\bibitem{Nishioka:2006gr}
T.~Nishioka and T.~Takayanagi,
``AdS Bubbles, Entropy and Closed String Tachyons,''
JHEP \textbf{01} (2007), 090
[arXiv:hep-th/0611035 [hep-th]].


\bibitem{Vidal:2002rm}
G.~Vidal, J.~I.~Latorre, E.~Rico and A.~Kitaev,
``Entanglement in quantum critical phenomena,''
Phys. Rev. Lett. \textbf{90} (2003), 227902
[arXiv:quant-ph/0211074 [quant-ph]].

\bibitem{Osborne:2002zz}
  T.~J.~Osborne and M.~A.~Nielsen,
  ``Entanglement in a simple quantum phase transition,''
  Phys.\ Rev.\ A {\bf 66}, 032110 (2002).

\bibitem{Bombelli:1986rw}
  L.~Bombelli, R.~K.~Koul, J.~Lee and R.~D.~Sorkin,
  ``A Quantum Source of Entropy for Black Holes,''
  Phys.\ Rev.\ D {\bf 34}, 373 (1986).

\bibitem{Srednicki:1993im}
  M.~Srednicki,
  ``Entropy and area,''
  Phys.\ Rev.\ Lett.\  {\bf 71}, 666 (1993)
  [hep-th/9303048].

\bibitem{Hoi}
H-K.~Lo, ``Classical Communication Cost in Distributed Quantum Information Processing - A generalization of Quantum Communication Complexity,'' Phys. Rev. A {\bf62} (2000) 012313 [arXiv:quant-ph/9912009].

\bibitem{Karpov}
E.~Karpov, D.~Daems and N.~J.~Cerf, ``Entanglement enhanced classical capacity of quantum communication channels with correlated noise in arbitrary dimensions,'' Phys. Rev. A {\bf74} (2006) 032320 [arXiv:quant-ph/0603286].

\bibitem{Ryu:2006bv}
S.~Ryu and T.~Takayanagi,
``Holographic derivation of entanglement entropy from AdS/CFT,''
Phys. Rev. Lett. \textbf{96} (2006), 181602
[arXiv:hep-th/0603001 [hep-th]].

\bibitem{Ryu:2006ef}
S.~Ryu and T.~Takayanagi,
``Aspects of Holographic Entanglement Entropy,''
JHEP \textbf{08} (2006), 045
[arXiv:hep-th/0605073 [hep-th]].

\bibitem{Pastawski:2015qua}
F.~Pastawski, B.~Yoshida, D.~Harlow and J.~Preskill,
``Holographic quantum error-correcting codes: Toy models for the bulk/boundary correspondence,''
JHEP \textbf{06} (2015), 149
[arXiv:1503.06237 [hep-th]].

\bibitem{Hayden:2016cfa}
P.~Hayden, S.~Nezami, X.~L.~Qi, N.~Thomas, M.~Walter and Z.~Yang,
``Holographic duality from random tensor networks,''
JHEP \textbf{11} (2016), 009
[arXiv:1601.01694 [hep-th]].

\bibitem{Johnson:2013dka}
C.~V.~Johnson,
``Large N Phase Transitions, Finite Volume, and Entanglement Entropy,''
JHEP \textbf{03} (2014), 047
[arXiv:1306.4955 [hep-th]].

\bibitem{Dey:2015ytd}
A.~Dey, S.~Mahapatra and T.~Sarkar,
``Thermodynamics and Entanglement Entropy with Weyl Corrections,''
Phys. Rev. D \textbf{94} (2016) no.2, 026006
[arXiv:1512.07117 [hep-th]].

\bibitem{Dey:2014voa}
A.~Dey, S.~Mahapatra and T.~Sarkar,
``Very General Holographic Superconductors and Entanglement Thermodynamics,''
JHEP \textbf{12}, 135 (2014)
[arXiv:1409.5309 [hep-th]].

\bibitem{VanRaamsdonk:2010pw}
  M.~Van Raamsdonk,
  ``Building up spacetime with quantum entanglement,''
  Gen.\ Rel.\ Grav.\  {\bf 42}, 2323 (2010)
  [Int.\ J.\ Mod.\ Phys.\ D {\bf 19}, 2429 (2010)]
  [arXiv:1005.3035 [hep-th]].

 \bibitem{Balasubramanian:2013lsa}
  V.~Balasubramanian, B.~D.~Chowdhury, B.~Czech, J.~de Boer and M.~P.~Heller,
  ``Bulk curves from boundary data in holography,''
  Phys.\ Rev.\ D {\bf 89}, no. 8, 086004 (2014)
  [arXiv:1310.4204 [hep-th]].

\bibitem{Balasubramanian:2011ur}
V.~Balasubramanian, A.~Bernamonti, J.~de Boer, N.~Copland, B.~Craps, E.~Keski-Vakkuri, B.~Muller, A.~Schafer, M.~Shigemori and W.~Staessens,
``Holographic Thermalization,''
Phys. Rev. D \textbf{84} (2011), 026010
[arXiv:1103.2683 [hep-th]].

\bibitem{Liu:2013iza}
H.~Liu and S.~J.~Suh,
``Entanglement Tsunami: Universal Scaling in Holographic Thermalization,''
Phys. Rev. Lett. \textbf{112} (2014), 011601
[arXiv:1305.7244 [hep-th]].

\bibitem{Dey:2015poa}
A.~Dey, S.~Mahapatra and T.~Sarkar,
``Holographic Thermalization with Weyl Corrections,''
JHEP \textbf{01} (2016), 088
[arXiv:1510.00232 [hep-th]].

\bibitem{Vidal:2002zz}
G.~Vidal and R.~F.~Werner,
``Computable measure of entanglement,''
Phys. Rev. A \textbf{65} (2002), 032314
[arXiv:0102117 [quant-ph]].

\bibitem{Horodecki1996}
R.~Horodecki, P.~Horodecki, M.~Horodecki and K.~Horodecki,
``Quantum entanglement,''
Rev. Mod. Phys. \textbf{81} (2009) 865 [arXiv:0702225 [quant-ph]].

\bibitem{Terhal}
B.~M.~Terhal, M.~Horodecki, D.~W.~Leung and D.~P.~DiVincenzo, ``The entanglement of purification,'' J. Math. Phys. {\bf 43} (2002) 4286, [arXiv:quant-ph/0202044].

\bibitem{Eisert}
J.~Eisert and M.~B.~Plenio, ``A comparison of entanglement measures,'' Journal of Modern Optics \textbf{46} (1999) 145 [arXiv:9807034 [quant-ph]].

\bibitem{Horodecki}
M.~Horodecki, P.~Horodecki and R.~Horodecki, ``Separability of mixed states: necessary and sufficient conditions,'' Phys. Lett. A \textbf{223} (1996) 1 [arXiv:9605038 [quant-ph]].

\bibitem{Peres:1996dw}
A.~Peres,
``Separability criterion for density matrices,''
Phys. Rev. Lett. \textbf{77} (1996), 1413-1415
[arXiv:9604005 [quant-ph]].

\bibitem{Takayanagi:2017knl}
T.~Takayanagi and K.~Umemoto,
``Entanglement of purification through holographic duality,''
Nature Phys. \textbf{14} (2018) no.6, 573-577
[arXiv:1708.09393 [hep-th]].

\bibitem{Nguyen:2017yqw}
P.~Nguyen, T.~Devakul, M.~G.~Halbasch, M.~P.~Zaletel and B.~Swingle,
``Entanglement of purification: from spin chains to holography,''
JHEP \textbf{01} (2018), 098
[arXiv:1709.07424 [hep-th]].

\bibitem{Kudler-Flam:2018qjo}
J.~Kudler-Flam and S.~Ryu,
``Entanglement negativity and minimal entanglement wedge cross sections in holographic theories,''
Phys. Rev. D \textbf{99} (2019) no.10, 106014
[arXiv:1808.00446 [hep-th]].

\bibitem{Kusuki:2019zsp}
Y.~Kusuki, J.~Kudler-Flam and S.~Ryu,
``Derivation of Holographic Negativity in AdS$_3$/CFT$_2$,''
Phys. Rev. Lett. \textbf{123} (2019) no.13, 131603
[arXiv:1907.07824 [hep-th]].

\bibitem{Chaturvedi:2016rft}
P.~Chaturvedi, V.~Malvimat and G.~Sengupta,
``Entanglement negativity, Holography and Black holes,''
Eur. Phys. J. C \textbf{78} (2018) no.6, 499
[arXiv:1602.01147 [hep-th]].

\bibitem{Chaturvedi:2016rcn}
P.~Chaturvedi, V.~Malvimat and G.~Sengupta,
``Holographic Quantum Entanglement Negativity,''
JHEP \textbf{05} (2018), 172
[arXiv:1609.06609 [hep-th]].

\bibitem{Jain:2017aqk}
P.~Jain, V.~Malvimat, S.~Mondal and G.~Sengupta,
``Holographic entanglement negativity conjecture for adjacent intervals in $AdS_3/CFT_2$,''
Phys. Lett. B \textbf{793} (2019), 104-109
[arXiv:1707.08293 [hep-th]].

\bibitem{Jain:2017xsu}
P.~Jain, V.~Malvimat, S.~Mondal and G.~Sengupta,
``Holographic entanglement negativity for adjacent subsystems in AdS$_{d+1}$/CFT$_{d}$,''
Eur. Phys. J. Plus \textbf{133} (2018) no.8, 300
[arXiv:1708.00612 [hep-th]].

\bibitem{Jain:2017uhe}
P.~Jain, V.~Malvimat, S.~Mondal and G.~Sengupta,
``Covariant holographic entanglement negativity for adjacent subsystems in AdS$_3$ /CFT$_2$,''
Nucl. Phys. B \textbf{945} (2019), 114683
[arXiv:1710.06138 [hep-th]].

\bibitem{Jain:2018bai}
P.~Jain, V.~Malvimat, S.~Mondal and G.~Sengupta,
``Holographic Entanglement Negativity for Conformal Field Theories with a Conserved Charge,''
Eur. Phys. J. C \textbf{78} (2018) no.11, 908
[arXiv:1804.09078 [hep-th]].

\bibitem{Malvimat:2018txq}
V.~Malvimat, S.~Mondal, B.~Paul and G.~Sengupta,
``Holographic entanglement negativity for disjoint intervals in $AdS_3/CFT_2$,''
Eur. Phys. J. C \textbf{79} (2019) no.3, 191
[arXiv:1810.08015 [hep-th]].

\bibitem{Malvimat:2018izs}
V.~Malvimat, H.~Parihar, B.~Paul and G.~Sengupta,
``Entanglement Negativity in Galilean Conformal Field Theories,''
Phys. Rev. D \textbf{100} (2019) no.2, 026001
[arXiv:1810.08162 [hep-th]].

\bibitem{Basak:2020bot}
J.~Kumar Basak, H.~Parihar, B.~Paul and G.~Sengupta,
``Holographic entanglement negativity for disjoint subsystems in $\mathrm{AdS_{d+1}/CFT_d}$,''
[arXiv:2001.10534 [hep-th]].

\bibitem{Malvimat:2018cfe}
V.~Malvimat, S.~Mondal and G.~Sengupta,
``Time Evolution of Entanglement Negativity from Black Hole Interiors,''
JHEP \textbf{05} (2019), 183
[arXiv:1812.04424 [hep-th]].

\bibitem{Dutta:2019gen}
S.~Dutta and T.~Faulkner,
``A canonical purification for the entanglement wedge cross-section,''
JHEP \textbf{03}, 178 (2021)
[arXiv:1905.00577 [hep-th]].

\bibitem{Tamaoka:2018ned}
K.~Tamaoka,
``Entanglement Wedge Cross Section from the Dual Density Matrix,''
Phys. Rev. Lett. \textbf{122}, no.14, 141601 (2019)
[arXiv:1809.09109 [hep-th]].

\bibitem{Akers:2019gcv}
C.~Akers and P.~Rath,
``Entanglement Wedge Cross Sections Require Tripartite Entanglement,''
JHEP \textbf{04} (2020), 208
[arXiv:1911.07852 [hep-th]].

\bibitem{Jain:2020rbb}
P.~Jain and S.~Mahapatra,
``Mixed state entanglement measures as probe for confinement,''
Phys. Rev. D \textbf{102}, 126022 (2020)
[arXiv:2010.07702 [hep-th]].

\bibitem{Caputa:2018xuf}
P.~Caputa, M.~Miyaji, T.~Takayanagi and K.~Umemoto,
``Holographic Entanglement of Purification from Conformal Field Theories,''
Phys. Rev. Lett. \textbf{122}, no.11, 111601 (2019)
[arXiv:1812.05268 [hep-th]].

\bibitem{Umemoto:2018jpc}
K.~Umemoto and Y.~Zhou,
``Entanglement of Purification for Multipartite States and its Holographic Dual,''
JHEP \textbf{10}, 152 (2018)
[arXiv:1805.02625 [hep-th]].

\bibitem{Bao:2018gck}
N.~Bao and I.~F.~Halpern,
``Conditional and Multipartite Entanglements of Purification and Holography,''
Phys. Rev. D \textbf{99}, no.4, 046010 (2019)
[arXiv:1805.00476 [hep-th]].

\bibitem{Espindola:2018ozt}
R.~Esp\'\i{}ndola, A.~Guijosa and J.~F.~Pedraza,
``Entanglement Wedge Reconstruction and Entanglement of Purification,''
Eur. Phys. J. C \textbf{78}, no.8, 646 (2018)
[arXiv:1804.05855 [hep-th]].

\bibitem{Jeong:2019xdr}
H.~S.~Jeong, K.~Y.~Kim and M.~Nishida,
``Reflected Entropy and Entanglement Wedge Cross Section with the First Order Correction,''
JHEP \textbf{12}, 170 (2019)
[arXiv:1909.02806 [hep-th]].

\bibitem{Jokela:2019ebz}
N.~Jokela and A.~P\"onni,
``Notes on entanglement wedge cross sections,''
JHEP \textbf{07}, 087 (2019)
[arXiv:1904.09582 [hep-th]].

\bibitem{Ghodrati:2022hbb}
M.~Ghodrati,
``Encoded information of mixed correlations: the views from one dimension higher,''
[arXiv:2209.04548 [hep-th]].

\bibitem{Bhattacharyya:2019tsi}
A.~Bhattacharyya, A.~Jahn, T.~Takayanagi and K.~Umemoto,
``Entanglement of Purification in Many Body Systems and Symmetry Breaking,''
Phys. Rev. Lett. \textbf{122}, no.20, 201601 (2019)
[arXiv:1902.02369 [hep-th]].

\bibitem{Camargo:2021aiq}
H.~A.~Camargo, L.~Hackl, M.~P.~Heller, A.~Jahn and B.~Windt,
``Long Distance Entanglement of Purification and Reflected Entropy in Conformal Field Theory,''
Phys. Rev. Lett. \textbf{127}, no.14, 141604 (2021)
[arXiv:2102.00013 [hep-th]].

\bibitem{Banuls:2022iwk}
M.~C.~Banuls, M.~P.~Heller, K.~Jansen, J.~Knaute and V.~Svensson,
``A quantum information perspective on meson melting,''
[arXiv:2206.10528 [hep-th]].

\bibitem{Asadi:2022mvo}
M.~Asadi, B.~Amrahi and H.~Eshaghi-Kenari,
``Probing Phase Structure of Strongly Coupled Matter with Holographic Entanglement Measures,''
[arXiv:2209.01586 [hep-th]].

\bibitem{Ali-Akbari:2021zsm}
M.~Ali-Akbari, M.~Asadi and B.~Amrahi,
``Non-conformal behavior of holographic entanglement measures,''
JHEP \textbf{04}, 014 (2022)
[arXiv:2112.02565 [hep-th]].

\bibitem{Vasli:2022kfu}
M.~J.~Vasli, M.~R.~Mohammadi Mozaffar, K.~Babaei Velni and M.~Sahraei,
``Holographic Study of Reflected Entropy in Anisotropic Theories,''
[arXiv:2207.14169 [hep-th]].

\bibitem{Liu:2021rks}
P.~Liu, C.~Niu, Z.~J.~Shi and C.~Y.~Zhang,
``Entanglement wedge minimum cross-section in holographic massive gravity theory,''
JHEP \textbf{08}, 113 (2021)
[arXiv:2104.08070 [hep-th]].

\bibitem{Liu:2020blk}
P.~Liu and J.~P.~Wu,
``Mixed state entanglement and thermal phase transitions,''
Phys. Rev. D \textbf{104}, no.4, 046017 (2021)
[arXiv:2009.01529 [hep-th]].

\bibitem{Saha:2021kwq}
A.~Saha and S.~Gangopadhyay,
``Holographic study of entanglement and complexity for mixed states,''
Phys. Rev. D \textbf{103}, no.8, 086002 (2021)
[arXiv:2101.00887 [hep-th]].

\bibitem{Chowdhury:2021idy}
A.~R.~Chowdhury, A.~Saha and S.~Gangopadhyay,
``Entanglement wedge cross-section for noncommutative Yang-Mills theory,''
JHEP \textbf{02}, 192 (2022)
[arXiv:2106.04562 [hep-th]].

\bibitem{Casalderrey-Solana:2011dxg}
J.~Casalderrey-Solana, H.~Liu, D.~Mateos, K.~Rajagopal and U.~A.~Wiedemann,
``Gauge/String Duality, Hot QCD and Heavy Ion Collisions,''
Cambridge University Press, 2014,
ISBN 978-1-139-13674-7
[arXiv:1101.0618 [hep-th]].

\bibitem{Gubser:2009md}
S.~S.~Gubser and A.~Karch,
``From gauge-string duality to strong interactions: A Pedestrian's Guide,''
Ann. Rev. Nucl. Part. Sci. \textbf{59}, 145-168 (2009)
[arXiv:0901.0935 [hep-th]].

\bibitem{Jarvinen:2021jbd}
M.~J\"arvinen,
``Holographic modeling of nuclear matter and neutron stars,''
Eur. Phys. J. C \textbf{82}, no.4, 282 (2022)
[arXiv:2110.08281 [hep-ph]].

\bibitem{Gursoy:2010fj}
U.~Gursoy, E.~Kiritsis, L.~Mazzanti, G.~Michalogiorgakis and F.~Nitti,
``Improved Holographic QCD,''
Lect. Notes Phys. \textbf{828}, 79-146 (2011)
[arXiv:1006.5461 [hep-th]].

\bibitem{Skokov:2009qp}
V.~Skokov, A.~Y.~Illarionov and V.~Toneev,
``Estimate of the magnetic field strength in heavy-ion collisions,''
Int. J. Mod. Phys. A \textbf{24}, 5925-5932 (2009)
[arXiv:0907.1396 [nucl-th]].

\bibitem{Bzdak:2011yy}
A.~Bzdak and V.~Skokov,
``Event-by-event fluctuations of magnetic and electric fields in heavy ion collisions,''
Phys. Lett. B \textbf{710}, 171-174 (2012)
[arXiv:1111.1949 [hep-ph]].

\bibitem{Voronyuk:2011jd}
V.~Voronyuk, V.~D.~Toneev, W.~Cassing, E.~L.~Bratkovskaya, V.~P.~Konchakovski and S.~A.~Voloshin,
``(Electro-)Magnetic field evolution in relativistic heavy-ion collisions,''
Phys. Rev. C \textbf{83}, 054911 (2011)
[arXiv:1103.4239 [nucl-th]].

\bibitem{Deng:2012pc}
W.~T.~Deng and X.~G.~Huang,
``Event-by-event generation of electromagnetic fields in heavy-ion collisions,''
Phys. Rev. C \textbf{85}, 044907 (2012)
[arXiv:1201.5108 [nucl-th]].

\bibitem{DElia:2010abb}
M.~D'Elia, S.~Mukherjee and F.~Sanfilippo,
``QCD Phase Transition in a Strong Magnetic Background,''
Phys. Rev. D \textbf{82}, 051501 (2010)
[arXiv:1005.5365 [hep-lat]].

\bibitem{DElia:2021tfb}
M.~D'Elia, L.~Maio, F.~Sanfilippo and A.~Stanzione,
``Confining and chiral properties of QCD in extremely strong magnetic fields,''
Phys. Rev. D \textbf{104}, no.11, 114512 (2021)
[arXiv:2109.07456 [hep-lat]].

\bibitem{Tuchin:2013ie}
K.~Tuchin,
``Particle production in strong electromagnetic fields in relativistic heavy-ion collisions,''
Adv. High Energy Phys. \textbf{2013}, 490495 (2013)
[arXiv:1301.0099 [hep-ph]].

\bibitem{Tuchin:2013apa}
K.~Tuchin,
``Time and space dependence of the electromagnetic field in relativistic heavy-ion collisions,''
Phys. Rev. C \textbf{88}, no.2, 024911 (2013)
[arXiv:1305.5806 [hep-ph]].

\bibitem{McLerran:2013hla}
L.~McLerran and V.~Skokov,
``Comments About the Electromagnetic Field in Heavy-Ion Collisions,''
Nucl. Phys. A \textbf{929}, 184-190 (2014)
[arXiv:1305.0774 [hep-ph]].

\bibitem{Bali:2011qj}
G.~S.~Bali, F.~Bruckmann, G.~Endrodi, Z.~Fodor, S.~D.~Katz, S.~Krieg, A.~Schafer and K.~K.~Szabo,
``The QCD phase diagram for external magnetic fields,''
JHEP \textbf{02}, 044 (2012)
[arXiv:1111.4956 [hep-lat]].

\bibitem{Bali:2012zg}
G.~S.~Bali, F.~Bruckmann, G.~Endrodi, Z.~Fodor, S.~D.~Katz and A.~Schafer,
``QCD quark condensate in external magnetic fields,''
Phys. Rev. D \textbf{86}, 071502 (2012)
[arXiv:1206.4205 [hep-lat]].

\bibitem{Ilgenfritz:2013ara}
E.~M.~Ilgenfritz, M.~Muller-Preussker, B.~Petersson and A.~Schreiber,
``Magnetic catalysis (and inverse catalysis) at finite temperature in two-color lattice QCD,''
Phys. Rev. D \textbf{89}, no.5, 054512 (2014)
[arXiv:1310.7876 [hep-lat]].

\bibitem{Bruckmann:2013oba}
F.~Bruckmann, G.~Endrodi and T.~G.~Kovacs,
``Inverse magnetic catalysis and the Polyakov loop,''
JHEP \textbf{04}, 112 (2013)
[arXiv:1303.3972 [hep-lat]].

\bibitem{Fukushima:2012kc}
K.~Fukushima and Y.~Hidaka,
``Magnetic Catalysis Versus Magnetic Inhibition,''
Phys. Rev. Lett. \textbf{110}, no.3, 031601 (2013)
[arXiv:1209.1319 [hep-ph]].

\bibitem{Ferreira:2014kpa}
M.~Ferreira, P.~Costa, O.~Louren\c{c}o, T.~Frederico and C.~Provid\^encia,
``Inverse magnetic catalysis in the (2+1)-flavor Nambu-Jona-Lasinio and Polyakov-Nambu-Jona-Lasinio models,''
Phys. Rev. D \textbf{89}, no.11, 116011 (2014)
[arXiv:1404.5577 [hep-ph]].

\bibitem{Mueller:2015fka}
N.~Mueller and J.~M.~Pawlowski,
``Magnetic catalysis and inverse magnetic catalysis in QCD,''
Phys. Rev. D \textbf{91}, no.11, 116010 (2015)
[arXiv:1502.08011 [hep-ph]].

\bibitem{Bali:2013esa}
G.~S.~Bali, F.~Bruckmann, G.~Endrodi, F.~Gruber and A.~Schaefer,
``Magnetic field-induced gluonic (inverse) catalysis and pressure (an)isotropy in QCD,''
JHEP \textbf{04}, 130 (2013)
[arXiv:1303.1328 [hep-lat]].

\bibitem{Fraga:2012fs}
E.~S.~Fraga and L.~F.~Palhares,
``Deconfinement in the presence of a strong magnetic background: an exercise within the MIT bag model,''
Phys. Rev. D \textbf{86}, 016008 (2012)
[arXiv:1201.5881 [hep-ph]].

\bibitem{Ayala:2014iba}
A.~Ayala, M.~Loewe, A.~J.~Mizher and R.~Zamora,
``Inverse magnetic catalysis for the chiral transition induced by thermo-magnetic effects on the coupling constant,''
Phys. Rev. D \textbf{90}, no.3, 036001 (2014)
[arXiv:1406.3885 [hep-ph]].

\bibitem{Ayala:2014gwa}
A.~Ayala, M.~Loewe and R.~Zamora,
``Inverse magnetic catalysis in the linear sigma model with quarks,''
Phys. Rev. D \textbf{91}, no.1, 016002 (2015)
[arXiv:1406.7408 [hep-ph]].

\bibitem{Fraga:2012ev}
E.~S.~Fraga, J.~Noronha and L.~F.~Palhares,
``Large $N_c$ Deconfinement Transition in the Presence of a Magnetic Field,''
Phys. Rev. D \textbf{87}, no.11, 114014 (2013)
[arXiv:1207.7094 [hep-ph]].

\bibitem{Bonati:2014ksa}
C.~Bonati, M.~D'Elia, M.~Mariti, M.~Mesiti, F.~Negro and F.~Sanfilippo,
``Anisotropy of the quark-antiquark potential in a magnetic field,''
Phys. Rev. D \textbf{89}, no.11, 114502 (2014)
[arXiv:1403.6094 [hep-lat]].

\bibitem{Bonati:2016kxj}
C.~Bonati, M.~D'Elia, M.~Mariti, M.~Mesiti, F.~Negro, A.~Rucci and F.~Sanfilippo,
``Magnetic field effects on the static quark potential at zero and finite temperature,''
Phys. Rev. D \textbf{94}, no.9, 094007 (2016)
[arXiv:1607.08160 [hep-lat]].

\bibitem{Fukushima:2008xe}
K.~Fukushima, D.~E.~Kharzeev and H.~J.~Warringa,
``The Chiral Magnetic Effect,''
Phys. Rev. D \textbf{78}, 074033 (2008)
[arXiv:0808.3382 [hep-ph]].

\bibitem{Kharzeev:2007jp}
D.~E.~Kharzeev, L.~D.~McLerran and H.~J.~Warringa,
``The Effects of topological charge change in heavy ion collisions: 'Event by event P and CP violation',''
Nucl. Phys. A \textbf{803}, 227-253 (2008)
[arXiv:0711.0950 [hep-ph]].

\bibitem{Kharzeev:2015znc}
D.~E.~Kharzeev, J.~Liao, S.~A.~Voloshin and G.~Wang,
``Chiral magnetic and vortical effects in high-energy nuclear collisions\textemdash{}A status report,''
Prog. Part. Nucl. Phys. \textbf{88}, 1-28 (2016)
[arXiv:1511.04050 [hep-ph]].

\bibitem{Johnson:2008vna}
C.~V.~Johnson and A.~Kundu,
``External Fields and Chiral Symmetry Breaking in the Sakai-Sugimoto Model,''
JHEP \textbf{12}, 053 (2008)
[arXiv:0803.0038 [hep-th]].

\bibitem{Callebaut:2011ab}
  N.~Callebaut, D.~Dudal and H.~Verschelde,
  ``Holographic rho mesons in an external magnetic field,''
  JHEP {\bf 1303}, 033 (2013)
  [arXiv:1105.2217 [hep-th]].

\bibitem{Callebaut:2013ria}
  N.~Callebaut and D.~Dudal,
  ``Transition temperature(s) of magnetized two-flavor holographic QCD,''
  Phys.\ Rev.\ D {\bf 87}, no. 10, 106002 (2013)
  [arXiv:1303.5674 [hep-th]].

\bibitem{Dudal:2015wfn}
  D.~Dudal, D.~R.~Granado and T.~G.~Mertens,
  ``No inverse magnetic catalysis in the QCD hard and soft wall models,''
  Phys.\ Rev.\ D {\bf 93} (2016) no.12,  125004
  [arXiv:1511.04042 [hep-th]].

  \bibitem{Dudal:2014jfa}
  D.~Dudal and T.~G.~Mertens,
  ``Melting of charmonium in a magnetic field from an effective AdS/QCD model,''
  Phys.\ Rev.\ D {\bf 91}, 086002 (2015)
  [arXiv:1410.3297 [hep-th]].

\bibitem{Dudal:2018rki}
  D.~Dudal and T.~G.~Mertens,
  ``Holographic estimate of heavy quark diffusion in a magnetic field,''
  Phys.\ Rev.\ D {\bf 97}, no. 5, 054035 (2018)
  [arXiv:1802.02805 [hep-th]].

  \bibitem{Gursoy:2017wzz}
U.~Gursoy, M.~Jarvinen and G.~Nijs,
``Holographic QCD in the Veneziano Limit at a Finite Magnetic Field and Chemical Potential,''
Phys. Rev. Lett. \textbf{120}, no.24, 242002 (2018)
[arXiv:1707.00872 [hep-th]].

\bibitem{Jokela:2013qya}
N.~Jokela, A.~V.~Ramallo and D.~Zoakos,
``Magnetic catalysis in flavored ABJM,''
JHEP \textbf{02}, 021 (2014)
[arXiv:1311.6265 [hep-th]].

\bibitem{Gursoy:2016ofp}
U.~G\"ursoy, I.~Iatrakis, M.~J\"arvinen and G.~Nijs,
``Inverse Magnetic Catalysis from improved Holographic QCD in the Veneziano limit,''
JHEP \textbf{03}, 053 (2017)
[arXiv:1611.06339 [hep-th]].

\bibitem{Li:2016gfn}
  D.~Li, M.~Huang, Y.~Yang and P.~H.~Yuan,
  ``Inverse Magnetic Catalysis in the Soft-Wall Model of AdS/QCD,''
  JHEP {\bf 1702}, 030 (2017)
  [arXiv:1610.04618 [hep-th]].

\bibitem{Critelli:2016cvq}
  R.~Critelli, R.~Rougemont, S.~I.~Finazzo and J.~Noronha,
  ``Polyakov loop and heavy quark entropy in strong magnetic fields from holographic black hole engineering,''
  Phys.\ Rev.\ D {\bf 94}, no. 12, 125019 (2016)
  [arXiv:1606.09484 [hep-ph]].

\bibitem{Giataganas:2017koz}
D.~Giataganas, U.~G\"ursoy and J.~F.~Pedraza,
``Strongly-coupled anisotropic gauge theories and holography,''
Phys. Rev. Lett. \textbf{121}, no.12, 121601 (2018)
[arXiv:1708.05691 [hep-th]].

\bibitem{Gursoy:2020kjd}
U.~G\"ursoy, M.~J\"arvinen, G.~Nijs and J.~F.~Pedraza,
``On the interplay between magnetic field and anisotropy in holographic QCD,''
JHEP \textbf{03}, 180 (2021)
[arXiv:2011.09474 [hep-th]].

\bibitem{Gursoy:2018ydr}
U.~G\"ursoy, M.~J\"arvinen, G.~Nijs and J.~F.~Pedraza,
``Inverse Anisotropic Catalysis in Holographic QCD,''
JHEP \textbf{04}, 071 (2019)
[erratum: JHEP \textbf{09}, 059 (2020)]
[arXiv:1811.11724 [hep-th]].

\bibitem{Rodrigues:2017cha}
D.~M.~Rodrigues, E.~Folco Capossoli and H.~Boschi-Filho,
``Deconfinement phase transition in a magnetic field in 2 + 1 dimensions from holographic models,''
Phys. Lett. B \textbf{780}, 37-40 (2018)
[arXiv:1709.09258 [hep-th]].

\bibitem{Rodrigues:2018pep}
D.~M.~Rodrigues, D.~Li, E.~Folco Capossoli and H.~Boschi-Filho,
``Chiral symmetry breaking and restoration in 2+1 dimensions from holography: Magnetic and inverse magnetic catalysis,''
Phys. Rev. D \textbf{98}, no.10, 106007 (2018)
[arXiv:1807.11822 [hep-th]].

\bibitem{Bohra:2019ebj}
H.~Bohra, D.~Dudal, A.~Hajilou and S.~Mahapatra,
``Anisotropic string tensions and inversely magnetic catalyzed deconfinement from a dynamical AdS/QCD model,''
Phys. Lett. B \textbf{801}, 135184 (2020)
[arXiv:1907.01852 [hep-th]].

\bibitem{Bohra:2020qom}
H.~Bohra, D.~Dudal, A.~Hajilou and S.~Mahapatra,
``Chiral transition in the probe approximation from an Einstein-Maxwell-dilaton gravity model,''
Phys. Rev. D \textbf{103}, no.8, 086021 (2021)
[arXiv:2010.04578 [hep-th]].

\bibitem{Dudal:2021jav}
D.~Dudal, A.~Hajilou and S.~Mahapatra,
``A quenched 2-flavour Einstein\textendash{}Maxwell\textendash{}Dilaton gauge-gravity model,''
Eur. Phys. J. A \textbf{57}, no.4, 142 (2021)
[arXiv:2103.01185 [hep-th]].

\bibitem{Arefeva:2022avn}
I.~Y.~Aref'eva, A.~Ermakov, K.~Rannu and P.~Slepov,
``Holographic model for light quarks in anisotropic hot dense QGP with external magnetic field,''
[arXiv:2203.12539 [hep-th]].

\bibitem{Arefeva:2020vae}
I.~Y.~Aref'eva, K.~Rannu and P.~Slepov,
``Holographic model for heavy quarks in anisotropic hot dense QGP with external magnetic field,''
JHEP \textbf{07}, 161 (2021)
[arXiv:2011.07023 [hep-th]].

\bibitem{Arefeva:2018cli}
I.~Aref'eva, K.~Rannu and P.~Slepov,
``Orientation Dependence of Confinement-Deconfinement Phase Transition in Anisotropic Media,''
Phys. Lett. B \textbf{792}, 470-475 (2019)
[arXiv:1808.05596 [hep-th]].

\bibitem{Arefeva:2018hyo}
I.~Aref'eva and K.~Rannu,
``Holographic Anisotropic Background with Confinement-Deconfinement Phase Transition,''
JHEP \textbf{05}, 206 (2018)
[arXiv:1802.05652 [hep-th]].

\bibitem{Jena:2022nzw}
S.~S.~Jena, B.~Shukla, D.~Dudal and S.~Mahapatra,
``Entropic force and real-time dynamics of holographic quarkonium in a magnetic field,''
Phys. Rev. D \textbf{105}, no.8, 086011 (2022)
[arXiv:2202.01486 [hep-th]].

\bibitem{Ballon-Bayona:2022uyy}
A.~Ballon-Bayona, J.~P.~Shock and D.~Zoakos,
``Magnetising the $ \mathcal{N} $ = 4 Super Yang-Mills plasma,''
JHEP \textbf{06}, 154 (2022)
[arXiv:2203.00050 [hep-th]].

\bibitem{Arefeva:2020bjk}
I.~Y.~Aref'eva, K.~Rannu and P.~Slepov,
``Energy Loss in Holographic Anisotropic Model for Heavy Quarks in External Magnetic Field,''
[arXiv:2012.05758 [hep-th]].

\bibitem{Ballon-Bayona:2020xtf}
A.~Ballon-Bayona, J.~P.~Shock and D.~Zoakos,
``Magnetic catalysis and the chiral condensate in holographic QCD,''
JHEP \textbf{10}, 193 (2020)
[arXiv:2005.00500 [hep-th]].

\bibitem{He:2020fdi}
S.~He, Y.~Yang and P.~H.~Yuan,
``Analytic Study of Magnetic Catalysis in Holographic QCD,''
[arXiv:2004.01965 [hep-th]].

\bibitem{Zhu:2019igg}
Z.~R.~Zhu, D.~f.~Hou and X.~Chen,
``Potential analysis of holographic Schwinger effect in the magnetized background,''
Eur. Phys. J. C \textbf{80}, no.6, 550 (2020)
[arXiv:1912.05806 [hep-ph]].

\bibitem{Braga:2020hhs}
N.~R.~F.~Braga and R.~da Mata,
``Configuration entropy description of charmonium dissociation under the influence of magnetic fields,''
Phys. Lett. B \textbf{811}, 135918 (2020)
[arXiv:2008.10457 [hep-th]].

\bibitem{Mamo:2015dea}
K.~A.~Mamo,
``Inverse magnetic catalysis in holographic models of QCD,''
JHEP \textbf{05}, 121 (2015)
[arXiv:1501.03262 [hep-th]].

\bibitem{Avila:2018sqf}
D.~\'Avila, V.~Jahnke and L.~Pati\~no,
``Chaos, Diffusivity, and Spreading of Entanglement in Magnetic Branes, and the Strengthening of the Internal Interaction,''
JHEP \textbf{09}, 131 (2018)
[arXiv:1805.05351 [hep-th]].

\bibitem{Giataganas:2012zy}
D.~Giataganas,
``Probing strongly coupled anisotropic plasma,''
JHEP \textbf{07}, 031 (2012)
[arXiv:1202.4436 [hep-th]].

\bibitem{STAR:2021mii}
 M.~Abdallah \textit{et al.} [STAR],
``Search for the chiral magnetic effect with isobar collisions at $\sqrt {s_{NN}}$=200 GeV by the STAR Collaboration at the BNL Relativistic Heavy Ion Collider,''
Phys. Rev. C \textbf{105}, no.1, 014901 (2022)
[arXiv:2109.00131 [nucl-ex]].

\bibitem{Duncan:1992hi}
R.~C.~Duncan and C.~Thompson,
``Formation of very strongly magnetized neutron stars - implications for gamma-ray bursts,''
Astrophys. J. Lett. \textbf{392}, L9 (1992).

\bibitem{Grasso:2000wj}
D.~Grasso and H.~R.~Rubinstein,
``Magnetic fields in the early universe,''
Phys. Rept. \textbf{348}, 163-266 (2001)
[arXiv:astro-ph/0009061 [astro-ph]].

\bibitem{Ecker:2019xrw}
C.~Ecker, M.~J\"arvinen, G.~Nijs and W.~van der Schee,
``Gravitational waves from holographic neutron star mergers,''
Phys. Rev. D \textbf{101}, no.10, 103006 (2020)
[arXiv:1908.03213 [astro-ph.HE]].

\bibitem{Kharzeev:2012ph}
D.~E.~Kharzeev, K.~Landsteiner, A.~Schmitt and H.~U.~Yee,
``'Strongly interacting matter in magnetic fields': an overview,''
Lect. Notes Phys. \textbf{871}, 1-11 (2013)
[arXiv:1211.6245 [hep-ph]].

\bibitem{Miransky:2015ava}
V.~A.~Miransky and I.~A.~Shovkovy,
``Quantum field theory in a magnetic field: From quantum chromodynamics to graphene and Dirac semimetals,''
Phys. Rept. \textbf{576}, 1-209 (2015)
[arXiv:1503.00732 [hep-ph]].

\bibitem{Buividovich:2008kq}
  P.~V.~Buividovich and M.~I.~Polikarpov,
  ``Numerical study of entanglement entropy in SU(2) lattice gauge theory,''
  Nucl.\ Phys.\ B {\bf 802} (2008) 458
  [arXiv:0802.4247 [hep-lat]].

\bibitem{Buividovich:2008gq}
  P.~V.~Buividovich and M.~I.~Polikarpov,
  ``Entanglement entropy in gauge theories and the holographic principle for electric strings,''
  Phys.\ Lett.\ B {\bf 670} (2008) 141
  [arXiv:0806.3376 [hep-th]].

\bibitem{Itou:2015cyu}
  E.~Itou, K.~Nagata, Y.~Nakagawa, A.~Nakamura and V.~I.~Zakharov,
  ``Entanglement in Four-Dimensional SU(3) Gauge Theory,''
  PTEP {\bf 2016} (2016) no.6,  061B01
  [arXiv:1512.01334 [hep-th]].

\bibitem{Rabenstein:2018bri}
A.~Rabenstein, N.~Bodendorfer, P.~Buividovich and A.~Schäfer,
``Lattice study of Rényi entanglement entropy in $SU(N_c)$ lattice Yang-Mills theory with $N_c = 2, 3, 4$,''
Phys. Rev. D \textbf{100} (2019) no.3, 034504
[arXiv:1812.04279 [hep-lat]].

\bibitem{Dudal:2016joz}
  D.~Dudal and S.~Mahapatra,
  ``Confining gauge theories and holographic entanglement entropy with a magnetic field,''
  JHEP {\bf 1704}, 031 (2017)
  [arXiv:1612.06248 [hep-th]].

\bibitem{Dudal:2018ztm}
  D.~Dudal and S.~Mahapatra,
  ``Interplay between the holographic QCD phase diagram and entanglement entropy,''
  JHEP {\bf 1807}, 120 (2018)
  [arXiv:1805.02938 [hep-th]].

\bibitem{Mahapatra:2019uql}
S.~Mahapatra,
``Interplay between the holographic QCD phase diagram and mutual \& $n$-partite information,''
JHEP \textbf{04} (2019), 137
[arXiv:1903.05927 [hep-th]].

\bibitem{Kola1403}
 U.~Kol, C.~Nunez, D.~Schofield, J.~Sonnenschein and M.~Warschawski,
  ``Confinement, Phase Transitions and non-Locality in the Entanglement Entropy,''
  JHEP {\bf 1406} (2014) 005
  [arXiv:1403.2721 [hep-th]].

  \bibitem{Ben-Ami:2014gsa}
O.~Ben-Ami, D.~Carmi and J.~Sonnenschein,
``Holographic Entanglement Entropy of Multiple Strips,''
JHEP \textbf{11} (2014), 144
[arXiv:1409.6305 [hep-th]].

\bibitem{Fujita0806}
M.~Fujita, T.~Nishioka and T.~Takayanagi,
``Geometric Entropy and Hagedorn/Deconfinement Transition,''
  JHEP {\bf 0809} (2008) 016
  [arXiv:0806.3118 [hep-th]].

\bibitem{Lewkowycz}
  A.~Lewkowycz,
  ``Holographic Entanglement Entropy and Confinement,''
  JHEP {\bf 1205} (2012) 032
  [arXiv:1204.0588 [hep-th]].

  \bibitem{Georgiou:2015pia}
G.~Georgiou and D.~Zoakos,
``Entanglement entropy of the Klebanov-Strassler model with dynamical flavors,''
JHEP \textbf{07} (2015), 003
[arXiv:1505.01453 [hep-th]].

\bibitem{Kim}
N.~Kim,
  ``Holographic entanglement entropy of confining gauge theories with flavor,''
  Phys.\ Lett.\ B {\bf 720} (2013) 232.

\bibitem{Ghodrati}
M.~Ghodrati,
  ``Schwinger Effect and Entanglement Entropy in Confining Geometries,''
  Phys.\ Rev.\ D {\bf 92} (2015) no.6,  065015
  [arXiv:1506.08557 [hep-th]].

\bibitem{Ali-Akbari:2017vtb}
  M.~Ali-Akbari and M.~Lezgi,
  ``Holographic QCD, entanglement entropy, and critical temperature,''
  Phys.\ Rev.\ D {\bf 96}, no. 8, 086014 (2017)
  [arXiv:1706.04335 [hep-th]].

\bibitem{Knaute:2017lll}
  J.~Knaute and B.~K\"ampfer,
  ``Holographic Entanglement Entropy in the QCD Phase Diagram with a Critical Point,''
  Phys.\ Rev.\ D {\bf 96}, no. 10, 106003 (2017)
  [arXiv:1706.02647 [hep-ph]].

\bibitem{Anber:2018ohz}
  M.~M.~Anber and B.~J.~Kolligs,
  ``Entanglement entropy, dualities, and deconfinement in gauge theories,''
  arXiv:1804.01956 [hep-th].

  \bibitem{Arefeva:2020uec}
I.~Y.~Aref'eva, A.~Patrushev and P.~Slepov,
``Holographic entanglement entropy in anisotropic background with confinement-deconfinement phase transition,''
JHEP \textbf{07} (2020), 043
[arXiv:2003.05847 [hep-th]].

\bibitem{Slepov:2019guc}
P.~Slepov,
``Entanglement entropy in strongly correlated systems with confinement/deconfinement phase transition and anisotropy,''
EPJ Web Conf. \textbf{222} (2019), 03024.

\bibitem{Liu:2019npm}
P.~Liu, C.~Niu and J.~P.~Wu,
``The Effect of Anisotropy on Holographic Entanglement Entropy and Mutual Information,''
Phys. Lett. B \textbf{796} (2019), 155-161
[arXiv:1905.06808 [hep-th]].

  \bibitem{Fujita:2020qvp}
M.~Fujita, S.~He and Y.~Sun,
``Thermodynamical property of entanglement entropy and deconfinement phase transition,''
[arXiv:2005.01048 [hep-th]].

\bibitem{Fu:2020oep}
G.~Fu, P.~Liu, H.~Gong, X.~M.~Kuang and J.~P.~Wu,
``Holographic informational properties for a specific Einstein-Maxwell-dilaton gravity theory,''
Phys. Rev. D \textbf{104}, no.2, 026016 (2021)
[arXiv:2007.06001 [hep-th]].

\bibitem{Jokela:2020wgs}
N.~Jokela and J.~G.~Subils,
``Is entanglement a probe of confinement?,''
JHEP \textbf{02}, 147 (2021)
[arXiv:2010.09392 [hep-th]].

\bibitem{DiNunno:2021eyf}
B.~S.~DiNunno, N.~Jokela, J.~F.~Pedraza and A.~P\"onni,
``Quantum information probes of charge fractionalization in large-$N$ gauge theories,''
JHEP \textbf{05}, 149 (2021)
[arXiv:2101.11636 [hep-th]].

\bibitem{Ghodrati:2021ozc}
M.~Ghodrati,
``Correlations of mixed systems in confining backgrounds,''
Eur. Phys. J. C \textbf{82}, no.6, 531 (2022)
[arXiv:2110.12970 [hep-th]].

\bibitem{Yadav:2021hmy}
G.~Yadav, V.~Yadav and A.~Misra,
``$\mathcal{M}$cTEQ ($\mathcal{M}$ chiral perturbation theory-compatible deconfinement Temperature and Entanglement Entropy up to terms Quartic in curvature) and FM (Flavor Memory),''
JHEP \textbf{10}, 220 (2021)
[arXiv:2108.05372 [hep-th]].

\bibitem{Chu:2019uoh}
C.~S.~Chu and D.~Giataganas,
``$c$-Theorem for Anisotropic RG Flows from Holographic Entanglement Entropy,''
Phys. Rev. D \textbf{101}, no.4, 046007 (2020)
[arXiv:1906.09620 [hep-th]].

\bibitem{Cartwright:2021hpv}
C.~Cartwright and M.~Kaminski,
``Inverted c-functions in thermal states,''
JHEP \textbf{01}, 161 (2022)
[arXiv:2107.12409 [hep-th]].

\bibitem{Fromm:2011qi}
M.~Fromm, J.~Langelage, S.~Lottini and O.~Philipsen,
``The QCD deconfinement transition for heavy quarks and all baryon chemical potentials,''
JHEP \textbf{01}, 042 (2012)
[arXiv:1111.4953 [hep-lat]].

\bibitem{Dudal:2017max}
D.~Dudal and S.~Mahapatra,
``Thermal entropy of a quark-antiquark pair above and below deconfinement from a dynamical holographic QCD model,''
Phys. Rev. D \textbf{96}, no.12, 126010 (2017)
[arXiv:1708.06995 [hep-th]].

\bibitem{Breitenlohner:1982bm}
P.~Breitenlohner and D.~Z.~Freedman,
``Positive Energy in anti-De Sitter Backgrounds and Gauged Extended Supergravity,''
Phys. Lett. B \textbf{115}, 197-201 (1982)

\bibitem{Gubser:2000nd}
S.~S.~Gubser,
``Curvature singularities: The Good, the bad, and the naked,''
Adv. Theor. Math. Phys. \textbf{4}, 679-745 (2000)
[arXiv:hep-th/0002160 [hep-th]].

\bibitem{Calabrese:2009qy}
P.~Calabrese and J.~Cardy,
``Entanglement entropy and conformal field theory,''
J. Phys. A \textbf{42}, 504005 (2009)
[arXiv:0905.4013 [cond-mat.stat-mech]].

\bibitem{Hayden:2011ag}
P.~Hayden, M.~Headrick and A.~Maloney,
``Holographic Mutual Information is Monogamous,''
Phys. Rev. D \textbf{87}, no.4, 046003 (2013)
[arXiv:1107.2940 [hep-th]].

\bibitem{Headrick:2010zt}
M.~Headrick,
``Entanglement Renyi entropies in holographic theories,''
Phys. Rev. D \textbf{82}, 126010 (2010)
[arXiv:1006.0047 [hep-th]].

\bibitem{Fischler:2012uv}
W.~Fischler, A.~Kundu and S.~Kundu,
``Holographic Mutual Information at Finite Temperature,''
Phys. Rev. D \textbf{87}, no.12, 126012 (2013)
[arXiv:1212.4764 [hep-th]].

\bibitem{Kundu:2016dyk}
S.~Kundu and J.~F.~Pedraza,
``Aspects of Holographic Entanglement at Finite Temperature and Chemical Potential,''
JHEP \textbf{08}, 177 (2016)
[arXiv:1602.07353 [hep-th]].

\bibitem{Casini:2015woa}
H.~Casini, M.~Huerta, R.~C.~Myers and A.~Yale,
``Mutual information and the F-theorem,''
JHEP \textbf{10}, 003 (2015)
[arXiv:1506.06195 [hep-th]].

\bibitem{Balasubramanian:2018qqx}
V.~Balasubramanian, N.~Jokela, A.~P\"onni and A.~V.~Ramallo,
``Information flows in strongly coupled ABJM theory,''
JHEP \textbf{01}, 232 (2019)
[arXiv:1811.09500 [hep-th]].

\bibitem{Cardy:2013nua}
J.~Cardy,
``Some results on the mutual information of disjoint regions in higher dimensions,''
J. Phys. A \textbf{46}, 285402 (2013)
[arXiv:1304.7985 [hep-th]].

\bibitem{Larkoski:2014pca}
A.~J.~Larkoski, J.~Thaler and W.~J.~Waalewijn,
``Gaining (Mutual) Information about Quark/Gluon Discrimination,''
JHEP \textbf{11}, 129 (2014)
[arXiv:1408.3122 [hep-ph]].

\bibitem{Agon:2022efa}
C.~A.~Ag\'on, P.~Bueno, O.~Lasso Andino and A.~Vilar L\'opez,
``Aspects of N-partite information in conformal field theories,''
[arXiv:2209.14311 [hep-th]].

\bibitem{Calabrese:2012ew}
P.~Calabrese, J.~Cardy and E.~Tonni,
``Entanglement negativity in quantum field theory,''
Phys. Rev. Lett. \textbf{109} (2012) 130502
[arXiv:1206.3092 [cond-mat.stat-mech]].

\bibitem{Calabrese:2012nk}
P.~Calabrese, J.~Cardy and E.~Tonni,
``Entanglement negativity in extended systems: A field theoretical approach,''
J. Stat. Mech. \textbf{1302} (2013) P02008
[arXiv:1210.5359 [cond-mat.stat-mech]].

\bibitem{Alba}
V.~Alba, ``Entanglement negativity and conformal field theory: a Monte Carlo study,'' Journal of Statistical Mechanics: Theory and Experiment 2013 (05) P05013 [arXiv:1302.1110 [cond-mat.stat-mech]].

\bibitem{Calabrese:2013mi}
P.~Calabrese, L.~Tagliacozzo and E.~Tonni,
``Entanglement negativity in the critical Ising chain,''
J. Stat. Mech. \textbf{1305} (2013), P05002
[arXiv:1302.1113 [cond-mat.stat-mech]].

\bibitem{Ruggiero:2016aqr}
P.~Ruggiero, V.~Alba and P.~Calabrese,
``Entanglement negativity in random spin chains,''
Phys. Rev. B \textbf{94} (2016) no.3, 035152
[arXiv:1605.00674 [cond-mat.str-el]].

\bibitem{Hoogeveen:2014bqa}
M.~Hoogeveen and B.~Doyon,
``Entanglement negativity and entropy in non-equilibrium conformal field theory,''
Nucl. Phys. B \textbf{898} (2015) 78-112
[arXiv:1412.7568 [cond-mat.stat-mech]].

\bibitem{Blondeau-Fournier:2015yoa}
O.~Blondeau-Fournier, O.~A.~Castro-Alvaredo and B.~Doyon,
``Universal scaling of the logarithmic negativity in massive quantum field theory,''
J. Phys. A \textbf{49} (2016) no.12, 125401
[arXiv:1508.04026 [hep-th]].

\bibitem{Castelnovo}
C.~Castelnovo, ``Negativity and topological order in the toric code,'' Phys.~Rev.~A \textbf{88} (2013) 042319 [arXiv:1306.4990 [cond-mat.stat-mech]].

\bibitem{Lee}
Y.~A.~Lee and G.~Vidal, ``Entanglement negativity and topological order,'' Phys. Rev. A \textbf{88} (2013) 042318 [arXiv:1306.5711 [cond-mat.stat-mech]].

\bibitem{Eisler}
V.~Eisler and Z.~Zimboras, ``Entanglement negativity in two-dimensional free lattice models,'' New Journal of Physics \textbf{16} (2015) 123020 [arXiv:1511.08819 [cond-mat.stat-mech]].

\bibitem{Wen:2015qwa}
X.~Wen, P.~Y.~Chang and S.~Ryu,
``Entanglement negativity after a local quantum quench in conformal field theories,''
Phys. Rev. B \textbf{92} (2015) no.7, 075109
[arXiv:1501.00568 [cond-mat.stat-mech]].

\bibitem{Wen:2016bla}
X.~Wen, P.~Y.~Chang and S.~Ryu,
``Topological entanglement negativity in Chern-Simons theories,''
JHEP \textbf{09} (2016), 012
[arXiv:1606.04118 [cond-mat.str-el]].

\bibitem{Rangamani:2014ywa}
M.~Rangamani and M.~Rota,
``Comments on Entanglement Negativity in Holographic Field Theories,''
JHEP \textbf{10} (2014), 060
[arXiv:1406.6989 [hep-th]].





\end{thebibliography}
\end{document}